\documentclass[lettersize,journal,onecolumn,comsoc,10pt]{IEEEtran}
\IEEEoverridecommandlockouts    

\usepackage{graphicx} % Required for inserting images
\usepackage{multirow}
\usepackage{makecell}
\usepackage{subcaption}
\usepackage{float}
\usepackage{bbm}
\newcommand{\diff}{\mathrm{d}}
\usepackage{cite}

\usepackage{amsthm}
\theoremstyle{definition}
\newtheorem{assumption}{\normalfont\bfseries Assumption}
\newtheorem{theorem}{\normalfont\bfseries Theorem}
\usepackage{color}
\newtheorem{definition}{\normalfont\bfseries Definition}

\usepackage[section]{placeins}

\begin{document}

\title{Physics-Informed Neural Networks for Nonlocal Flow Modeling of Connected Automated Vehicles}

\author{
Chenguang Zhao,
Huan Yu*\thanks{* Corresponding author. Email: huanyu@ust.hk} \\
\IEEEauthorblockA{Thrust of Intelligent Transportation, The Hong Kong University of Science and Technology (Guangzhou), Nansha, Guangzhou, 511400, Guangdong, China}
}

\maketitle

\begin{abstract}
Connected automated vehicles (CAVs) cruising control strategies have been extensively studied at the microscopic level. CAV controllers sense and react to traffic both upstream and downstream, yet most macroscopic models still assume locality, where the desired speed only depends on local density. The nonlocal macroscopic traffic flow models that explicitly capture the ``look ahead'' and ``look behind'' nonlocal CAV dynamics remain underexplored. 
In this paper, we propose a Physics-informed Neural Network framework to directly learn a macroscopic non-local flow model from a generic  looking-ahead looking-behind vehicle motion model, which bridges the micro-macro modeling gap. We reconstruct macroscopic traffic states from synthetic CAV trajectories generated by the proposed microscopic control designs, and then learn a  non-local traffic flow model that embeds a non-local conservation law to capture the resulting look-ahead/look-behind dynamics. To analyze how CAV control parameters affect nonlocal traffic flow, we conduct high-fidelity driving simulator experiments to collect human drivers' trajectory data with varying downstream and upstream visibility, which serves as a baseline for tuning CAV control gains.  Our analysis validates that the learned non-local flow model predicts CAV traffic dynamics more accurately than local models, and the fundamental diagram exhibits far less scatter in the speed–density relation. We further show that the looking-ahead/looking-behind control gains mainly reshape the non-local kernels, while the macroscopic speed and non-local density relation mainly depends on the desired speed function  choice of the CAV controller.  Our results 
 provide a systematic approach for learning non-local macroscopic traffic-flow models directly from generic CAV control designs. 
\end{abstract}

\section{Introduction}

The deployment of connected and automated vehicles (CAVs) is expected to enhance traffic systems in many aspects, such as reducing speed perturbation and alleviating traffic congestion~\cite{wang2021leading,stern2018dissipation},  and avoiding collisions and improving safety~\cite{zhao2023safety,alan2023integrating}. Both theoretical analyses and field experiments have investigated the trade-offs among these performance improvements, which are brought by automation and connectivity~\cite{li2022trade,zhao2024leveraging,barth2024co,sun2020optimal}.  Automation enables CAVs to achieve shorter reaction times, more accurate perception of the traffic environment, and more efficient deceleration strategies. These benefits can be further amplified through connectivity, which allows CAVs to share information and coordinate with surrounding vehicles and infrastructure.  With vehicle-to-vehicle communication, the CAV can measure not only the leader vehicle's speed and position, but also downstream and upstream traffic~\cite{li2023information}.  For example, with a ``looking-ahead'' strategy, CAVs can detect downstream traffic congestion and decelerate in advance to reduce future jerk~\cite{orosz2016connected,zhao2024leveraging}. Some research has also proposed the ``looking-behind'' strategy, where the CAV adjusts its speed based on the states of upstream vehicles to lead the following vehicles and alleviate upstream traffic congestion~\cite{wang2021leading,zhao2024leveraging}. In our previous work, we have shown that via looking-behind, the CAV achieves safe stabilization of traffic flow~\cite{zhao2023safety}.

Despite these benefits brought by looking-ahead and looking-behind behaviors of CAVs, most existing research has mainly focused on evaluating CAV controller performance at the microscopic level, such as stability and safety~\cite{li2022trade,barth2024co,zhao2023safety} — as a result of optimizing vehicle motions.  To fully understand and evaluate the impact of CAV design on traffic systems, it is crucial to model and analyze macroscopic traffic dynamics. Some studies have analyzed how CAVs will affect macroscopic characteristics, such as capacity, flow stability, and congestion propagation~\cite{jiang2024dynamic,yao2022modeling,ma2021analysis}. However, these analyses are typically based on local flow models, where the desired speed depends solely on local traffic density.  With connectivity, the CAV reacts to downstream and upstream traffic dynamics, as shown in Fig.~\ref{fig:intro}. A natural expectation is that the desired speed is also affected by surrounding density. It remains underexplored how the CAV, especially its looking-ahead/looking-behind behaviors designed from the microscopic motion level, will affect the macroscopic-level traffic flow dynamics.

\subsection{Look-ahead and look-behind cruising control strategies}

In the most basic adaptive cruise control (ACC), the vehicle is controlled using three components: its own speed, the gap to the immediate leader, and the leader’s speed.  The ACC vehicle decelerates mainly under three scenarios: when the ego vehicle speed increases, when the gap decreases, and when the leader vehicle decelerates. By tuning how ACC reacts to these scenarios, the ACC helps to stabilize traffic, i.e., its speed has a smaller perturbation. Research has shown that connectivity helps automated vehicles to obtain upstream and downstream traffic information, which can be integrated into the autonomous driving strategy to further enhance traffic system performance~\cite{li2023information,orosz2016connected,wang2021leading}. 
In~\cite{orosz2016connected}, connected cruise control is designed, which assumes that the CAV measures the speed and gap information of multiple downstream leading vehicles and responds to their motions.  For example, when congestion occurs downstream, the CAV will begin to decelerate as soon as a more distant leading vehicle slows down, even if the immediately preceding vehicle maintains its speed. This allows the CAV to decelerate proactively, reducing the need for abrupt braking when passing through congested traffic.  Theoretical analysis and field experiments have both validated that downstream traffic information helps to stabilize traffic~\cite{li2023information,orosz2016connected,jin2018experimental}.

In addition to looking ahead to respond to downstream traffic,  the CAV can also monitor upstream following vehicles and react correspondingly, a cruising strategy known as leading cruise control~\cite{wang2021leading}. The main idea is that when the CAV decelerates, it also considers how its deceleration affects the vehicles behind, aiming to avoid inducing abrupt deceleration in followers and to achieve ``head-to-tail" string stability.  The stability ensures that speed perturbations attenuate along the vehicle chain—that is, the disturbance experienced by the last connected following vehicle is smaller than that of the first leading vehicle. The effectiveness of smoothing upstream traffic through this looking-behind design has been demonstrated in related research~\cite{wang2021leading,zhao2024leveraging}.

\subsection{Look-ahead and look-behind behaviors of human drivers }

Compared with adaptive cruising control, the looking-ahead/looking-behind controller design involves more control parameters, such as the looking-ahead and looking-behind gains. And it is challenging to provide a comprehensive analysis of all parameters. In this paper, we conduct experiments to collect human driving behavior during looking-ahead and looking-behind scenarios, and use it as a baseline to analyze the looking-ahead/looking-behind CAV controllers. This approach is motivated by human-inspired controller design~\cite{nie2024human,zhang2022human,hang2020human,liu2023safe}. Human driving strategies often serve as templates for designing and optimizing autonomous control algorithms~\cite{plebe2024human}. If autonomous vehicles have human-like behaviors, it will be easier for human drivers to predict CAV's motion and interact with them. On the other hand, an unnatural autonomous driving algorithm that lacks interpretability for human drivers and passengers may hinder the deployment of CAVs. 

In car-following cruising, human drivers also have looking-ahead and looking-behind behaviors. The looking-ahead anticipation behavior means that the human driver observes the downstream leading vehicle's motion and adjusts the driving strategy correspondingly. For example, when the driver observes congestion, he/she may decelerate in advance to avoid abrupt braking. To describe such behaviors, multi-vehicle anticipation car-following models have been proposed and calibrated from trajectory data~\cite{nirmale2024multi,nie2024human}.  Besides looking ahead, human drivers also look-behind and consider the upstream following vehicles' motion when making driving decisions. This is particularly evident in the case where the following vehicle accelerates. For example, when the following vehicle suddenly accelerates, the driver may also accelerate to avoid collisions. Car-following models that describe such looking-behind behaviors have been proposed and calibrated using NGSIM trajectory data~\cite{li2024modular}. But in these trajectory data-sets, the visibility of leading and following vehicles is unknown. For example, there may be trucks, foggy weather, or a long distance, and it is difficult to identify whether the human drivers can observe surrounding traffic. Therefore, it is insufficient to conclude from these analyses whether human drivers look-ahead/look-behind when they drive.  In this paper, we design simulator experiments to collect and analyze human drivers' reactions under three traffic scenarios: 1) when they can only observe the leader vehicle, 2) when they can look-ahead and observe more leading vehicles, and 3) when the following vehicle accelerates aggressively to nudge the driver. The collected human driving behavior is then used as a baseline for analyzing CAV controller design and analysis.

\begin{figure}[!t]
    \centering
    \includegraphics[width=0.8\linewidth]{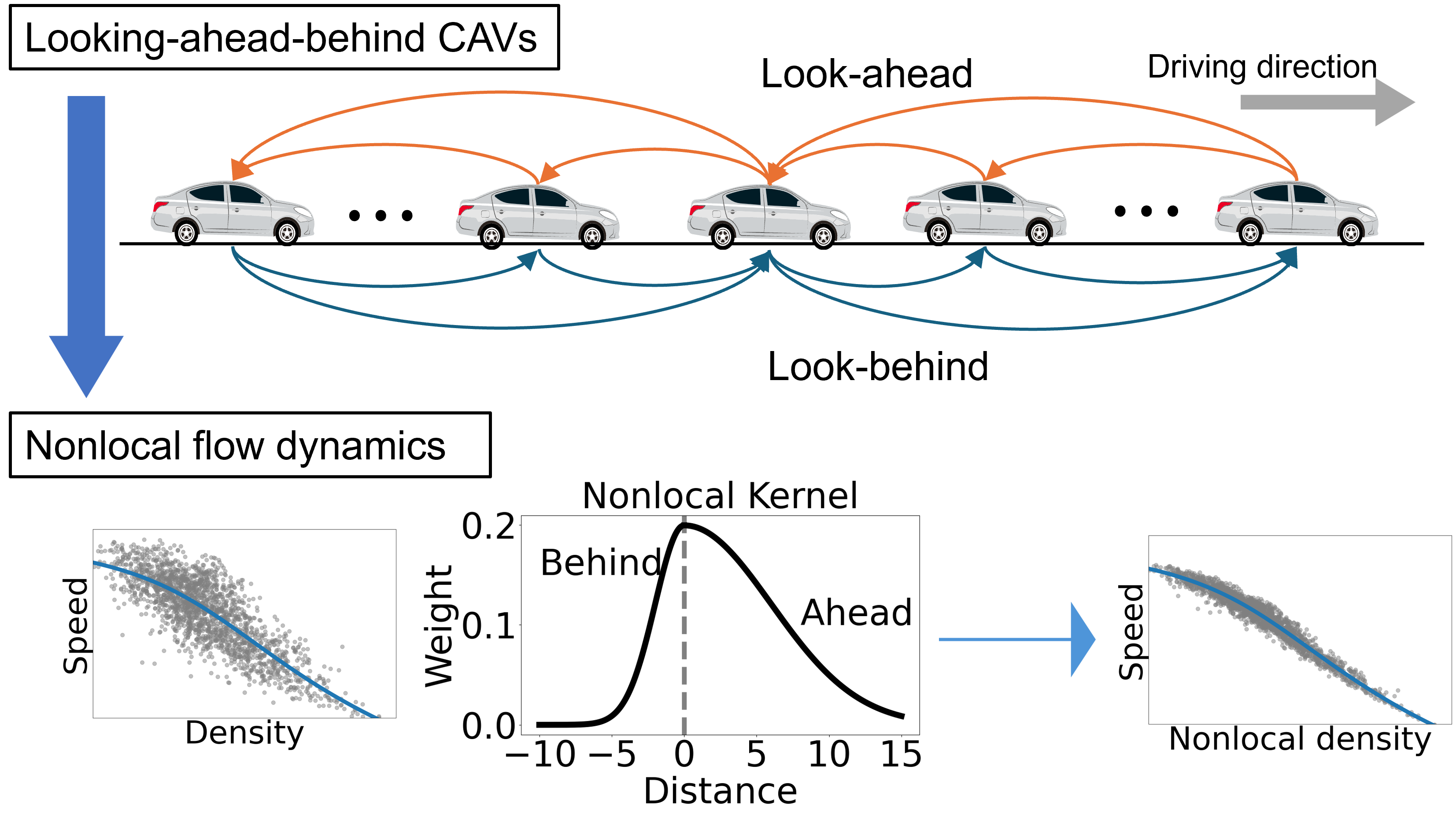}
    \caption{The cruising CAVs “look ahead” at leading vehicles (red arrows) and “look behind” at following vehicles (blue arrows), introducing nonlocal dynamics into the traffic flow modeling. The classical speed–local density fundamental diagram, showing considerable scatter. The nonlocal dynamics are modeled by nonlocal kernel, and the speed versus nonlocal density (i.e. the weighted average density using nonlocal kernel) exhibits a much convergent relationship.}
    \label{fig:intro}
\end{figure}

\subsection{Traffic flow modeling for CAVs}

Traffic flow models describe aggregated traffic dynamics, including density, flow, and average speed, and are used to capture macroscopic characteristics such as capacity and congested wave propagation. The basic traffic flow model is the conservation equation, which is derived through the conservation law of mass in human-driven vehicular traffic. By further assuming that the speed is determined by the density via the fundamental diagram (FD), we have the first-order Lighthill–Whitham–Richards (LWR) model~\cite{lighthill1955kinematic,richards1956shock}. The LWR model has been adopted for various traffic management applications, such as moving bottlenecks~\cite{yu2020bilateral}  and varying speed limit~\cite{han2021linear}. 
By further assuming that traffic speed has its own dynamics and evolves toward the desired speed determined by the fundamental diagram from density, the second order Aw–Rascle–Zhang (ARZ) model is proposed~\cite{aw2000resurrection,zhang2002non}, which has a conservation PDE and an advection equation to describe the dynamics of traffic speed.

With the deployment of CAVs, future traffic systems can be viewed as artificial fluids, where the macroscopic traffic dynamics are governed by the driving behaviors of CAV control designs. A major challenge lies in deriving macroscopic traffic flow models from various microscopic CAV motion control designs. A number of theoretical works treat the traffic system as a multi-agent system and derive the aggregated dynamics as the number of agents approaches infinity, including representative approaches such as continuum limit~\cite{chiarello2021statistical,tordeux2018traffic,ancona2025continuum} and mean-field theory~\cite{huang2020game,mo2024game}. While these methods offer theoretical rigor, they often assume certain forms of vehicle dynamics. For example, only relatively simple microscopic models have been analyzed using continuum limit theory, such as  a simplified follow-the-leader model where the  vehicle instantaneously adopts a desired speed determined by the gap. Extending these derivations to encompass more complex vehicle dynamics remains an open question.

Besides the analytical complexity in mapping from the micro dynamics of CAVs to a macroscopic model, direct modifications of classical models such as the LWR or ARZ  are sometimes insufficient for accurately capturing connected traffic behavior~\cite{do2025nonlocal}.  In the LWR model, the traffic speed instantaneously adapts to the desired speed decided by the fundamental diagram from local density. In the ARZ model, while a kinematic equation is incorporated to describe the evolution of speed towards the desired speed, the fundamental diagram still describes the relationship between the desired speed and local density. However, traffic systems also have non-local dynamics, where the speed depends not only on local density, but also the downstream or upstream density~\cite{zhao2024learning,do2025nonlocal,kachroo2023nonlocal}. For example, in our previous work~\cite{zhao2024learning}, we have shown that human drivers exhibit looking-ahead behavior in a ring-road experiment, and a flow model incorporating downstream density into the desired speed yields a more accurate representation of traffic dynamics compared to the local model.
Other non-local models have also been formulated and validated on various datasets, such as  PeMS \cite{do2025nonlocal} and  NGSIM~\cite{kachroo2023nonlocal}. 
This non-local phenomenon is expected to  become more evident in connected traffic, especially when the vehicles are controlled using feedback from downstream leading vehicles or upstream following vehicles as shown in Fig.~\ref{fig:intro}. When CAVs look ahead or look behind, their desired speed is influenced by the behavior of nearby vehicles in both directions. This suggests that the macroscopic fundamental diagram should account for a relationship between speed and non-local density, rather than relying solely on local measurements.

To describe the non-local effects at the macroscopic level, the non-local LWR model has been proposed, where the desired speed is determined by a non-local fundamental diagram that describes the relationship between traffic speed and non-local downstream/upstream density. Related research has discussed several aspects of the non-local model, such as  well-posedness~\cite{blandin2016well}, stability~\cite{huang2022stability}, controllability~\cite{bayen2021boundary}, the vanishing nonlocality limit \cite{colombo2019singular,keimer2019approximation}, numerical schemes~\cite{goatin2016well}, and calibration using real data~\cite{zhao2024learning}.  A few studies have discussed the continuum limit theory to derive non-local models from vehicle dynamics~\cite{holden2024continuum,chiarello2020micro,helbing2009derivation,karafyllis2022constructing}. These approaches typically rely on simplified forms of vehicle behaviors. As a result, a general and systematic modeling framework for capturing non-local traffic flow dynamics—especially in the context of connected traffic—remains an open research problem.

The above modeling methods are physics-based and use vehicle dynamics equations to derive flow dynamics formulations.  Another category of modeling approaches is data-driven methods, which directly learn traffic dynamics from data. Representative data-driven methods include statistical approaches, such as Gaussian processes~\cite{liu2023gaussian}, and machine learning approaches, such as k-Nearest Neighbors, Random Forests, and neural networks~\cite{parsa2021data,rahman2023data}. However, these methods operate as black-box models, as they directly provide traffic states without offering an explicit formulation of traffic wave propagation. As a result, they offer little insight into how CAV controller designs affect flow dynamics. Another limitation is that they require a large amount of mixed-traffic data to analyze the effect of CAVs, which is still unavailable in current real-world traffic.

To reduce the dependence on data and incorporate  prior knowledge of traffic flow dynamics, physics-data hybrid modeling has been proposed~\cite{bezgin2021data,raissi2019physics}.  There are two main approaches to combining physics knowledge with data-driven methods. The first approach treats physical dynamics as hard constraints and designs a neural network architecture that automatically satisfies this prior knowledge~\cite{bezgin2021data}. The second approach, referred to as the physics-informed neural network (PINN), treats physical knowledge as a soft constraint and includes it in the loss function~\cite{raissi2019physics}. For traffic flow modeling, PINN is more appropriate, as physical knowledge provides only an approximate description of traffic dynamics. PINN-related methods have been used with local flow models~\cite{shi2021physics, zhao2023observer,wilkman2025online} and non-local models~\cite{huang2020game,zhao2024learning,singh2025non} as the physics knowledge. Specifically, in our previous work, we showed that PINN using non-local traffic flow as prior knowledge provides a more accurate description of traffic dynamics than local models. But these analyses are restricted to human driver behavior. In this paper, we propose a modeling framework to analyze how CAV controller design affects non-local flow dynamics.

\subsection{Contribution}

To summarize, a key research gap lies in the absence of macroscopic non-local traffic flow models that explicitly capture the designed looking-ahead and looking-behind behaviors of CAVs. To address this micro–macro gap, we utilize PINN and propose a data–physics  hybrid modeling framework to derive a macroscopic non-local flow model from general microscopic vehicle motion dynamics.  Given a vehicle motion model, we first run simulations to generate trajectories, which are then used to reconstruct macroscopic traffic states. We use neural networks to learn a data-physics hybrid model, i.e., a mapping from spatial-temporal points to macroscopic states. The neural networks are trained to minimize two losses, the data loss that reflects the difference between the estimated state and reconstructed state, and the physics loss that regulates the learned flow dynamics using a non-local flow model.  Based on the proposed modeling method, we further analyze how microscopic design parameters of  CAV controllers will affect macroscopic traffic flow dynamics. 
We first conduct simulator experiments to collect human drivers' data when they can observe downstream/upstream traffic, and calibrate a microscopic model from collected trajectory data. The calibrated model serves as a baseline to tune and analyze how CAV controllers affect flow models. The objective is to develop CAV cruising control that optimizes traffic flow while 
retaining human-like driving characteristics, making its motions intuitive for human drivers and fostering seamless HV-CAV interaction in mixed traffic.

{The main contribution of this paper is two-fold. 
From a methodological perspective, we bridge the micro–macro gap by proposing a hybrid learning framework to formulate a non-local traffic flow model for looking-ahead looking-behind controllers and to analyze how these controllers affect flow dynamics.  Unlike previous models that rely on fixed-form continuum assumptions, our framework learns a macroscopic non-local traffic flow model directly from  microscopic vehicle dynamics with look-ahead and look-behind behaviors. This enables systematic analysis of how CAV control strategies influence macroscopic traffic flow patterns.
From a data perspective,  we design driving-simulator experiments that record human-driver trajectories under varying downstream and upstream visibility, yielding a dataset for calibrating baseline microscopic models and tuning CAV controllers. This enriches traffic data with looking-ahead/looking-behind behaviors, and facilitates development of interpretable CAV cruising control.}

This paper is organized as follows. We introduce a general looking-ahead and looking-behind CAV controller in Section~\ref{sec:controller}. In Section~\ref{sec:macro}, we introduce the proposed modeling method to get a non-local flow model from the general CAV controller. We conduct simulator experiments and analyze flow models corresponding to  different vehicle dynamics in Section~\ref{sec:analysis}.

\section{CAV Cruising Control design}\label{sec:controller}

In this paper, we consider longitudinal motion and control CAVs traveling in a platoon as in Fig.~\ref{fig:intro}. We index the vehicle chain in a descending order with the driving direction, i.e., vehicle $i$ follows vehicle $i-1$ and leads vehicle $i+1$. We assume that the CAV looks-ahead of $n$ leading vehicles and looks-behind of $m$ following vehicles. For CAV $i$, its longitudinal dynamics are:
\begin{align}
    \dot{s}_i &= v_{i-1} - v_i, \label{eq:CAV s}\\
    \dot{v}_i &=  u_i, \label{eq:CAV v}
\end{align}
where $v_i$ represent its speed, $s_i$ represents is gap to the leader vehicle $i-1$, and $v_{i-1}$ is the leader vehicle speed. The CAV controller $u_i$ takes a feedback form:
\begin{align}\label{eq:controller}
    u_i = \underbrace{\alpha_0(V_{\mathrm{opt}}(s_i) -v_i)+\beta_0 (v_{i-1}-v_i)}_{\mathrm{ACC}} + \underbrace{\sum_{j=-1}^{-n} \alpha_{j} (V_{\mathrm{opt}}(s_{i+j}) - v_i) + \beta_j (v_{i+j-1} - v_i)}_{\mathrm{look-ahead}}  + \underbrace{\sum_{j=1}^{m} \alpha_{j} (V_{\mathrm{opt}}(s_{i+j}) - v_i) + \beta_j (v_{i+j} - v_i)}_{\mathrm{look-behind}}.
\end{align}

The controller contains three parts, ACC, look-ahead, and look-behind. 

In the ACC part, the CAV adjusts its speed based on three components: its own speed $v_0$, its gap with the leader $s_0$, and  the leader's speed $v_{-1}$. The CAV acceleration is a weighted sum of its reaction to gap and speed difference, with $\alpha_0>0$ and $\beta_0>0$ being the sensitivity parameters to  gap and speed difference respectively. The $V_{\mathrm{opt}}(s)$ is the optimal speed function, which specifies the desired speed dependent on the gap $s$. In this paper, we choose the piece-wise linear desired speed function:
\begin{align}\label{eq:micro Vopt}
	V_{\mathrm{opt}}(s) = \left\{ \begin{array}{ll}
		0, & s\le s_{\mathrm{st}} \\
		v_{\max} \frac{s- s_{\mathrm{st}}}{s_{\mathrm{go}} - s_{\mathrm{st}}}, & s_{\mathrm{st}} \le s \le s_{\mathrm{go}} \\
		v_{\max}, & s\ge s_{\mathrm{go}}
	\end{array} \right.,
\end{align}
with $v_{\max}$ being the free speed, $s_{\mathrm{rm}}$ being the stopping gap, and $s_{\mathrm{go}}$ being the free-flow gap.

In the looking-ahead part, the CAV also adjusts its speed to downstream traffic. Similar to the ACC part, the CAV's reaction to downstream traffic is also a weighted average of its reaction to leading vehicles' gap and speed. And $\alpha_j>0$ and $\beta_j>0$ with $j=-1,\cdots,-n$ denote the CAV's sensitivity parameter to leading vehicles' gap and speed.  With the looking-ahead terms, the CAV anticipates downstream traffic and takes actions in advance when traffic oscillates. For example, consider the scenario where the vehicle platoon drives at a constant speed and the first leading vehicle is traveling through a bottleneck. If the CAV is controlled via only ACC, the CAV begins to decelerate only when the leading vehicle $i-1$ begins to decelerate. But if the CAVs look-ahead, it will begin to decelerate when the first leading vehicle $i-n$ begins to decelerate. This brings smoother speed perturbations in the CAV motion, which implies more stable traffic.

In the looking-behind part, the CAV uses feedback from upstream following vehicles, with $\alpha_j$ and $\beta_j$ being CAV's sensitivity parameters to the gap and speed of the following vehicles. By adding these looking-behind terms, the CAV also considers how its deceleration process will affect upstream traffic. For example, considering the same scenario of traveling through a congestion, when the CAV begins to decelerate, the following vehicles are still at the original speed. Therefore $v_i<v_j$ ($i>j$), and the looking-behind term makes CAV's deceleration become smaller, which also makes the following vehicles have a smaller deceleration. By looking-behind, the CAV acts as a leader of following vehicles and helps to stabilize upstream traffic. 

The controller parameters, including the controller gain and parameters in the desired speed function, should be designed so that the system~\eqref{eq:CAV s}-\eqref{eq:CAV v} is stable. In vehicle platooning control, there are two types of commonly adopted stability, plant stability and string stability. 
\begin{definition}[System stability~\cite{zhao2024leveraging}]
    The system is plant stable if all vehicles can cruise at a given constant equilibrium speed $v^*>0$, i.e., $\lim_{t\to\infty} v_i(t) = v^*$ holds for all $i$. The system is string stable if the speed perturbations of the following vehicle is smaller than the leading vehicle, i.e., $\Vert v_i(t)\Vert_\infty \le \Vert v_i(t) \Vert_\infty$ with $\Vert \cdot \Vert_\infty$ being the $L_\infty$ norm of a function. 
\end{definition}

To analyze the constraints on the control parameters under which the system is stable, we use the transfer function $G(\mathrm{s})$, which is derived from the vehicle dynamics of~\eqref{eq:CAV s}-\eqref{eq:controller} in the Laplace domain $\mathrm{s}$. The property of the  transfer function depends on the controller parameters, and the system stability can be achieved by choosing control parameters that make the transfer function satisfy certain conditions. We refer readers to related study for more theoretical details on derivation and analysis of the transfer function~\cite{zhao2024learning}.
\begin{theorem}[System stability]
    The system is plant stable if the controller parameters, including the feedback gains $\alpha$, $\beta$, and parameters in the desired speed function, i.e., $s_{\mathrm{go}}$, $v_{\max}$, and  $s_{\mathrm{st}}$,  are chosen such that all solutions of $D(\mathrm{s}) = 0$ have negative real parts, where $D$ is the denominator of the transfer function $G$. The system is string stable if $ |G(\mathrm{j} \omega) | < 1 $  holds for all $\omega > 0$, where $\mathrm{j}^2 = -1$ is the imaginary unit.
\end{theorem}

\section{Non-local traffic flow modeling}\label{sec:macro}

In this section, we give the designed approach to get a non-local flow model from the microscopic vehicle dynamics. The designed method contains two steps: a) from micro model to macro data: we simulate the microscopic vehicle motion dynamics to generate trajectory data, from which macroscopic traffic states are reconstructed; and b) from macro data to macro model: using the reconstructed macroscopic states, we learn a model that best captures the underlying macroscopic dynamics.

\subsection{From micro model to macro data}

Given the microscopic model~\eqref{eq:controller}, we run simulations of $N$ vehicles on a ring road of length $L$. All vehicles adopt the given microscopic model dynamics. The vehicle length is set as $l=5$ m. Given the road length $L$  and the vehicle number $N$, the equilibrium gap of each vehicle is $ s^* = L/N - l$. The equilibrium speed is then decided from the $V_{\mathrm{opt}}$ as $v^* = V_{\mathrm{opt}}(s^*)$. We run simulations for a time period of $T$ and collect the position  $x_i(t)$ and speed $v_i(t)$ data of these $N$ vehicles.

We reconstruct the macroscopic state using the kernel density estimation. The reconstructed density, flow, and speed at position $x$ and time $t$ is:
\begin{align}
	\rho(x,t) &= \sum_{i=1}^{N} K(d(x-x_i(t)),h),\\
	q(x,t) &= \sum_{i=1}^N v_i(t) K(d(x-x_i(t)),h),\\
	v(x,t) & = q(x,t)/\rho(x,t),
\end{align}
where  $d(x,y)$ is the distance of two positions $x,y$ on the ring road, i.e., 
\begin{align}
	d(x,y) = \left\{ 
	\begin{array}{ll}
		y-x, & x<L/2, x\le y<x+L/2,\\
		L- y + x, & x<L/2, y\ge x+L/2, \\
		x-y, & x<L/2, y< x, \\
		y-x, & x\ge L/2, y\ge x,\\
		x-y, & x\ge L/2, x-L/2 \le y < x,\\
		L-x + y, & x\ge L/2, y < x-L/2,
	\end{array}
	\right. ,
\end{align}
and $K$ is a smooth kernel function. We use the Gaussian kernel as:
\begin{align}
	K(x,h) = \frac{1}{\sqrt{2\pi} h} \exp \left(-\frac{x^2}{2h^2}\right),
\end{align}
with $h>0$ being a smooth parameter.

\begin{figure}[!t]
    \centering
    \includegraphics[width=1\linewidth]{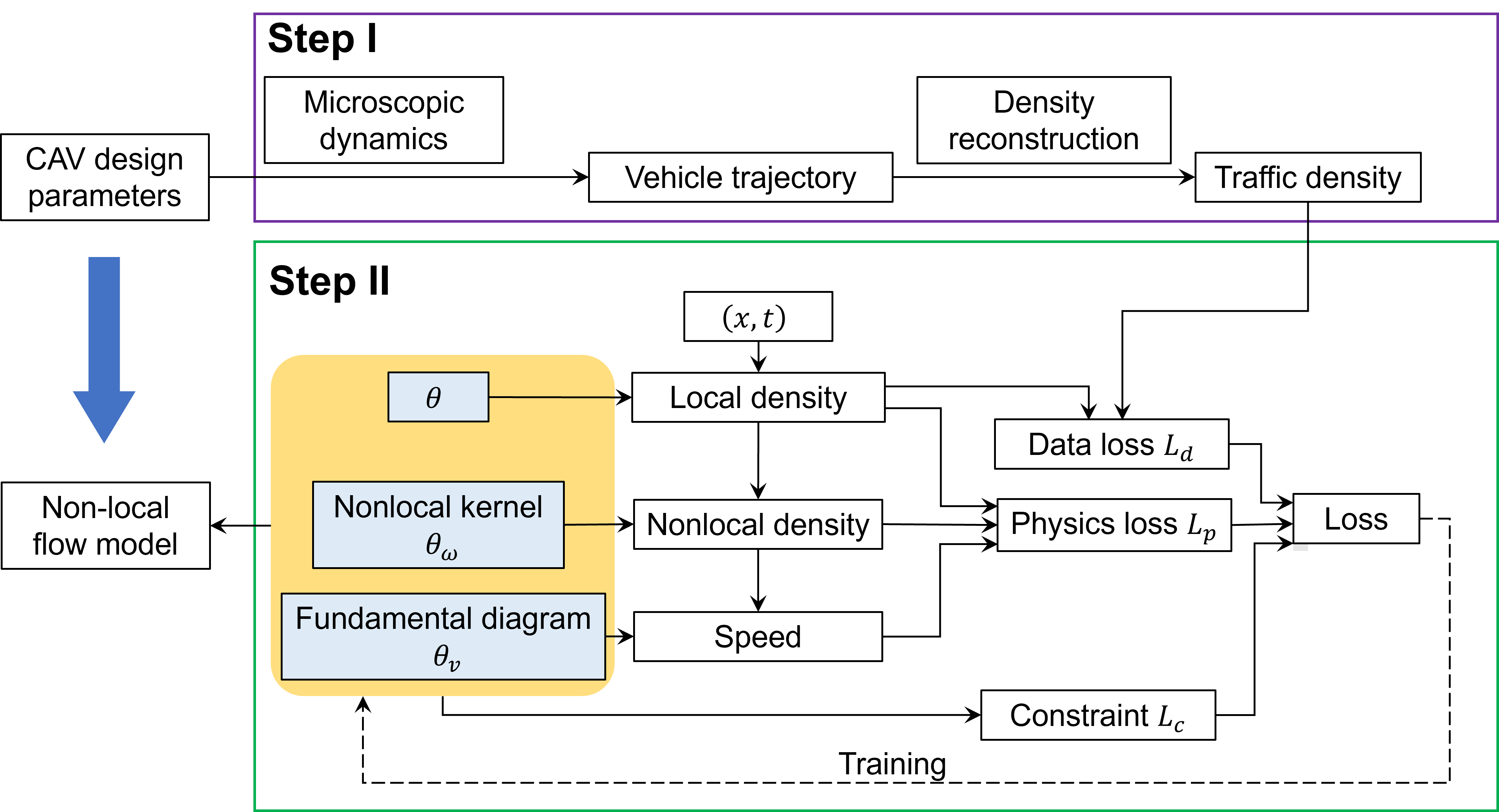}
    \caption{The designed modeling structure. We get a non-local macroscopic model from a given microscopic model. The proposed approach contains two steps: I) Generate vehicle trajectories from the given microscopic model and reconstruct macroscopic values; II) Lean a macroscopic model from the macroscopic data to minimize a weighted loss function of data error and physics regularization.  }
    \label{fig:NN}
\end{figure}

\subsection{From macro data to macro model}

In this paper, we bring the ideas of PINN and design a model-data hybrid modeling method. Fig.~\ref{fig:NN} gives the diagram  of the designed method. We aim to find a mapping from the spatial-temporal points $(x,t)$ that accurately estimates the macroscopic data and follows an underlying non-local flow dynamics. In the designed structure, there are three trainable parameters: $\theta$ to learn the mapping from spatial-temporal locations $(x,t)$ to density $\rho(x,t)$, $\theta_{\omega}$ to learn non-local weights in the non-local model, and $\theta_v$ to learn the non-local fundamental diagram. To optimize these trainable parameters, there are three types of loss functions: data-loss $L_{\mathrm{d}}$ to minimize the difference between the learned density and the reconstructed ground-truth density, the physics loss $L_{\mathrm{p}}$ to minimize the difference between the learned dynamics and a non-local flow model, and the constraint loss $L_{\mathrm{c}}$ to ensure that the learned non-local kernel and fundamental diagram satisfy well-posedness constraints. The overall loss function is a weighted sum of these three loss functions:
\begin{align}\label{eq:loss}
	L(\theta,\theta_v,\theta_{\omega}) = L_{\mathrm{d}}(\theta) +  L_{\mathrm{p}} (\theta,\theta_v,\theta_{\omega}) +L_{\mathrm{c}} (\theta_v,\theta_{\omega}).
\end{align}
We introduce the calculation of each loss function as follows.

\subsubsection{Data loss}
We run $N_s$ cases of microscopic simulations, each with different vehicle numbers. For each simulated density $\rho_k$, we use a neural network with trainable parameter $\theta_k$ to learn the traffic density at location $x$ and  time $t$ as $\hat{\rho}(x,t;\theta_k)$. The data loss is designed to minimize the difference between the learned density $\hat{\rho}(x,t;\theta_k)$ and the ground-truth density $\rho_k(x,t)$. 
We select some spatial-temporal points $\mathcal{D}$ and assume that the ground-truth density is available as the training data. The density values at other spatial-temporal points are considered as test data. The data loss is then calculated as
\begin{equation}\label{eq:loss data}
	L_{\mathrm{d}}(\theta) = c_d \sum_{k=1}^{N_s} \frac{1}{N_D} \sum_{(x,t)\in \mathcal{D}}\left( \rho_k(x,t) 
	- \hat{\rho}(x,t;\theta_k)\right)^2,
\end{equation}
where $N_D = |\mathcal{D}|$  represents the total number of available ground-truth density data, and $c_d>0$ is a weight coefficient.

\subsubsection{Physics loss}
The physics-loss is designed as a regulator to guide the training process and ensure that the learned dynamics agree with basic physics laws.   To capture the looking-ahead and looking-behind strategies in the microscopic controller~\eqref{eq:controller}, we use the non-local LWR model  as the underlying flow dynamics:
\begin{align}
	&\partial_t \rho(x,t) + \partial_x (\rho(x,t) v(x,t)) = 0, \label{eq:nonlocal LWR}\\
	&v(x,t) = V_{\eta}(\rho_{\eta}),\\
	&\rho_{\eta}(x,t) =   \int_{-\eta^b}^{\eta^a} \rho(x+y,t)\omega(y) \diff y, \label{eq:nonlocal rho look-ahead-behind}
\end{align}
where the desired speed is a function of non-local density $\rho_{\eta}$ via the non-local fundamental diagram $V_{\eta}$. The non-local density is a weighted average of downstream and upstream density with the non-local kernel $\omega(x):[-\eta_b,\eta_a]\to \mathbb{R}$ with  $\eta_a>0$ being the looking-ahead distance and  $\eta_b>0$ being the looking-behind distance. And for the non-local kernel, we have:
\begin{align}
	\int_{-\eta^b}^{\eta^a} \omega(x)\diff x = 1.
\end{align}
For the well-posedness of the non-local model~\eqref{eq:nonlocal LWR}, we make the following assumptions on the fundamental diagram and non-local kernel function:
\begin{assumption}\label{assumption:FD}
	The fundamental diagram $V_{\eta}(\rho)$ satisfies $V_{\eta}(\rho)\ge 0$ and $\frac{\diff V_{\eta}}{\diff \rho} \le 0$ for all $\rho\le \rho_{\max}$ with $\rho_{\max}$ being the jam density.
\end{assumption}

\begin{assumption}\label{assumption:look-ahead-behind kernel}
	The looking-ahead looking-behind kernel in~\eqref{eq:nonlocal rho look-ahead-behind} satisfies: 1) $\omega(x)\ge 0$ for all $x\in[-\eta^b,\eta^a]$, 2) $\diff \omega /\diff x \le 0$ for all $x\in [0,\eta]$, and 3) $\diff \omega /\diff x \ge 0$ for all $x\in [-\eta^b,0]$.
\end{assumption}

Since the NN only outputs local density $\rho(x_i,t_j)$, to evaluate the learned dynamics, we need to first calculate the  non-local density.  We adopt a numerical approximation scheme~\cite{karafyllis2022analysis} to approximate the non-local density $\rho_{\eta}(x_i,t_j)$ as
\begin{align}\label{eq:nonlocal rho discrete}
	\bar{\rho}(x_i,t_j) = \sum_{k=-N_{\eta}^b}^{N_{\eta}^a-1}\rho(x_i+k\Delta x,t_j) \bar{\omega}_k,
\end{align}
with $\Delta x$ being the discrete cell length, $N_{\eta}^a = \eta^a/\Delta x$, $N_{\eta}^b = \eta^b/ \Delta x$, and $\bar{\omega}_k = \int_{k\Delta x}^{(k+1)\Delta x} \omega(s) ds $ is the discretized non-local weights with
\begin{align}\label{eq:omega bar constraint  sum1}
	\sum_{k=0}^{N_{\eta}-1} \bar{\omega}_k = 1.
\end{align}
We use a trainable vector $\theta_{\omega} \in \mathbb{R}^{N_{\eta}^a+N_{\eta}^b-1}$ to learn the non-local weights. To ensure that the condition in~\eqref{eq:omega bar constraint  sum1} is always met, we  construct the learned kernel weight $\hat{\omega}(\theta_\omega)$  as:
\begin{align}\label{eq:omega discrete learned}
	\hat{\omega}_k(\theta_\omega) = \frac{\theta_{\omega,k}}{\sum_{k=0}^{N_{\eta}-1} \theta_{\omega,k}}.
\end{align} 
The learned non-local density at a spatial-temporal location $(x_i,t_j)$ is then a weighted sum of the local density $\hat{\rho}(x_i,t_j;\theta)$ weighted by the kernel weight $\hat{\omega}$ as:
\begin{align}
	\hat{\rho}_{\eta}(x_i,t_j;\theta,\theta_\omega) = \sum_{k=0}^{N_{\eta}^a-1} \hat{\rho}(x_i+k\Delta x,t_j;\theta) \hat{\omega}_k(\theta_\omega) + \sum_{k=1}^{N_{\eta}^b} \hat{\rho}(x_i-k\Delta x,t_j;\theta) \hat{\omega}_{k-1+N_{\eta}^a}(\theta_\omega),
\end{align}
where the first $N_a-1$ values in $\hat{\omega}_k$ are the looking-ahead weights and the last $N_b$ values in $\hat{\omega}_k$   are the looking-behind weights.

We use a neural network with trainable parameter $\theta_v$ to learn the fundamental diagram $V_{\eta}(\rho_{\eta})$, i.e., the mapping from non-local density to speed, as $\hat{V}_{\eta}(\rho_{\eta};\theta_v)$. 
To evaluate the discrepancy between the learned dynamics and the nonlocal LWR model, we define a residual value at the spatial-temporal location $(x_i,t_j)$ as
\begin{equation}
	\begin{aligned}
		f(x_i,t_j;\theta,\theta_v,\theta_{\omega}) = &\partial_t \hat{\rho}(x_i,t_j;\theta) + \partial_x \left( \hat{\rho} (x_i,t_j;\theta) \hat{V}_{\eta}(\hat{\rho}_{\eta}(x_i,t_j;\theta,\theta_{\omega});\theta_v) \right)\\
		= & \partial_t \hat{\rho}(x_i,t_j;\theta) + \partial_x \hat{\rho} (x_i,t_j;\theta) \cdot \hat{V}_{\eta}(\hat{\rho}_{\eta}(x_i,t_j;\theta,\theta_{\omega});\theta_v)   \\
		&+  \hat{\rho} (x_i,t_j;\theta)  \cdot \partial_{\rho}\hat{V}_{\eta}(\hat{\rho}_{\eta}(x_i,t_j;\theta,\theta_{\omega});\theta_v)   \cdot \partial_x \hat{\rho}_{\eta} (x_i,t_j;\theta,\theta_{\omega}), 
	\end{aligned}
\end{equation}
where the partial derivative of the non-local density $ \partial_x \hat{\rho}_{\eta} (x_i,t_j;\theta,\theta_{\omega})$ is given by the same numerical discretization method in~\eqref{eq:nonlocal rho discrete} as:
\begin{equation}
	\partial_x \hat{\rho}_{\eta} (x_i,t_j;\theta,\theta_{\omega})   = \sum_{k=0}^{N_{\eta}^a-1}\partial_x \hat{\rho}(x_i+k\Delta x,t_j;\theta) \hat{\omega}_k(\theta_\omega) + \sum_{k=1}^{N_{\eta}^b}\partial_x \hat{\rho}(x_i-k\Delta x,t_j;\theta) \hat{\omega}_{k-1+N_{\eta}^a}(\theta_\omega), 
\end{equation}
and the partial derivatives $\partial_t \hat{\rho} (x_i,t_j;\theta)$, $\partial_x \hat{\rho} (x_i,t_j;\theta)$, and $\partial_{\rho}\hat{V}_{\eta}(\hat{\rho}_{\eta}(x_i,t_j;\theta,\theta_{\omega});\theta_v)$ are calculated via automatic differentiation in Tensorflow. We select
some spatial-temporal points $\mathcal{P}$ from the whole spatial-temporal domain and calculate the physic loss as:
\begin{equation}\label{eq:loss physics}
	L_{\mathrm{p}}(\theta,\theta_v,\theta_{\omega}) = \sum_{k=1}^{N_s} \frac{1}{N_p}\sum_{(i,j)\in P} \left( f(x_i,t_j;\theta_k,\theta_v,\theta_{\omega}) \right)^2,
\end{equation}
where $N_p = |\mathcal{P}|$ is the number of selected spatial-temporal points. For the mapping from the spatial-temporal points $(x,t)$ to density  $\rho$, each simulated density use a separate NN $\theta_k$. This is because that the initial data is different for different simulations. But for the kernel  and fundamental diagram , all simulations share the same $\theta_{\omega}$ and $\theta_{v}$. This aims to learn a model that covers a wide range of density values.

\subsubsection{Constraint loss}

The constraint loss is included to ensure that the learned non-local kernel and fundamental diagram meet the model well-posedness assumptions. 
Given the conditions on $\omega(x)$ in Assumption~\ref{assumption:look-ahead-behind kernel}, we have three constraints as:  
\begin{align}
	&\bar{\omega}_k \ge 0, \forall k=0,1,\cdots,N_{\eta}^a + N_{\eta}^b-1,\\
	&\bar{\omega}_{k+1}\le \bar{\omega}_k, \forall k=0,1,\cdots,N_{\eta}^a-2,\\
	&\bar{\omega}_{k+1}\le \bar{\omega}_k, \forall k = N_{\eta}^a,\cdots,N_{\eta}^a+N_{\eta}^b-2, \\
    & \bar{\omega}_{N_{\eta}^a} \le \bar{\omega}_0.
\end{align}
To satisfy these constraints, we define a constraint loss on the non-local kernel as:
\begin{align}
	L_{c,\omega} (\theta_\omega) &= \sum_{i=0}^{N_{\eta}^a+N_{\eta}^b-1} \left(\min \{\hat{\omega}_i(\theta_\omega),0\}\right)^2  + \sum_{i=0}^{N_{\eta}^a-2} \left(\max \{\hat{\omega}_{i+1}(\theta_\omega)-\hat{\omega}_i(\theta_\omega),0\}\right)^2 + \sum_{i=N_{\eta}^a}^{N_{\eta}^a+N_{\eta}^b-2} \left(\max \{\hat{\omega}_{i+1}(\theta_\omega)-\hat{\omega}_i(\theta_\omega),0\}\right)^2 \notag\\
	& + \left(\max \{\hat{\omega}_{N_{\eta}^a}(\theta_\omega)-\hat{\omega}_{0}(\theta_\omega),0\}\right)^2. 
\end{align}

For the fundamental diagram, given the constraints in Assumption~\ref{assumption:FD}, to ensure that $\hat V_{\eta}(\rho;\theta_v)\ge 0$, we design a constraint penalty as
\begin{equation}\label{eq:loss FD>0}
	L_{c,v,1} = \sum_{i=0}^{N_{\rho}-1}   \left(\min\left\{ \hat{V}_{\eta} (i\Delta\rho;\theta_v),0\right\}\right)^2 , 
\end{equation}
where $N_{\rho} = \rho_m/\Delta \rho$ with $\rho_m$ being the maximum density.  To ensure that the fundamental diagram is a non-increasing function with respect to the density, we  design the second constraint penalty as:
\begin{equation}\label{eq:loss dFD/drho<0}
	L_{c,v,2} = \sum_{i=0}^{N_{\rho}-1}   \left( \max\left\{
	\partial_{\rho} \hat V_{\eta}(i \Delta \rho;\theta_v),0
	\right\} \right)^2,
\end{equation}
where $\partial_{\rho} \hat V_{\eta}$ is calculated using automatic differentiation provided in Tensorflow. The constraint loss is a weighted sum of penalties for the non-local kernel and fundamental diagram:
\begin{align}\label{eq:loss constraint}
	L_{\mathrm{c}} = p_{\omega} L_{c,\omega} (\theta_\omega)+  + p_{v} ( L_{c,v,1}(\theta_v)  L_{c,v,2}(\theta_v)),  
\end{align}
where $p_{\omega}>0$  and $p_{v}>0$  are penalty weight coefficients.

\section{Look-ahead and look-behind behavior analysis}\label{sec:analysis}

In this section, we first conduct simulator experiments to collect human drivers' data when they can look-ahead and look-behind, and  calibrate looking-ahead/looking-behind car-following models for human drivers in Section~\ref{subsec:human}. These models are used as baselines to analyze how the controller~\eqref{eq:controller} affects the macroscopic flow model. We analyze how the feedback gains and the optimal speed function affect the flow dynamics in Section~\ref{subsec:gain} and Section~\ref{subsec:vs} respectively. In Section~\ref{subsec:mixed}, we analyze the mixed traffic flow with CAV and human-driven vehicles. And we conduct sensitivity analysis to demonstrate how hyper-parameters in the learning algorithm affect the learned flow model.

\subsection{Human drivers' behavior analysis} \label{subsec:human}

\subsubsection{Human drivers' data collection}

\begin{figure}[!t]
    \centering
    \includegraphics[width=0.6\linewidth]{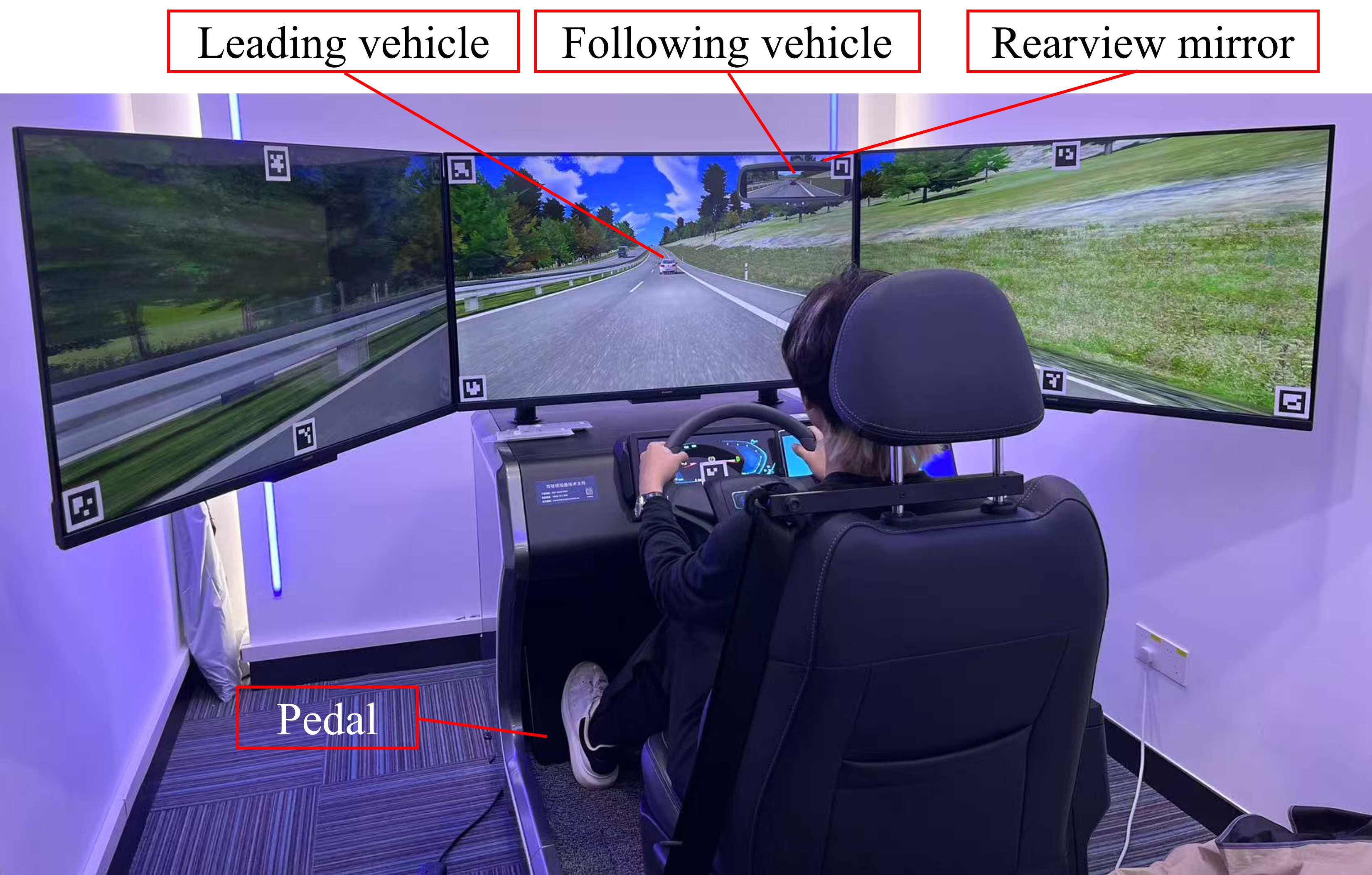}
    \caption{Simulator settings.  The screens shows the leading vehicles and the rearview mirror  shows the following vehicle.} \label{fig:simulator}
\end{figure}
	
We collect human drivers' trajectories via a simulator experiment. The collected human drivers' data is used as a baseline to analyze CAV controllers. Fig.~\ref{fig:simulator} gives a picture taken during one experiment. The participant drives in a vehicle chain and is indexed as $i=0$. The participant is told to cruise without lane-changing. There are three leading vehicles: $i=-3,-2,-1$, and one following vehicle $i=1$. The leading vehicle $i=-2$ follows $i=-3$, and the vehicle $i=-1$ follows $i=-2$. In each experiment, the participant goes through three scenarios:
\begin{itemize}
    \item Normal cruising. The participant can observe the leading vehicle $i=-1$ on the screen and the following vehicle $i=1$ through the rearview mirror. The three leading vehicles cruise at a pre-decided speed $v^*$, and the following vehicle follows the participant.  In this scenario, the leading vehicle $i=-3$ is set to have two deceleration-acceleration processes to cover a wider range of driving scenarios. For the first deceleration process, the leading vehicle decelerates at a deceleration of $-a_1$ ($a_1>0$) with a duration of $t_1$ and then accelerates back to $v^*$ with the  acceleration of $a_1$.  For the second deceleration, the leading vehicle $i=-3$ has a deceleration of $-a_2$ ($a_2>0$) and duration of $t_2$, then it accelerates back to $v^*$ with an acceleration of $a_2$.  
    \item Cruising when the following vehicle has a nudging acceleration. Participants can observe the leading vehicle $i=-1$ and the following vehicle. The three leading vehicles keep the pre-set speed $v^*$, and the following vehicle $i=1$ takes three nudging accelerations: with the acceleration and duration being $(a_3,t_3)$, $(a_4,t_4)$, $(a_5,t_5)$ respectively. After each nudging acceleration, the following vehicle $i=1$ takes the IDM model to decelerate. 
    \item Cruising when they can see the leading vehicles $i=-2$, $i=-3$. Participants can observe all three leading vehicles on the screen. The leading vehicle $i=-3$ will repeat the two deceleration processes as in the normal cruising scenario.
\end{itemize}
Each participant takes three experiments, with high, low, and medium pre-set speed $v^*$. The parameter settings for different experiments are given in Table~\ref{tab:exp}.
	
\begin{table}[!t]
    \centering
    \caption{Experiment settings}
    \label{tab:exp}
    \begin{tabular}{c|c|c|c|c}
        &  & High speed & Medium speed & Low speed\\ \hline
        & Pre-set speed $v^*$ (km/h) & 70 &  50 & 30 \\ \hline 
        \multirow{4}{*}{Leader  decelerates} & Deceleration $a_1$ (m/s$^2$) & 5.5 & 4.5 & 3.1 \\ 
        & Duration $t_1$ (s) & 3 & 3 & 3 \\
        & Deceleration $a_2$ (m/s$^2$) & 7.5 & 6.5 & 5.0  \\ 
        & Duration $t_2$ (s) &  3 & 3 &  3 \\ \hline
        \multirow{6}{*}{Follower  accelerates} & Acceleration $a_3$ (m/s$^2$) & 1.25 & 0.8 & 0.56 \\ 
        & Duration $t_3$ (s) & 6 & 4 & 2 \\
        & Acceleration $a_4$ (m/s$^2$) & 2.81 & 1.8 & 1.25  \\ 
        & Duration $t_4$ (s) & 6 & 4& 2 \\
        & Acceleration $a_5$ (m/s$^2$) & 8 & 7.2 & 5.0  \\ 
        & Duration $t_5$ (s) & 6 & 4 & 2\\ \hline
    \end{tabular}
\end{table}
	
Since the collected data only covers a relatively short spatial-temporal period, to analyze macroscopic attributes,  we first calibrate car-following models from the collected data, and then run microscopic simulations to get more trajectories on a larger spatial-temporal range. We calibrate the following three microscopic models from participants' trajectory data:
\begin{itemize}
    \item Car-following model: We use the OVM model:
    \begin{align}\label{eq:micro CF}
        \dot{v}_0 = \alpha_0(V_{\mathrm{opt}}(s_0) -v_0) + \beta_0(v_{-1} - v_0),
    \end{align}
    and we use the piece-wise linear desired speed function as in~\eqref{eq:micro Vopt}.
    \item Car-following model with looking-ahead speed:
    \begin{align}\label{eq:micro look ahead}
        \dot{v}_0 = \alpha_0 (V_{\mathrm{opt}}(s_0) - v_0) + \beta_{0}(v_{-1} - v_0) + \beta_{-1} (v_{-2} - v_0) + \beta_{-2} (v_{-3}-v_0),
    \end{align}
    where $\beta_{-1}$ and $\beta_{-2}$ are the sensitive parameters to the speed of leading vehicles $i=-2$ and $i=-3$ respectively. 
    \item Car-following model with nudging effect:
    \begin{align}\label{eq:micro nudging pos}
        \dot{v}_0 = \alpha_0 (V_{\mathrm{opt}}(s_0) - v_0) + \beta_{0}(v_{-1} - v_0) + \beta_1 (v_{1} - v_0) \mathbbm{1}({v_1>v_0})
    \end{align}
    where 
    \begin{align}
        \mathbbm{1}({v_1>v_0}) = \left \{ \begin{array}{ll}
            1, & v_1>v_0 \\
            0, & v_1\le v_0
        \end{array} 
        \right. ,
    \end{align}
    is an indicator function. We add this indicator function because human drivers only accelerate when the following vehicle accelerates, but will not decelerate when the following vehicle decelerates.
\end{itemize}
We run genetic algorithms to get the model parameters by minimizing the relative calibration error of speed and gap:
\begin{align}
    \sum_i \sum_t \left( \frac{\hat{v}_0(i,t) - v_0(i,t)}{v_0(i,t)} \right)^2 + \left( \frac{\hat{s}_0(i,t) - s_0(i,t)}{s_0(i,t)} \right)^2
\end{align}
with $v_0(i,t)$ and $\hat{v}_0(i,t)$ being the recorded and predicted speed in the $i$-th experiment at time-step $t$, $s_0(i,t)$ and $\hat{s}_0(i,t)$ being the recorded and predicted gap in the $i$-th experiment at time-step $t$. We give in Table~\ref{tab:micro parameter} the calibrated microscopic model parameters.

\begin{table}[!t]
    \centering
    \caption{The calibrated microscopic model parameters}
    \label{tab:micro parameter}
    \begin{tabular}{c|c|c|c}
        Parameter  & Car-following~\eqref{eq:micro CF} & Look-ahead~\eqref{eq:micro look ahead} & Look-behind~\eqref{eq:micro nudging pos} \\ \hline
        $\alpha_0$  & 0.011 & 0.027 & 0.014 \\ \hline 
        $\beta_0$  & 0.718 & 0.965 & 0.789 \\ \hline 
        $v_{\max}$ (m/s) & 17.08 & 26.53 & 16.24 \\ \hline
        $s_{\mathrm{st}}$ (m) & 1.53 & 1.12 &  1.00 \\ \hline 
        $s_{\mathrm{go}}$ (m) & 24.96 & 27.97& 21.40\\ \hline 
        $\beta_{-1}$ & $\backslash$ &  0.0500 & $\backslash$\\ \hline 
        $\beta_{-2}$  & $\backslash$ & 0.0082 & $\backslash$\\ \hline 
        $\beta_{1}$ & $\backslash$ & $\backslash$ & 0.092 \\ \hline 
    \end{tabular}
\end{table}

\subsubsection{Macroscopic modeling for human drivers}

Now we run simulations to get macroscopic non-local flow models corresponding to different microscopic cruising models. For the normal car-following model in~\eqref{eq:micro CF} and the car-following model with looking-ahead behaviors in~\eqref{eq:micro look ahead}, since there is no looking-behind behaviors, we set the looking-behind kernel length as $\eta_b = 0$. We use the looking-ahead kernel length as $\eta_a = 30 $ m. For the car-following model with looking-behind behaviors in~\eqref{eq:micro nudging pos}, we set the kernel length as  $\eta_a = 30 $ m and $\eta_b = 30 $ m.  For the coefficient weights in the loss function, we set the coefficient weight for the  data loss in~\eqref{eq:loss data}  as $c_d = 0.1$, and the penalty for the physics constraint loss in~\eqref{eq:loss constraint} as $p_v = 10^6$ and $p_{\omega} = 10^6$. We set the neural network as a fully connected feedforward network with 6 hidden layers and 64 neurons in each hidden layer. For the numerical scheme to get the non-local density in~\eqref{eq:nonlocal rho discrete}, we set the discrete cell length as $\Delta x = 1$ m. To get the training dataset, we first discrete  the original spatial-temporal by small cells with the cell length being 1 meter and 1 second. Then we get $N_x\times N_t$ grids.  For the training data,  we assume that the initial data at $t=0$, i.e., $\rho(x,0)$ is given. Besides, there are $N_l=5$ loop detectors evenly distributed along the road at  $x=l_j$ to measure the ground-truth density $\rho_k(l_i,t)$. The spatial-temporal locations where the ground-truth training density is available are as $\mathcal{D} = \{(x_i,0)|i=0,\cdots N_x-1\} \cup \{(l_i,t_j)|i=0,\cdots N_l-1;j=0,\cdots,N_t-1\}$.

\begin{figure}[!t]
    \centering
    \subcaptionbox{Ground truth}{\includegraphics[width=0.3\linewidth]{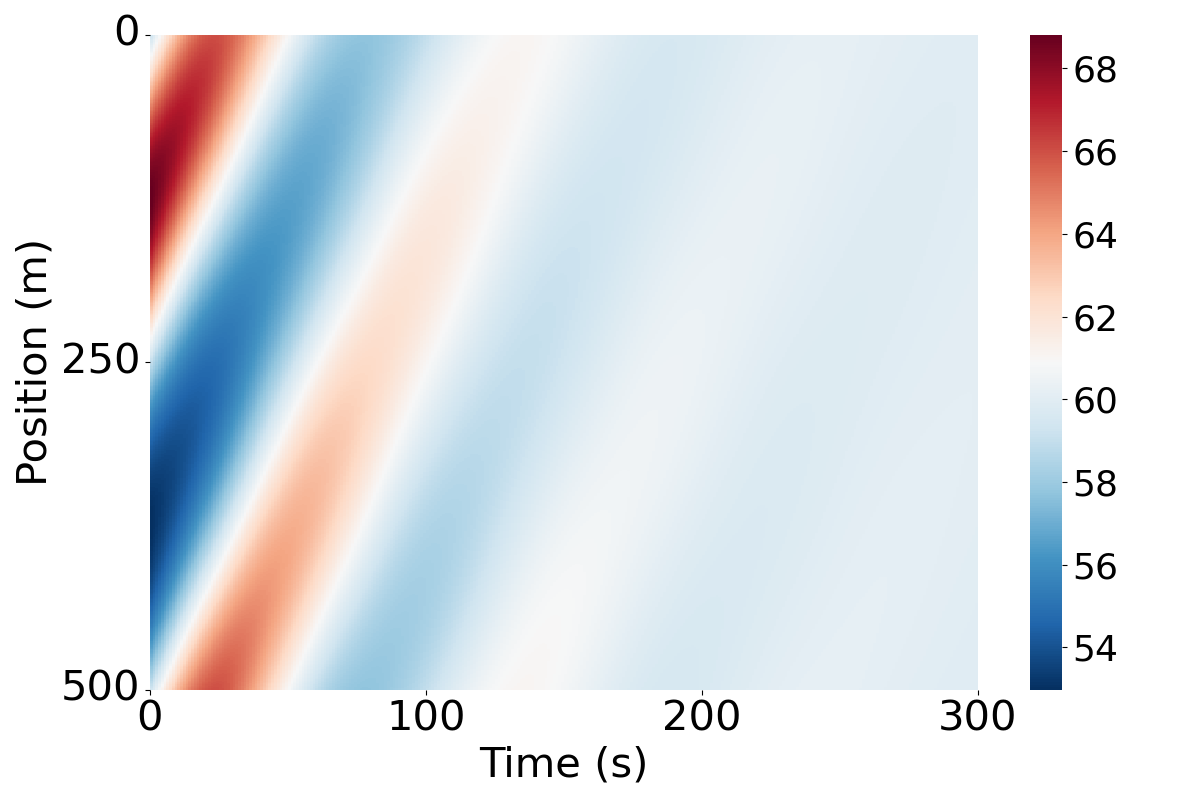}}
    \subcaptionbox{PINN with looking-ahead LWR}{\includegraphics[width=0.3\linewidth]{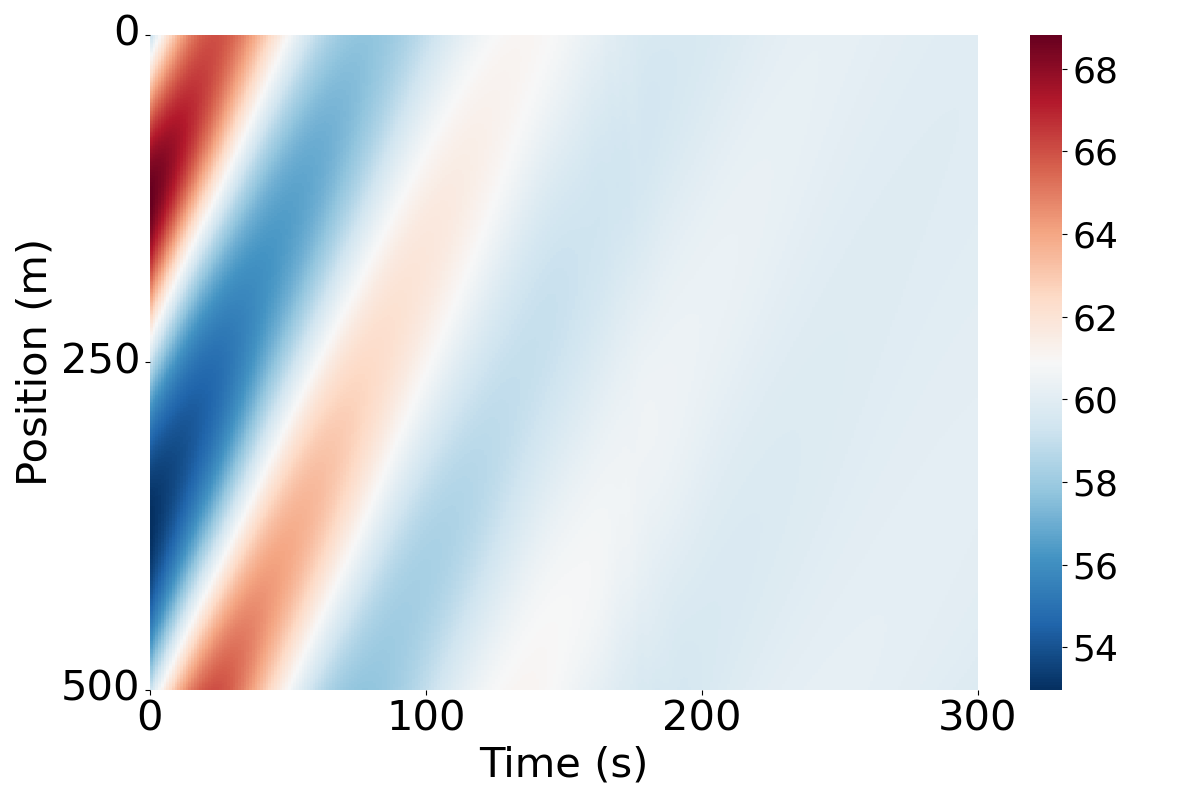}}
    \subcaptionbox{PINN with LWR}{\includegraphics[width=0.3\linewidth]{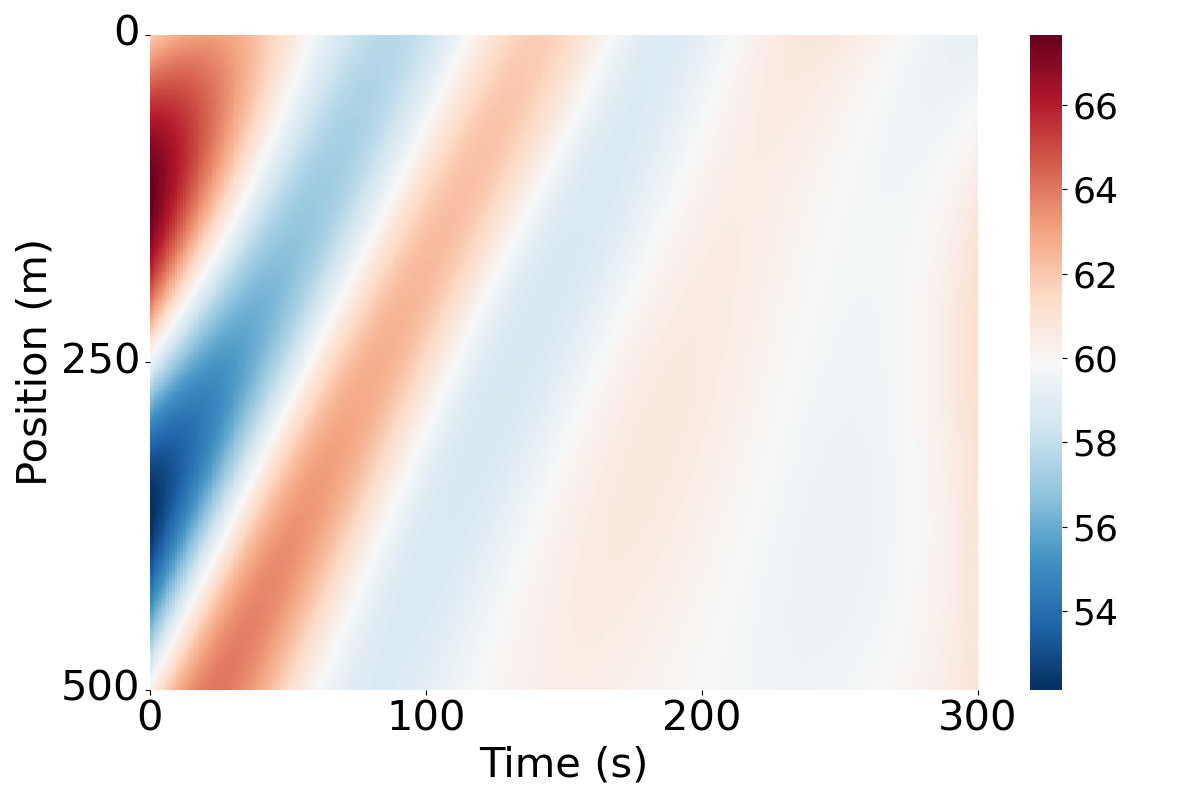}}
    \caption{The traffic density corresponding to the car-following model~\eqref{eq:micro CF}.}
    \label{fig:rho HV acc}
\end{figure}

For the normal car-following model in~\eqref{eq:micro CF}, we compare the accuracy when the non-local LWR model and local LWR model are adopted as physics regularization Fig.~\ref{fig:rho HV acc}. Fig.~\ref{fig:rho HV acc}(a) gives the ground-truth traffic density, and Figs.~\ref{fig:rho HV acc}(b),(c) plot the density simulated with the looking-ahead LWR model and LWR model respectively. We see that the looking-ahead LWR model provides a more accurate description of the traffic density wave. To quantitatively measure the modeling accuracy, we calculate the estimation error over the whole spatial-temporal domain :
\begin{align}
    \sqrt{\frac{1}{N_xN_t}\sum_{x,t}\left(\frac{\rho(x,t)-\hat\rho(x,t)}{\rho(x,t)}\right)^2} \times 100\%
\end{align}
with $\rho(x,t)$ and $\hat \rho(x,t)$ being the ground-truth and estimated density at $(x,t)$ respectively. From the  estimation error  given in Table~\ref{tab:rho error HV}, we see that the non-local LWR model provides a more accurate description of the macroscopic dynamics.
	
\begin{figure}[!t]
    \centering
    \subcaptionbox{Speed vs Density}{\includegraphics[width=0.3\linewidth]{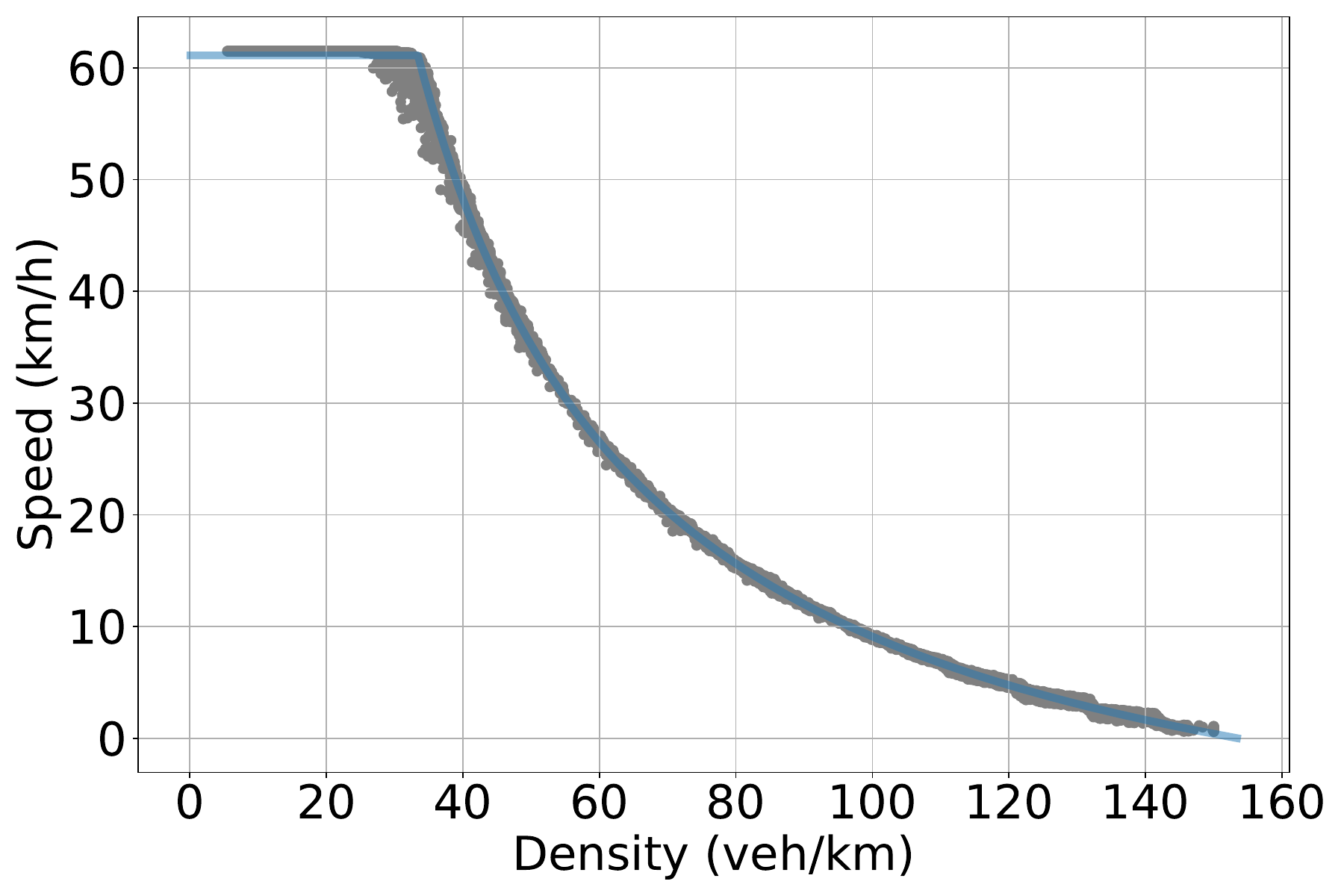}}
    \subcaptionbox{Speed vs Nonlocal density}{\includegraphics[width=0.3\linewidth]{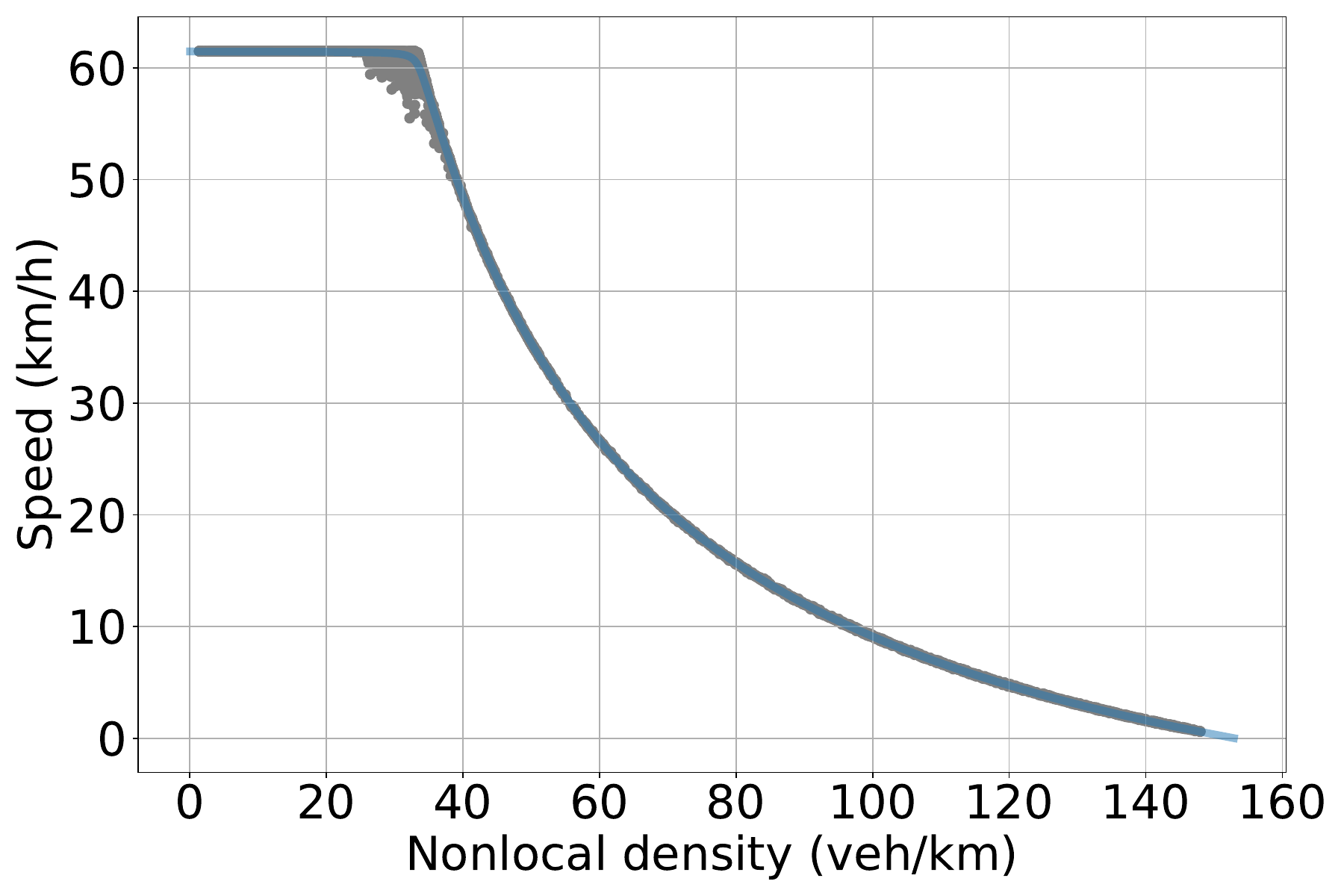}}
    \subcaptionbox{Looking-ahead weights}{\includegraphics[width=0.3\linewidth]{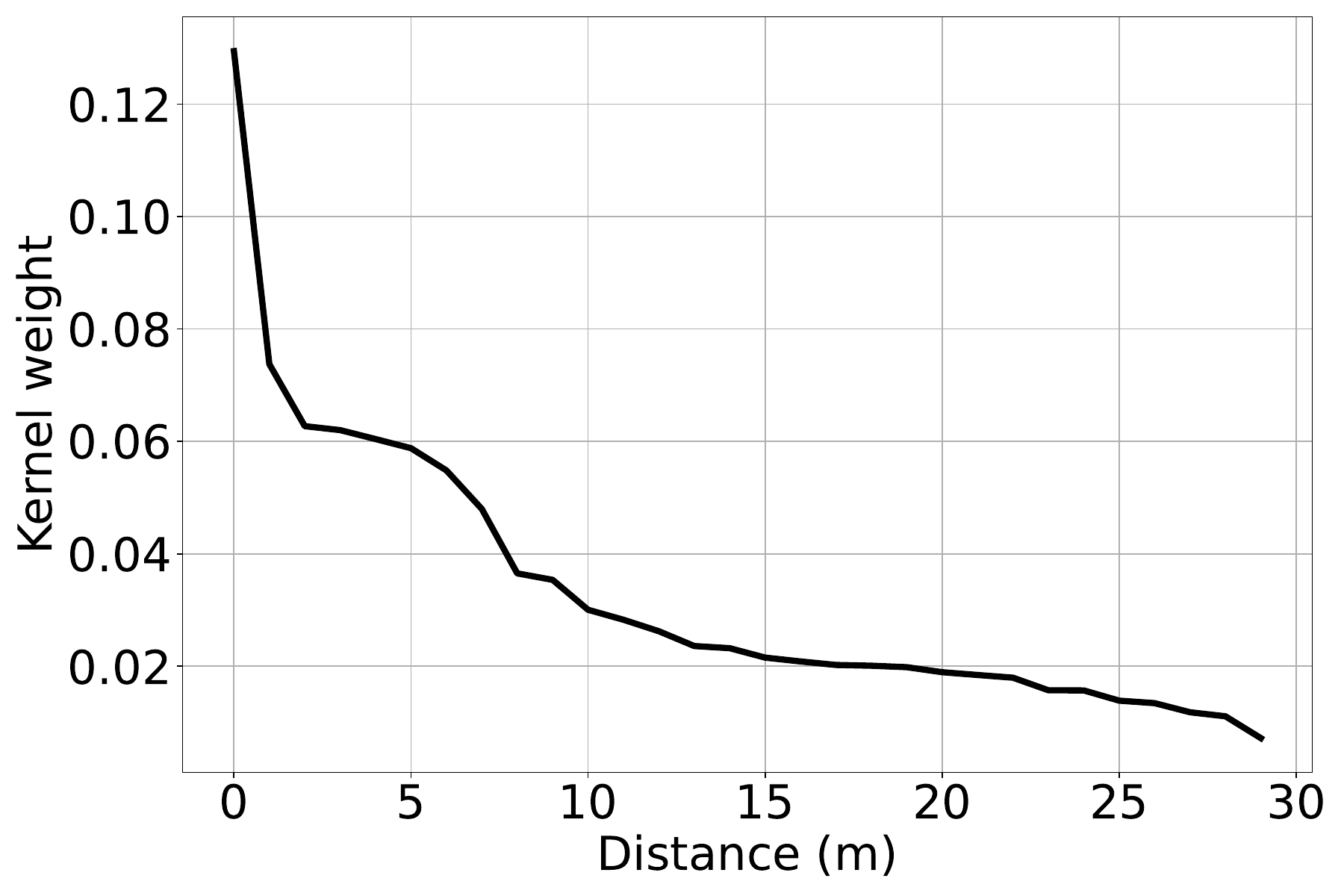}}
    \caption{Fundamental diagram (a)(b) and the learned looking-ahead weights (c) for the car-following model~\eqref{eq:micro CF}.}
    \label{fig:FD HV acc}
\end{figure}

We plot the fundamental diagram in Fig.~\ref{fig:FD HV acc}. Fig.~\ref{fig:FD HV acc}(a) plots the speed $v$ vs density $\rho$ scatter plot and Fig.~\ref{fig:FD HV acc}(b) gives the speed $v$ vs non-local density $\rho_{\eta}$ scatter plot. We see that the speed  vs non-local density presents a smaller scatter range around the fundamental diagram, which means the desired speed $V_{\eta}(\rho_{\eta})$ based on non-local density provides a more accurate calibration than the local-density-desired speed $V(\eta)$. Fig.~\ref{fig:FD HV acc}(c) plots the learned looking-ahead weights. We see that a majority of the looking-ahead weights are within 5 meters.

\begin{figure}[!t]
    \centering
    \subcaptionbox{Ground truth}{\includegraphics[width=0.3\linewidth]{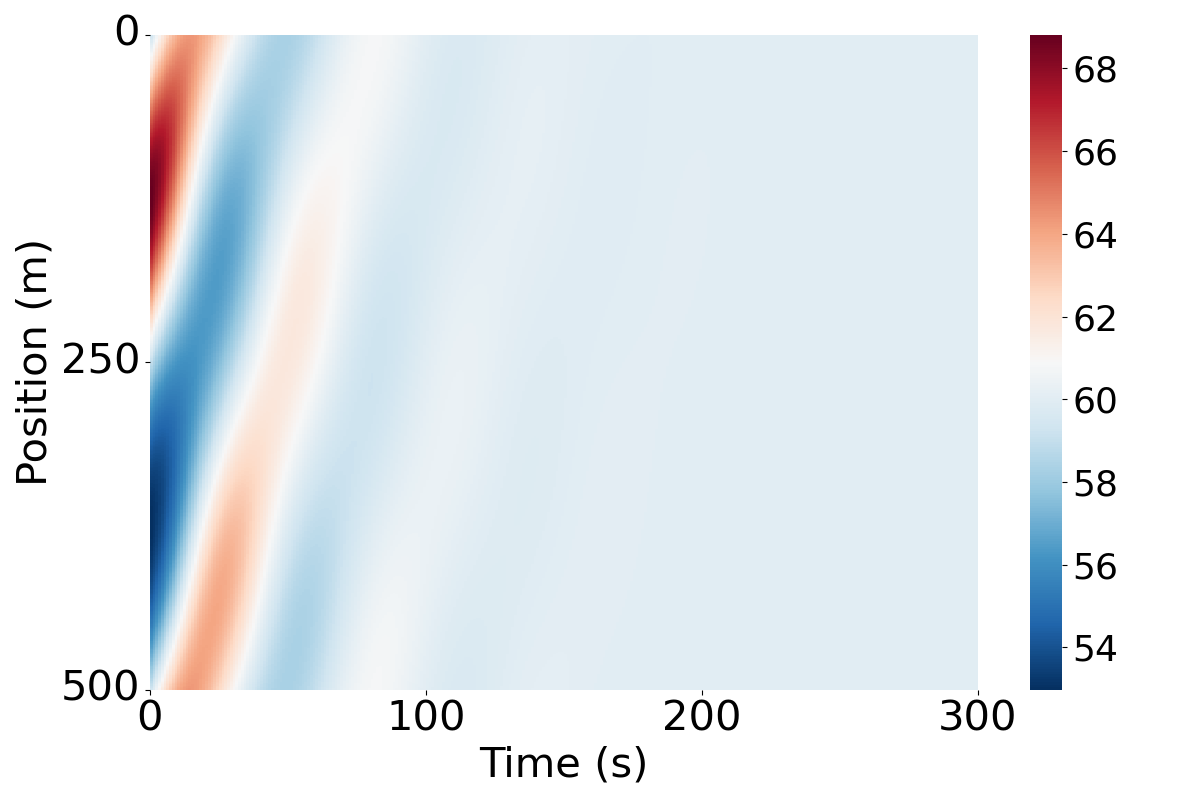}}
    \subcaptionbox{PINN with looking-ahead LWR}{\includegraphics[width=0.3\linewidth]{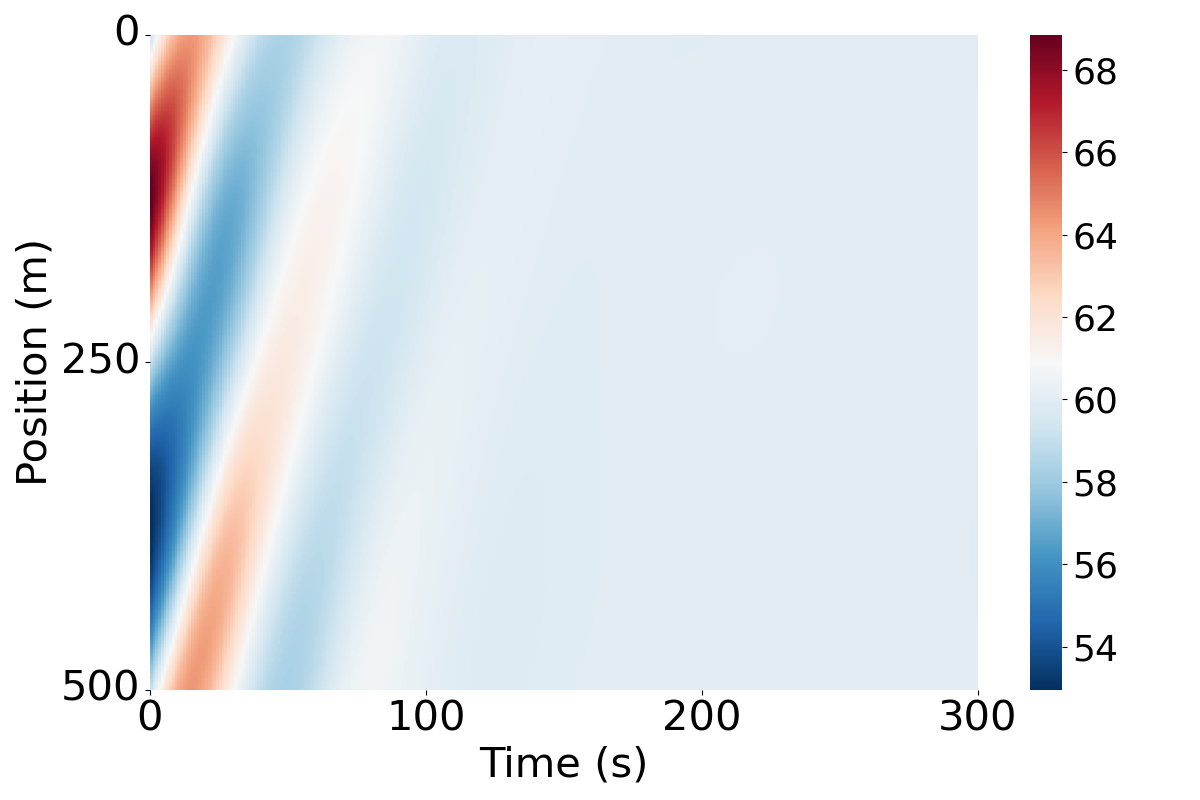}}
    \subcaptionbox{PINN with LWR}{\includegraphics[width=0.3\linewidth]{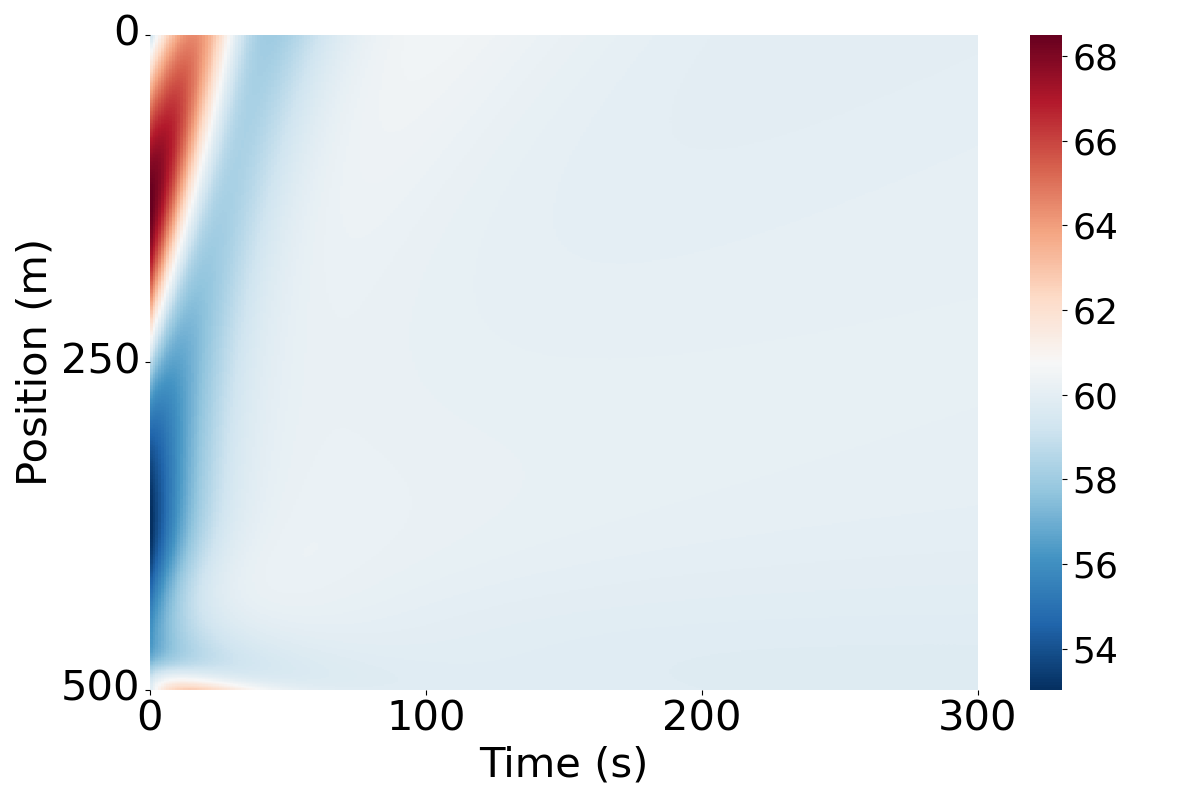}}
    \caption{The traffic density corresponding to the car-following model with looking-ahead speed~\eqref{eq:micro look ahead}.}
    \label{fig:rho HV look ahead}
\end{figure}

\begin{figure}[!t]
    \centering
    \subcaptionbox{Speed vs Density}{\includegraphics[width=0.3\linewidth]{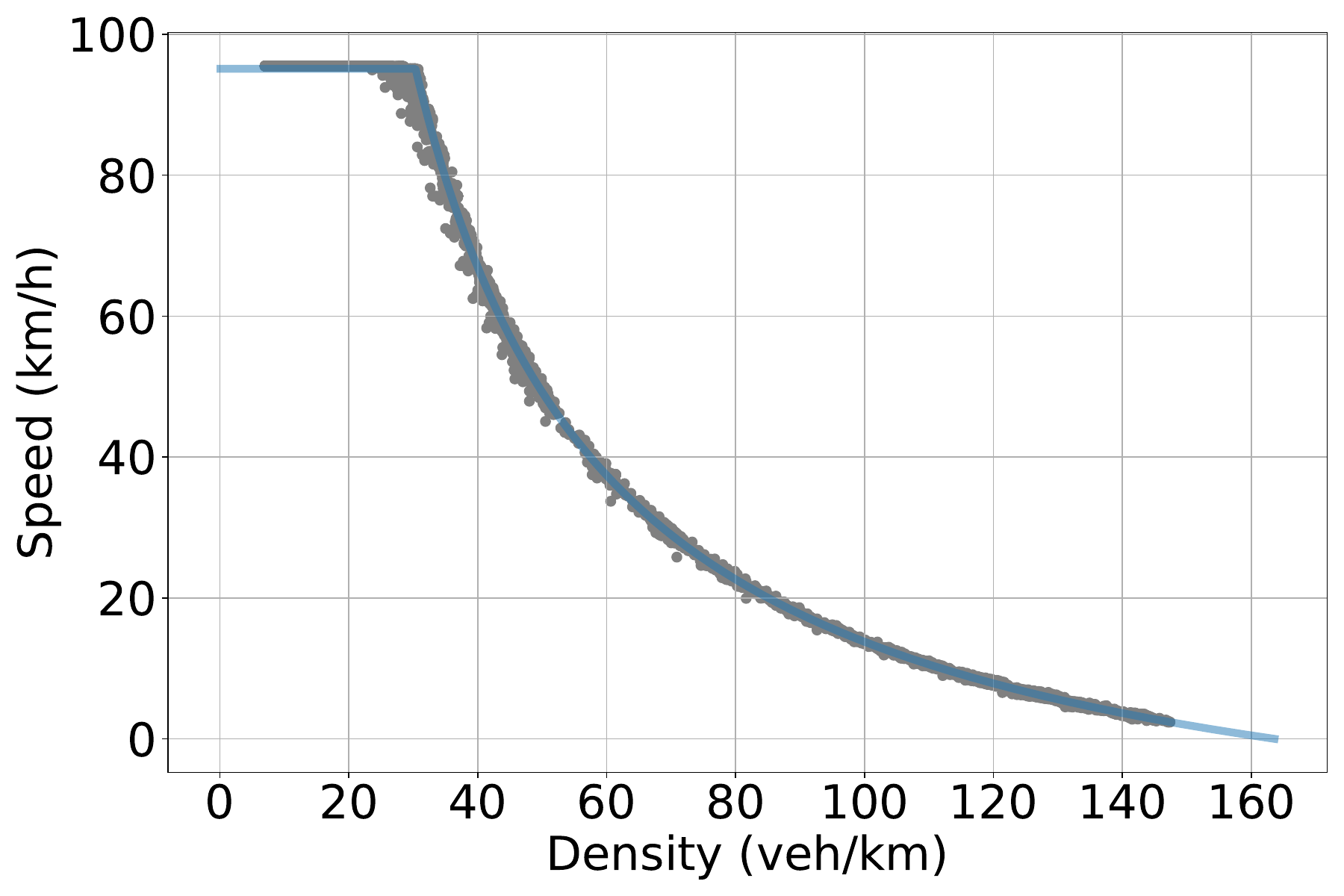}}
    \subcaptionbox{Speed vs Nonlocal density}{\includegraphics[width=0.3\linewidth]{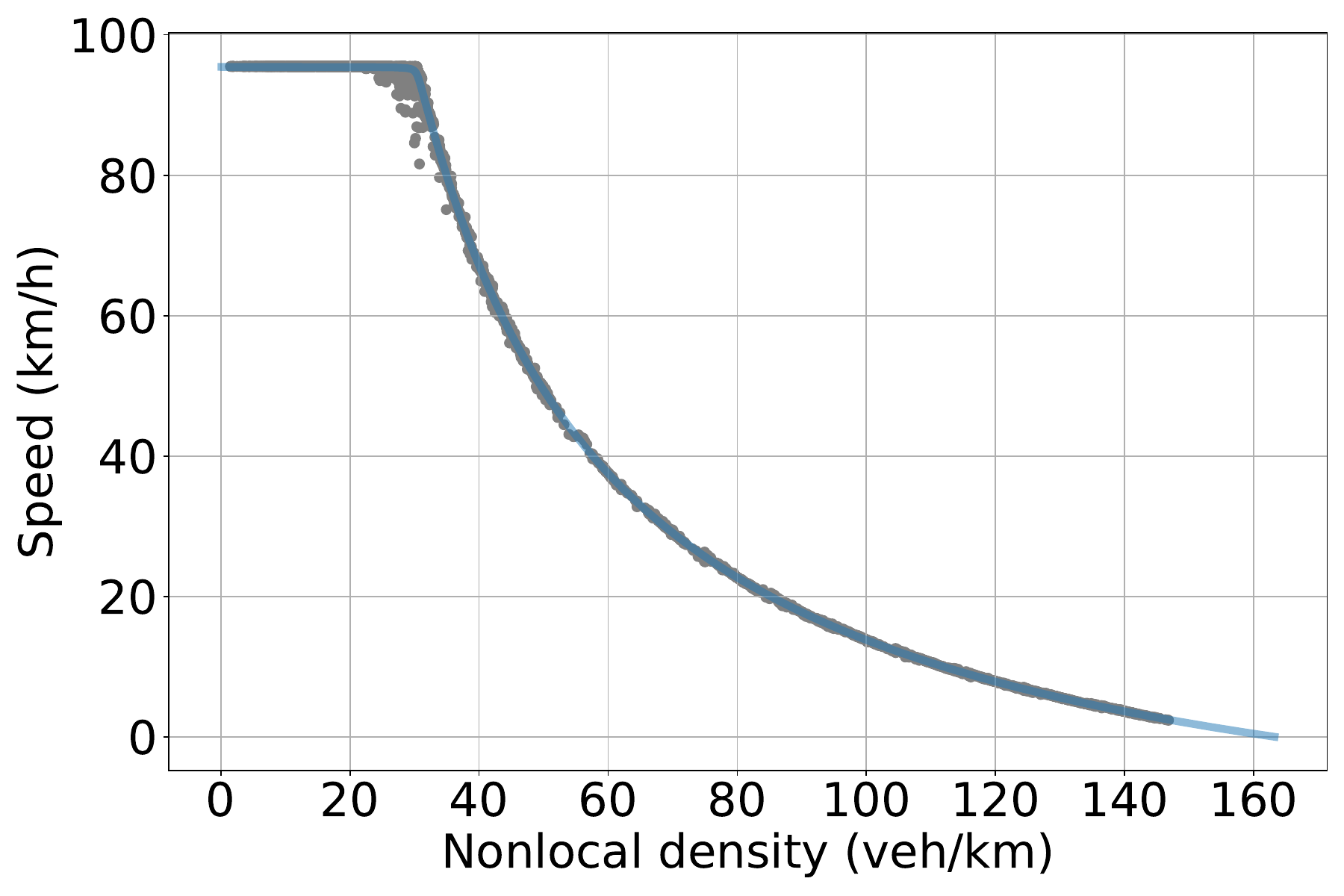}}
    \subcaptionbox{Looking-ahead weights}{\includegraphics[width=0.3\linewidth]{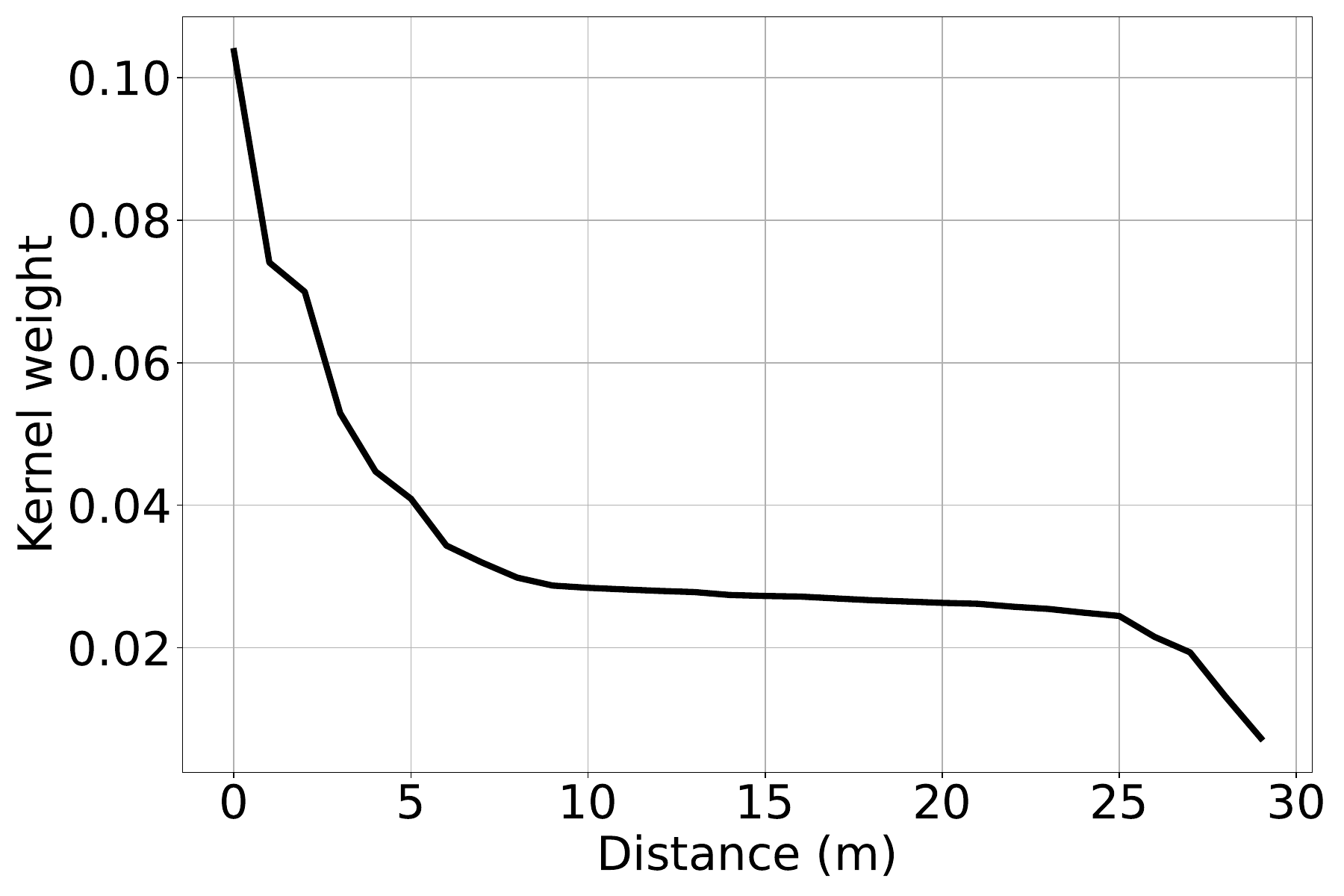}}
    \caption{Fundamental diagram (a)(b) and the learned looking-ahead weights (c) for the car-following model~\eqref{eq:micro look ahead}.} \label{fig:FD HV look-ahead}
\end{figure}

Fig.~\ref{fig:rho HV look ahead} compares the corresponding macroscopic dynamics for the calibrated looking-ahead car-following model~\eqref{eq:micro look ahead}. Similarly, the non-local LWR also provides a more accurate prediction of the wave speed. Fig.~\ref{fig:FD HV look-ahead} gives the fundamental diagram and the learned looking-ahead kernels. We see that the  speed vs looking-ahead non-local density in Fig.~\ref{fig:FD HV look-ahead}(b) provides a better calibration than the speed vs local density in Fig.~\ref{fig:FD HV look-ahead}(a). For the looking-ahead weights in Fig.~\ref{fig:FD HV look-ahead}(c), a majority of the looking-ahead kernel is also within 5 meters. But compared with the car-following case in Fig.~\ref{fig:FD HV acc}(c) where the weights decelerate to 0.02 around 15 meters, the weights in Fig.~\ref{fig:FD HV look-ahead}(c) still keep around 0.3 within 5 to 25 meters. This shows that if the microscopic dynamics model includes leader's information, the looking-ahead non-local kernel also includes more weights on the downstream traffic information. 

\begin{figure}[!t]
    \centering
    \subcaptionbox{Ground truth}{\includegraphics[width=0.3\linewidth]{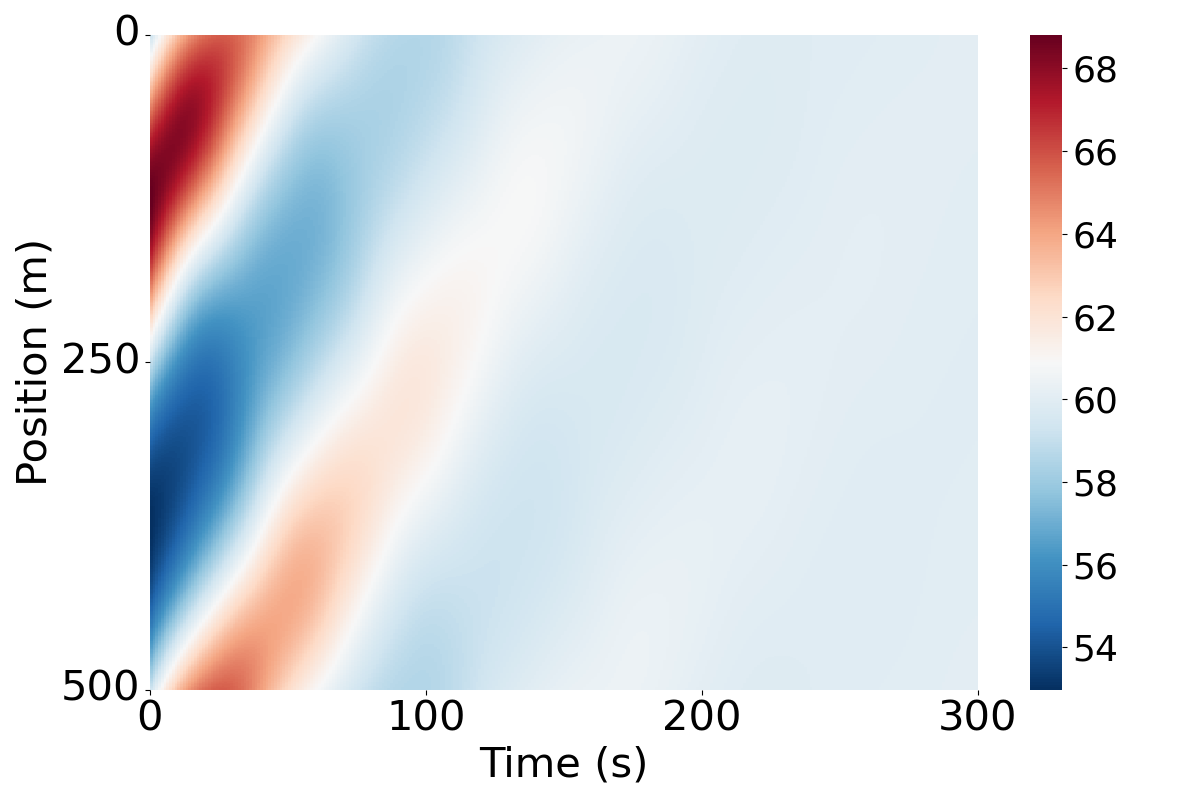}}
    \subcaptionbox{PINN with looking-ahead LWR}{\includegraphics[width=0.3\linewidth]{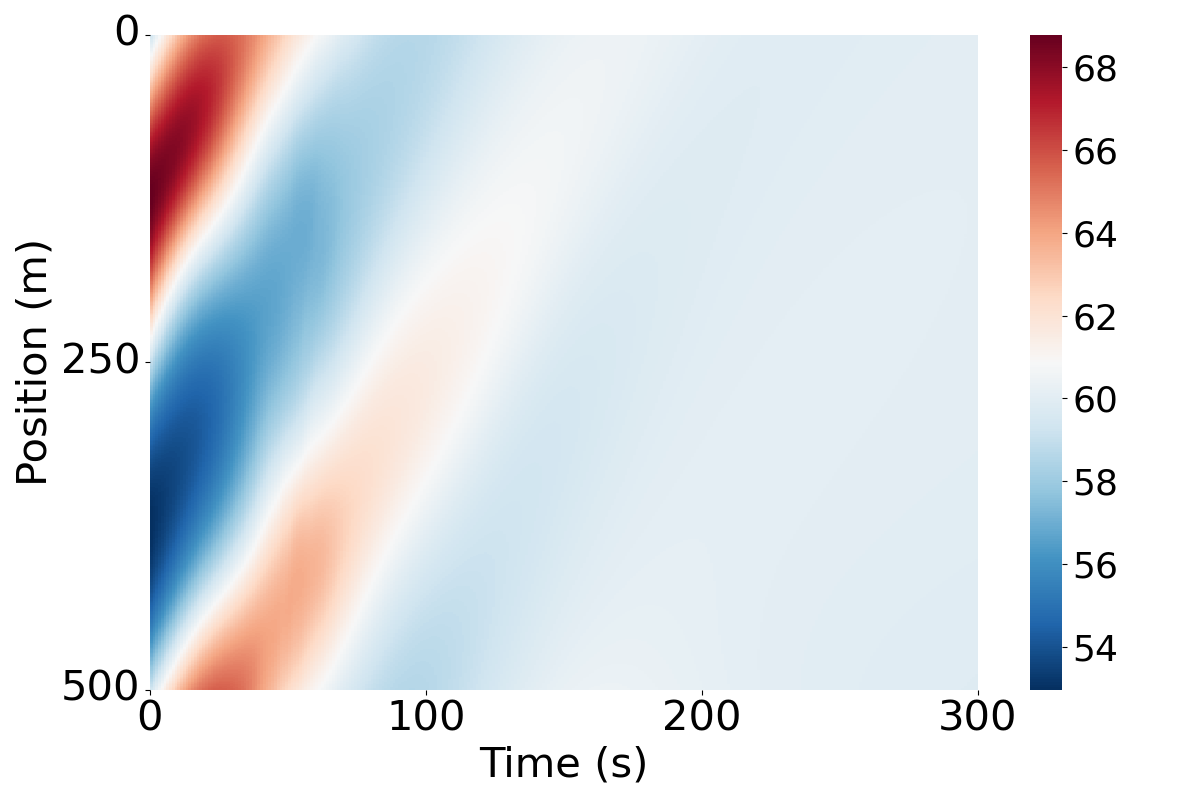}}
    \subcaptionbox{PINN with LWR}{\includegraphics[width=0.3\linewidth]{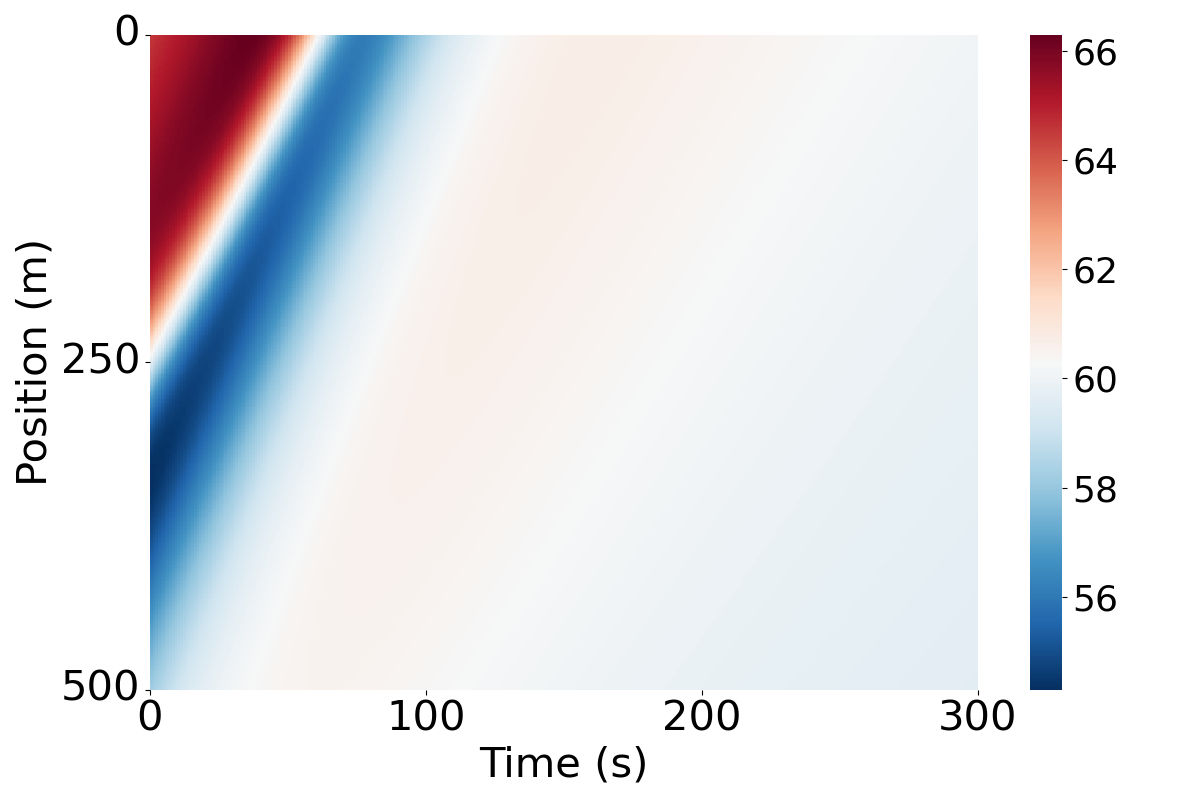}}
    \caption{The traffic density corresponding to the car-following model with looking-behind speed~\eqref{eq:micro nudging pos}.}
    \label{fig:rho HV nudging}
\end{figure}

\begin{figure}[!t]
    \centering
    \subcaptionbox{Speed vs Density}{\includegraphics[width=0.3\linewidth]{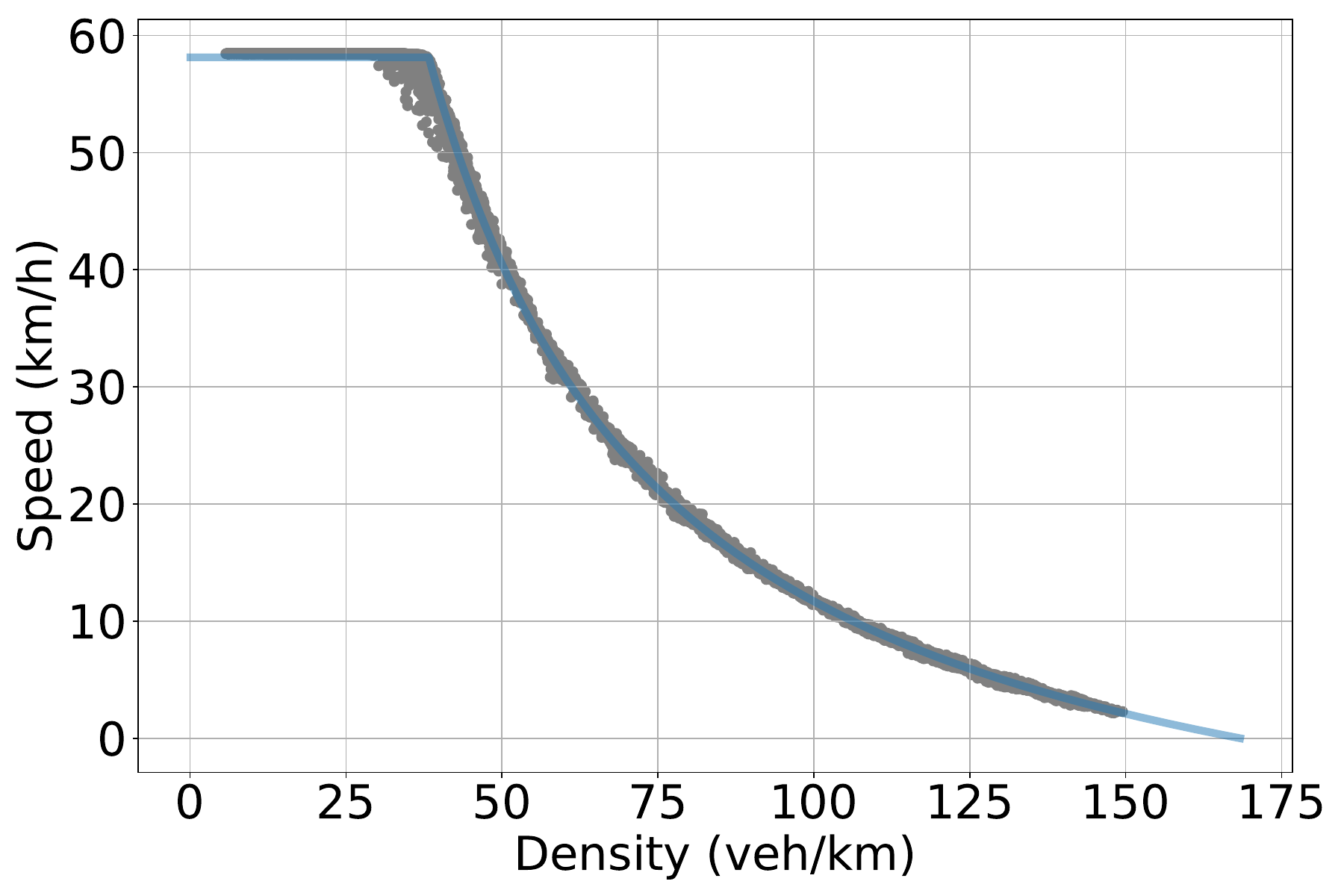}}
    \subcaptionbox{Speed vs Nonlocal density}{\includegraphics[width=0.3\linewidth]{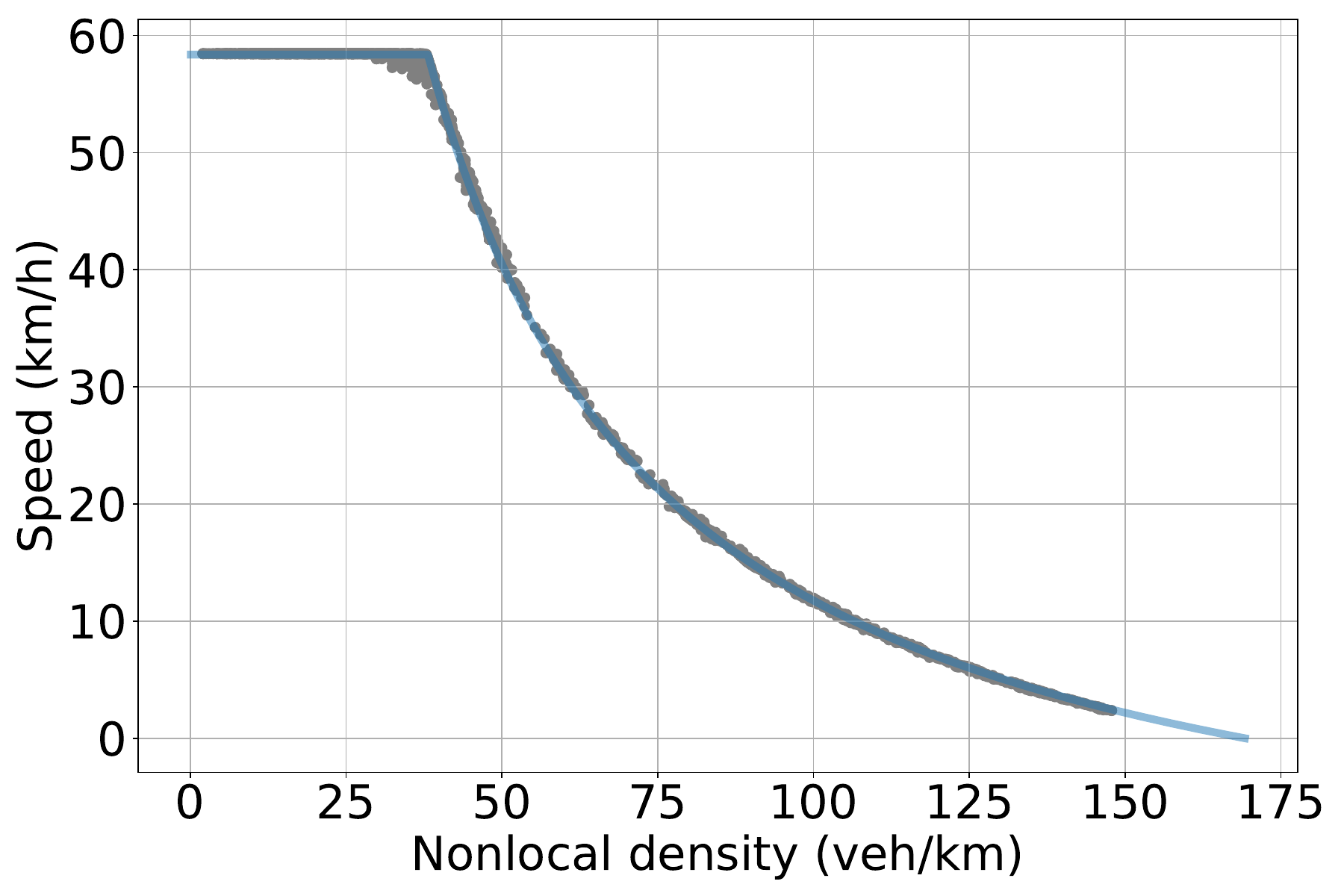}}
    \subcaptionbox{Nonlocal weights}{\includegraphics[width=0.3\linewidth]{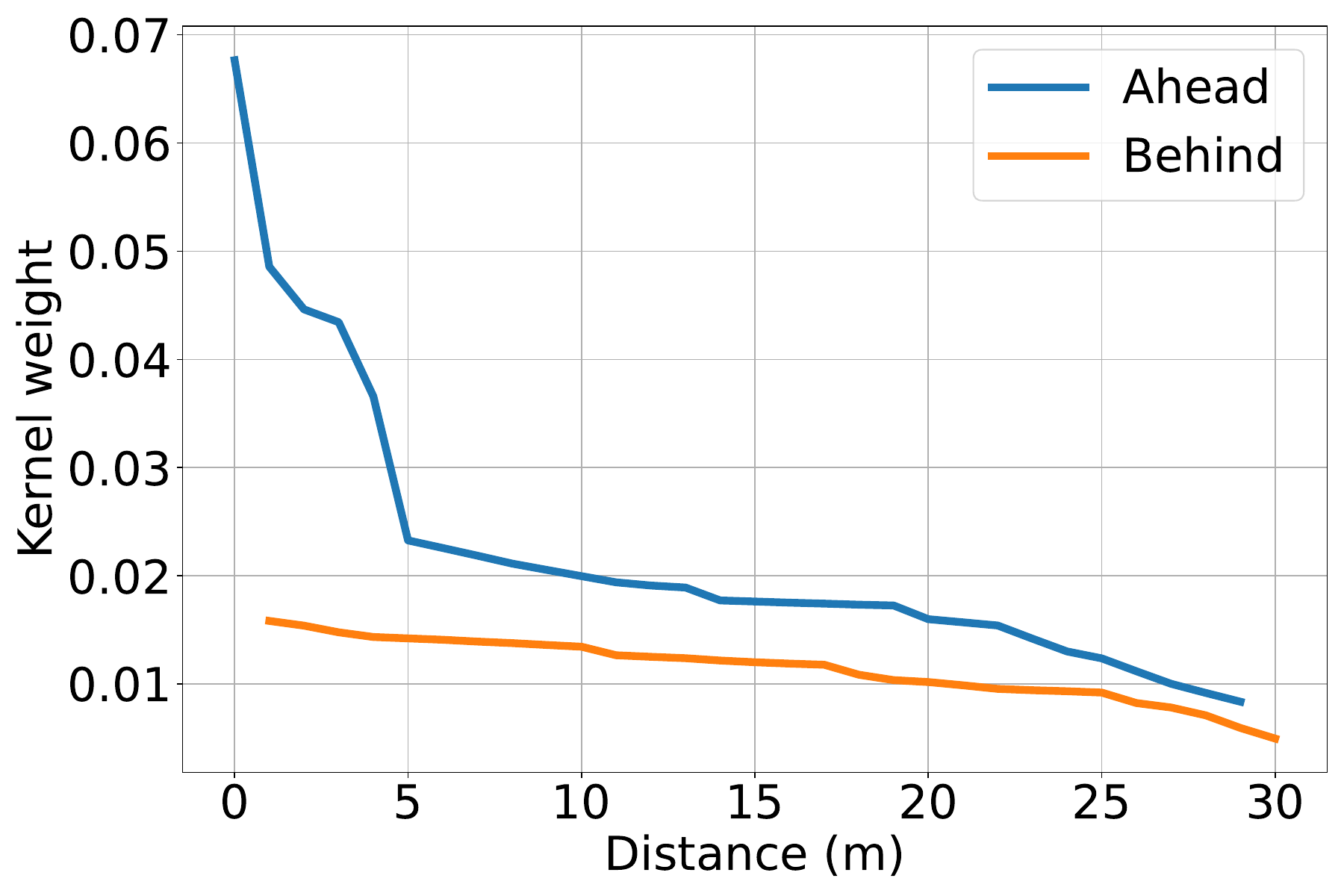}}
    \caption{Fundamental diagram (a)(b) and the learned looking-ahead looking-behind weights (c) for the car-following model~\eqref{eq:micro nudging pos}.}  \label{fig:FD HV look-behind}
\end{figure}

In Fig.~\ref{fig:rho HV nudging}, we compare the results for the macroscopic dynamics corresponding to the car-following model with looking-behind effects in~\eqref{eq:micro nudging pos}. From the wave propagation in Fig.~\ref{fig:rho HV nudging} and the estimation error in Table~\ref{tab:rho error HV}, the looking-ahead looking-behind non-local model still provides a more accurate estimation of traffic flow than the local LWR model. From the fundamental diagram in Fig.~\ref{fig:FD HV look-behind}, the speed vs non-local density in Fig.~\ref{fig:FD HV look-behind}(b) presents a smaller scatter range than the speed vs local density in Fig.~\ref{fig:FD HV look-behind}(a). For the looking-ahead weights in Fig.~\ref{fig:FD HV look-behind}(c), we see that a majority of the looking-ahead weights are within 5 meters, and the looking-ahead weights decelerate quickly after 5 meters. This is similar to the trend for the car-following result in Fig.~\ref{fig:FD HV acc}(c), since in the looking-behind model~\eqref{eq:micro nudging pos}, the vehicle only considers one preceding vehicle's speed. The looking-behind kernel weights account for only a small proportion, as shown in Fig.~\ref{fig:FD HV look-behind}(c). This is because the feedback gain corresponding to the follower speed's $\beta_1$ is relatively small compared with the feedback gain of the leading vehicle $\beta$, $\alpha$.

\begin{table}[!t]
    \centering
    \caption{The error (\%) when non-local LWR and LWR is adopted as the physics constraint}
    \label{tab:rho error HV}
    \begin{tabular}{c|c|c|c}
        & Car-following~\eqref{eq:micro CF} & Look-ahead~\eqref{eq:micro look ahead} & Look-behind~\eqref{eq:micro nudging pos}  \\ \hline
        PINN with non-local LWR  &  0.05 & 0.13 & 0.11 \\ \hline
        PINN with LWR  & 0.90 & 1.20 & 1.61 \\ \hline
    \end{tabular}
\end{table}

\subsection{Feedback gain analysis}\label{subsec:gain}

\subsubsection{ACC gain}
\begin{figure}[!t]
    \centering
    \subcaptionbox{Estimation error}{\includegraphics[width=0.3\linewidth]{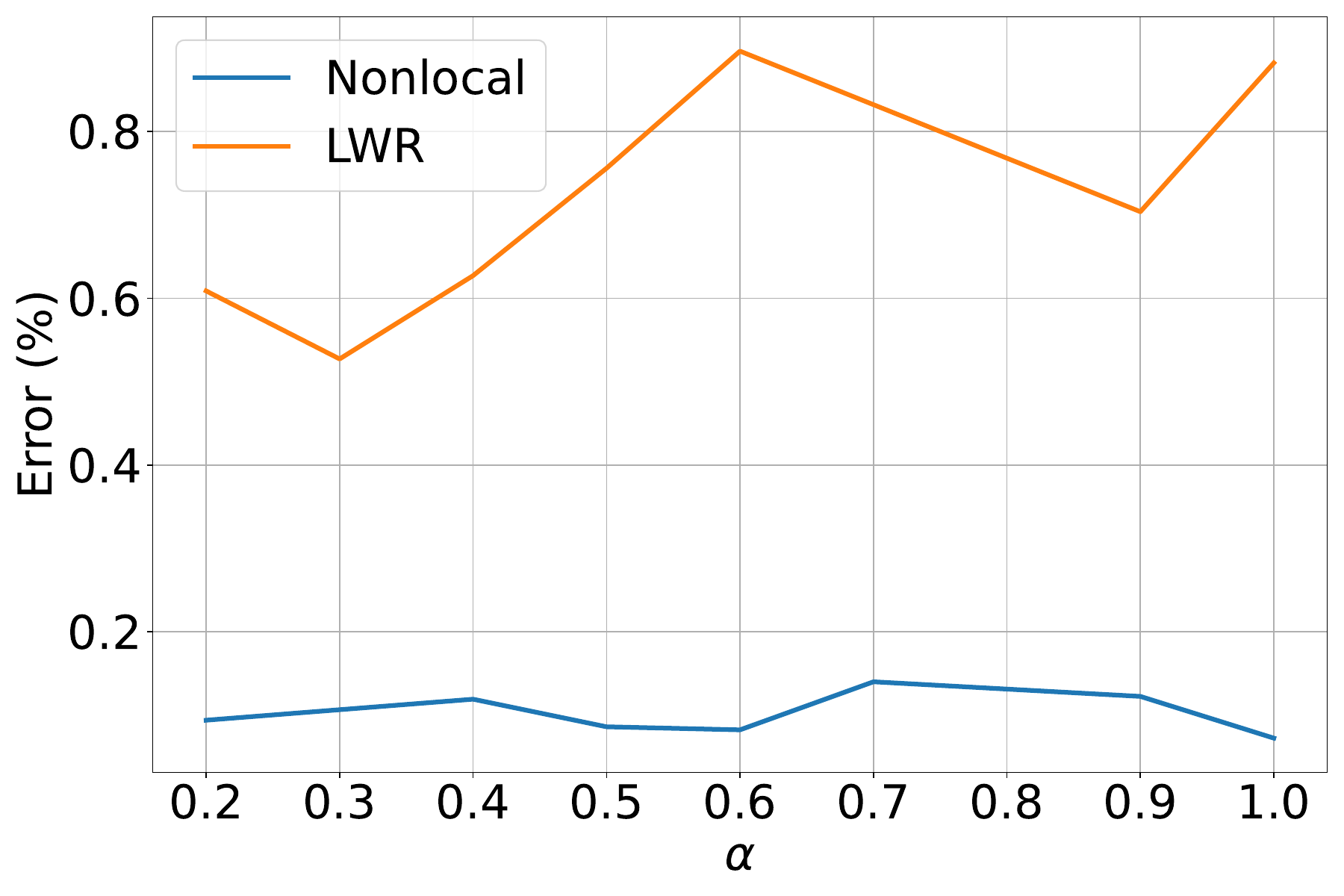}}
    \subcaptionbox{Non-local kernel}{\includegraphics[width=0.3\linewidth]{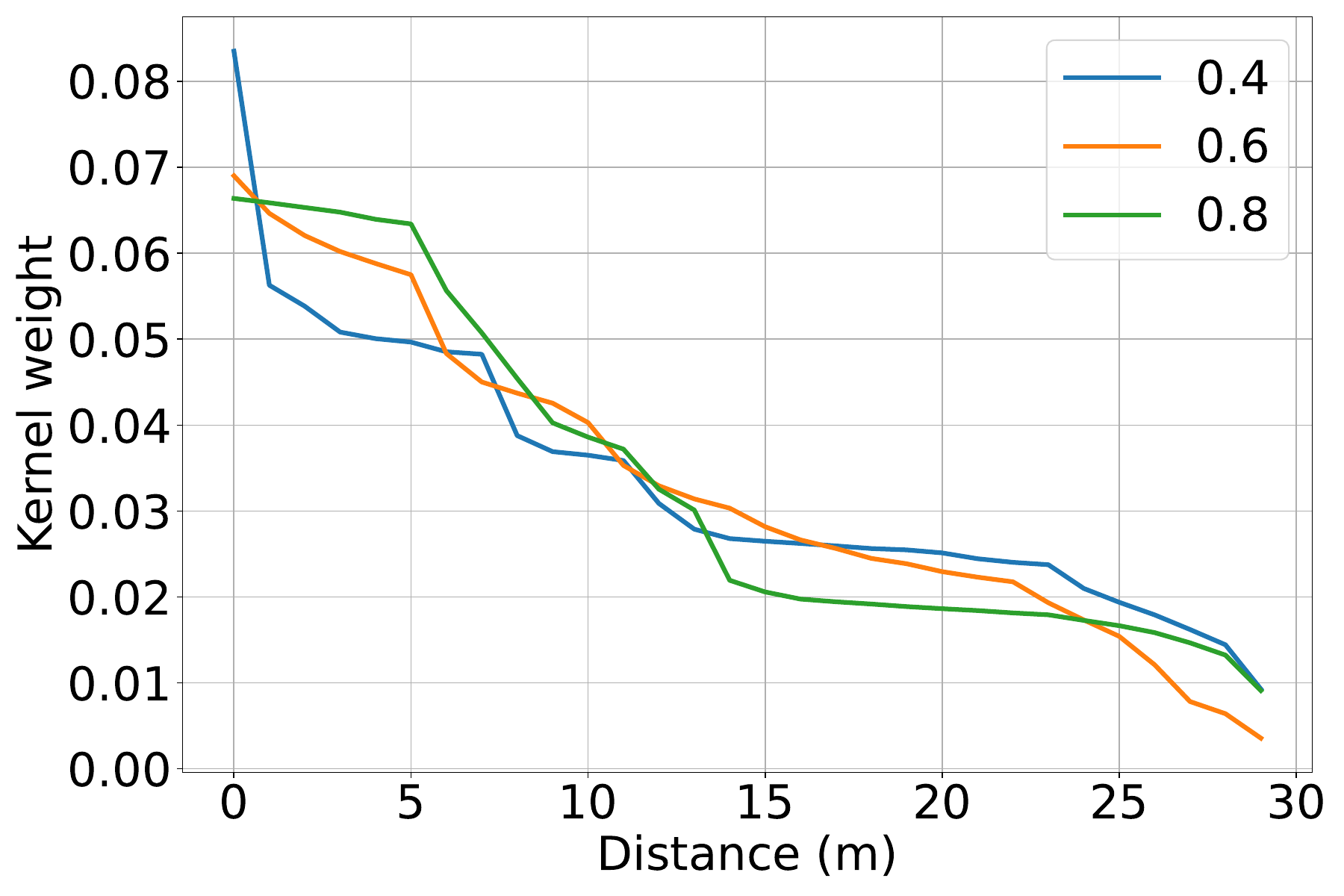}}
    \caption{Simulation results for different speed feedback gain $\alpha_0$ values. (a): The estimation error with LWR model and non-local model. (b): the looking-ahead kernel comparison for different $\alpha_0$.}
    \label{fig:AV acc alpha}
\end{figure}

\begin{figure}[!t]
    \centering
    \subcaptionbox{Local, $\alpha_0=0.4$}{\includegraphics[width=0.24\linewidth]{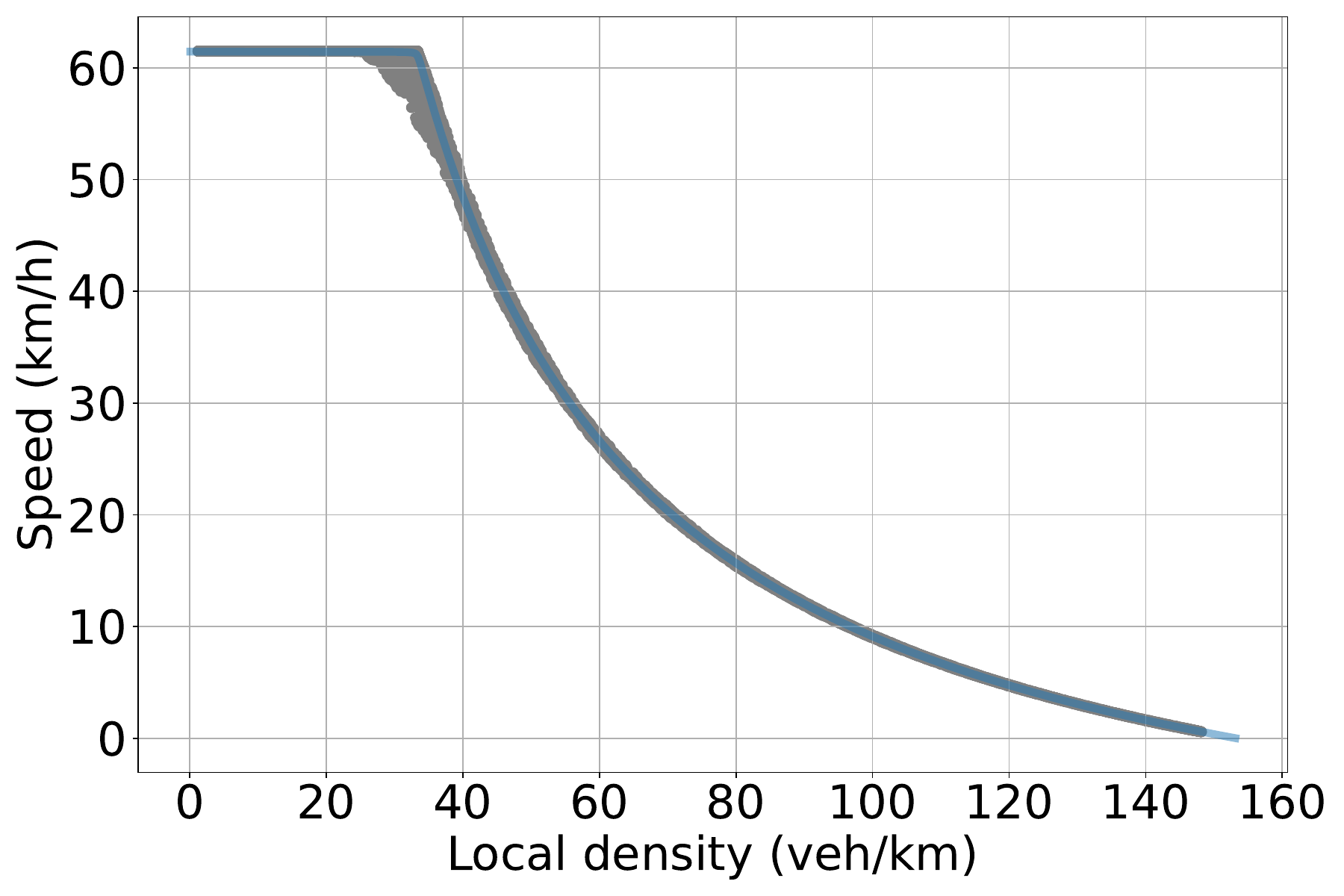}}
    \subcaptionbox{
    Non-local, $\alpha_0=0.4$}{\includegraphics[width=0.24\linewidth]{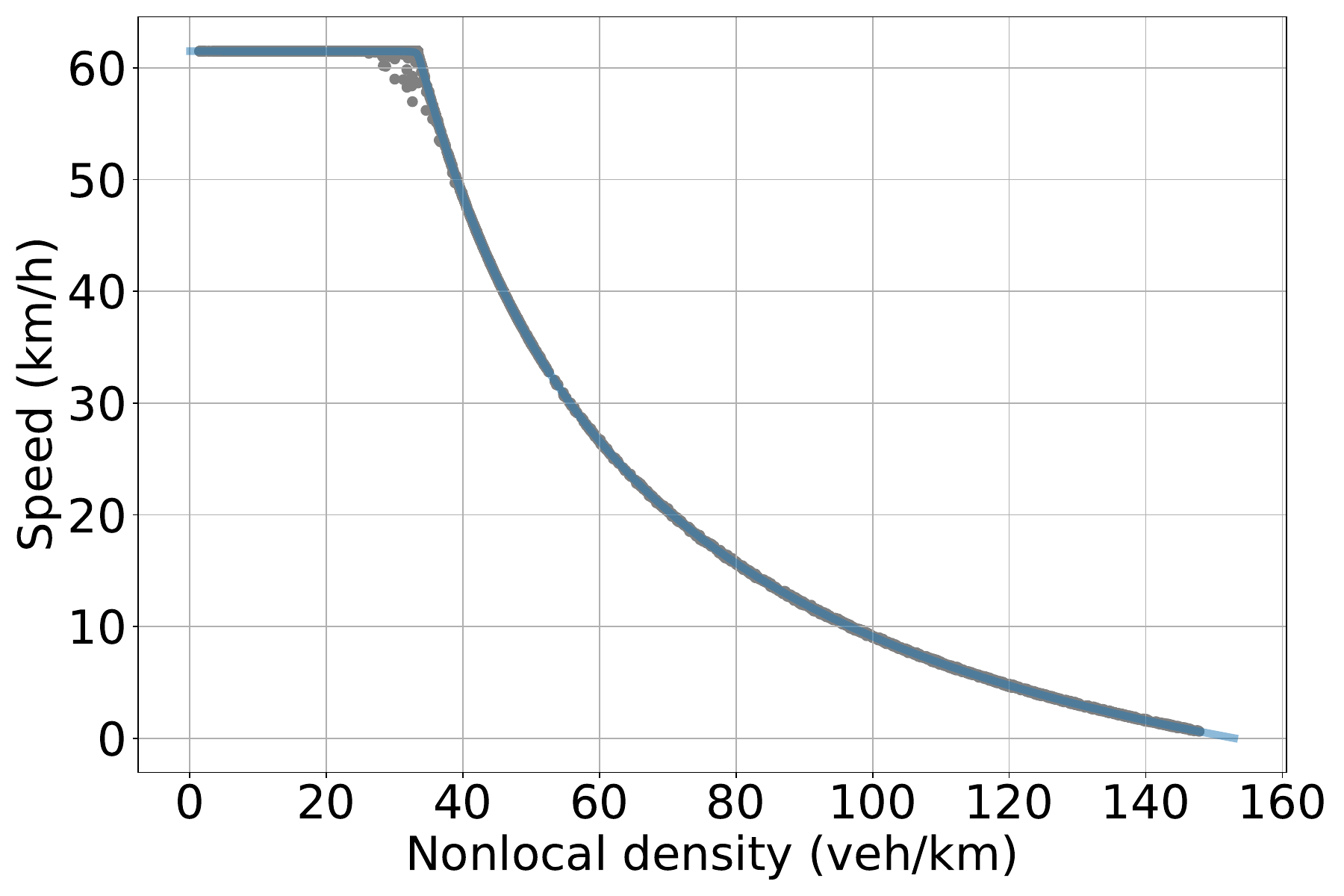}}
    \subcaptionbox{Local, $\alpha_0=0.8$}{\includegraphics[width=0.24\linewidth]{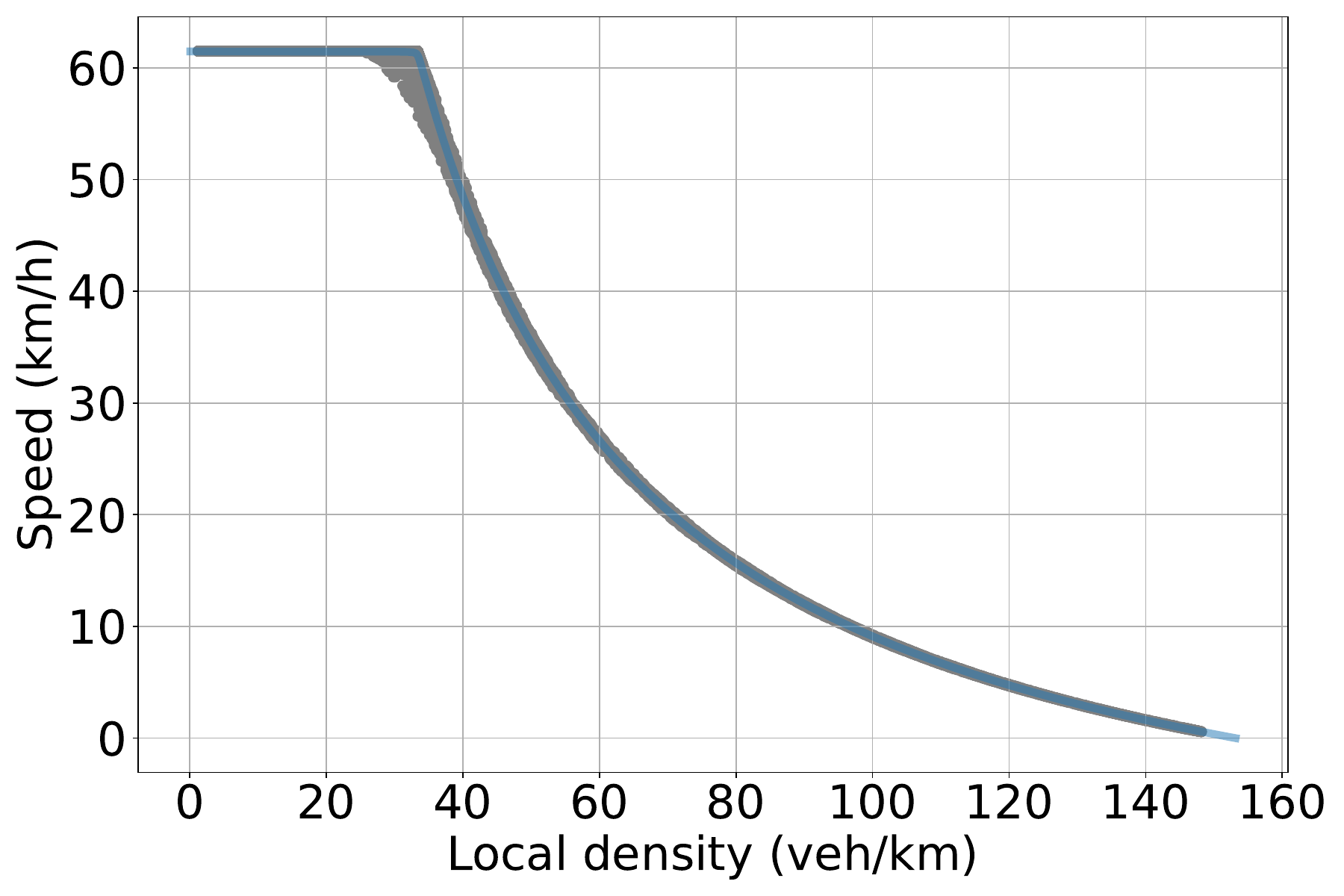}}
    \subcaptionbox{Nonlocal, $\alpha_0=0.8$}{\includegraphics[width=0.24\linewidth]{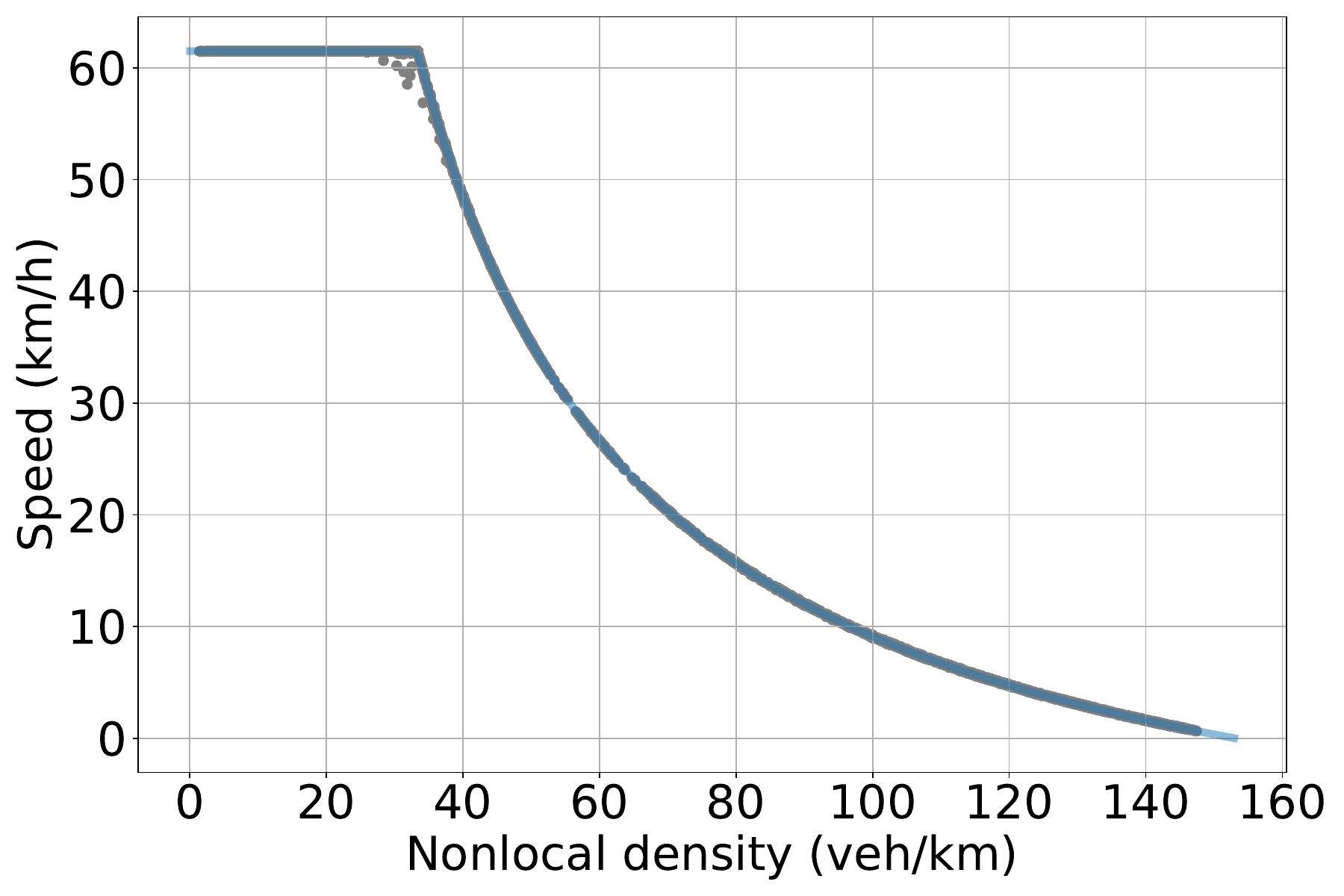}}
    \caption{The fundamental diagram for different speed feedback gain $\alpha_0$.}\label{fig:FD AV acc alpha}
\end{figure}

For the gap feedback $\alpha_0$, we run simulations with $\alpha_0$ varying from 0.2 to 1. We give the estimation error with different $\alpha_0$ values as in Fig.~\ref{fig:AV acc alpha}(a). For all considered $\alpha_0$ values, the looking-ahead non-local model gives an error less than 0.2\%. This indicates that the non-local model provides an accurate description of the traffic dynamics.  Fig.~\ref{fig:AV acc alpha}(b) compares looking-ahead kernel weights for different $\alpha_0$ values.  We see that for all $\alpha_0$ values, a majority of the weights still lie within the $0 \le x \le 5$ meters range. A second trend is that with the increase of $\alpha_0$, more weights are allocated to downstream traffic. Comparing the three kernels, with the increase of $\alpha_0$, the weights within 5 meters decrease, and the weights for $5\le x\le 20$~m increase. For example, the weight at $x=0$ is 0.085 for $\alpha_0=0.4$, and decreases to 0.07 and 0.065 for $\alpha_0=0.6$ and $\alpha_0=0.8$ respectively. And for the weights at $x=5$ meters, $\alpha_0=0.8$ has the highest weight, and $\alpha_0=0.4$ corresponds to the smallest value. We give in Fig.~\ref{fig:FD AV acc alpha} the non-local fundamental diagram for different $\alpha_0$ values. The results show that the non-local fundamental diagram remains approximately the same for different $\alpha_0$ values.

\begin{figure}[!t]
    \centering
    \subcaptionbox{Estimation error}{\includegraphics[width=0.3\linewidth]{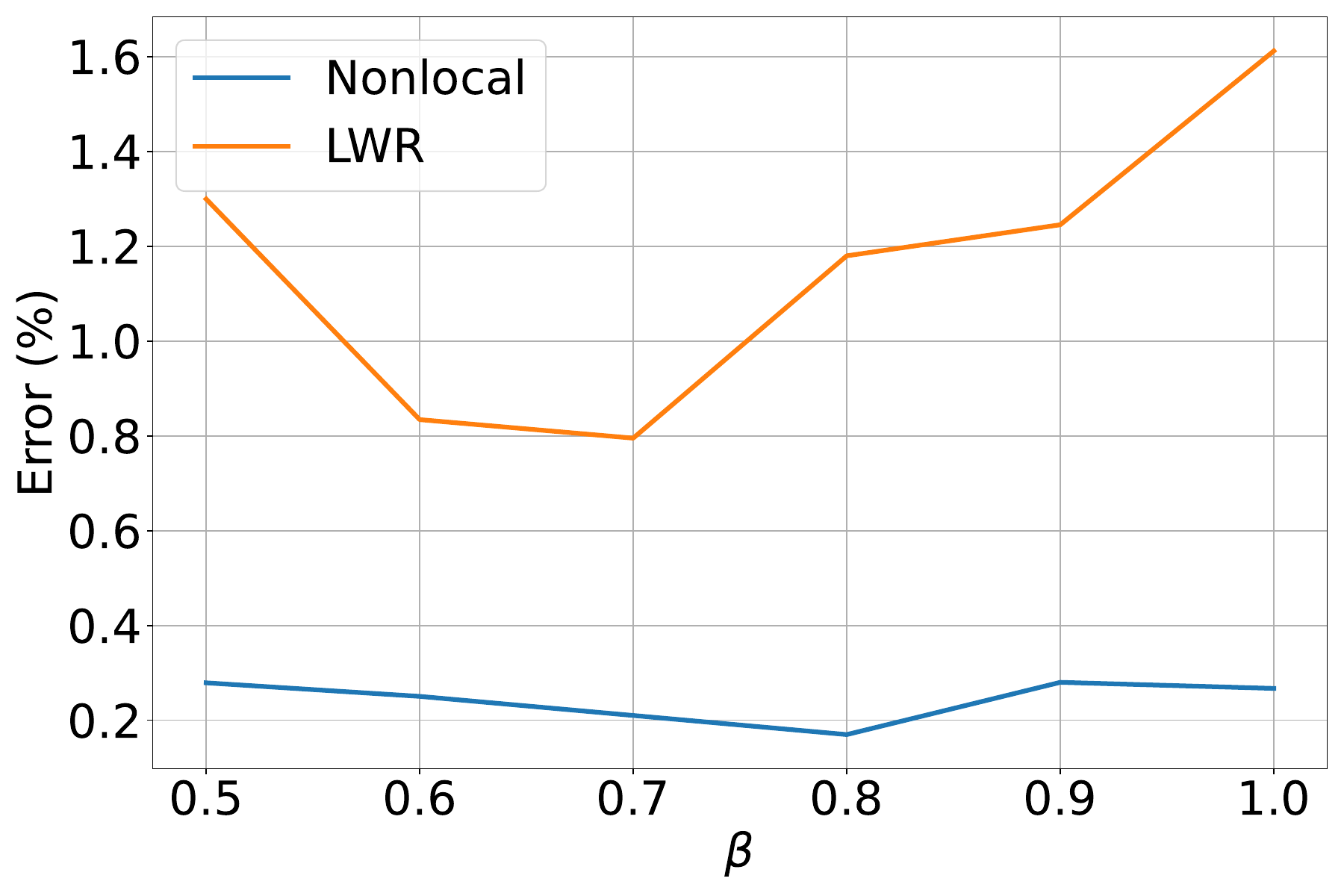}}
    \subcaptionbox{Non-local kernel}{\includegraphics[width=0.3\linewidth]{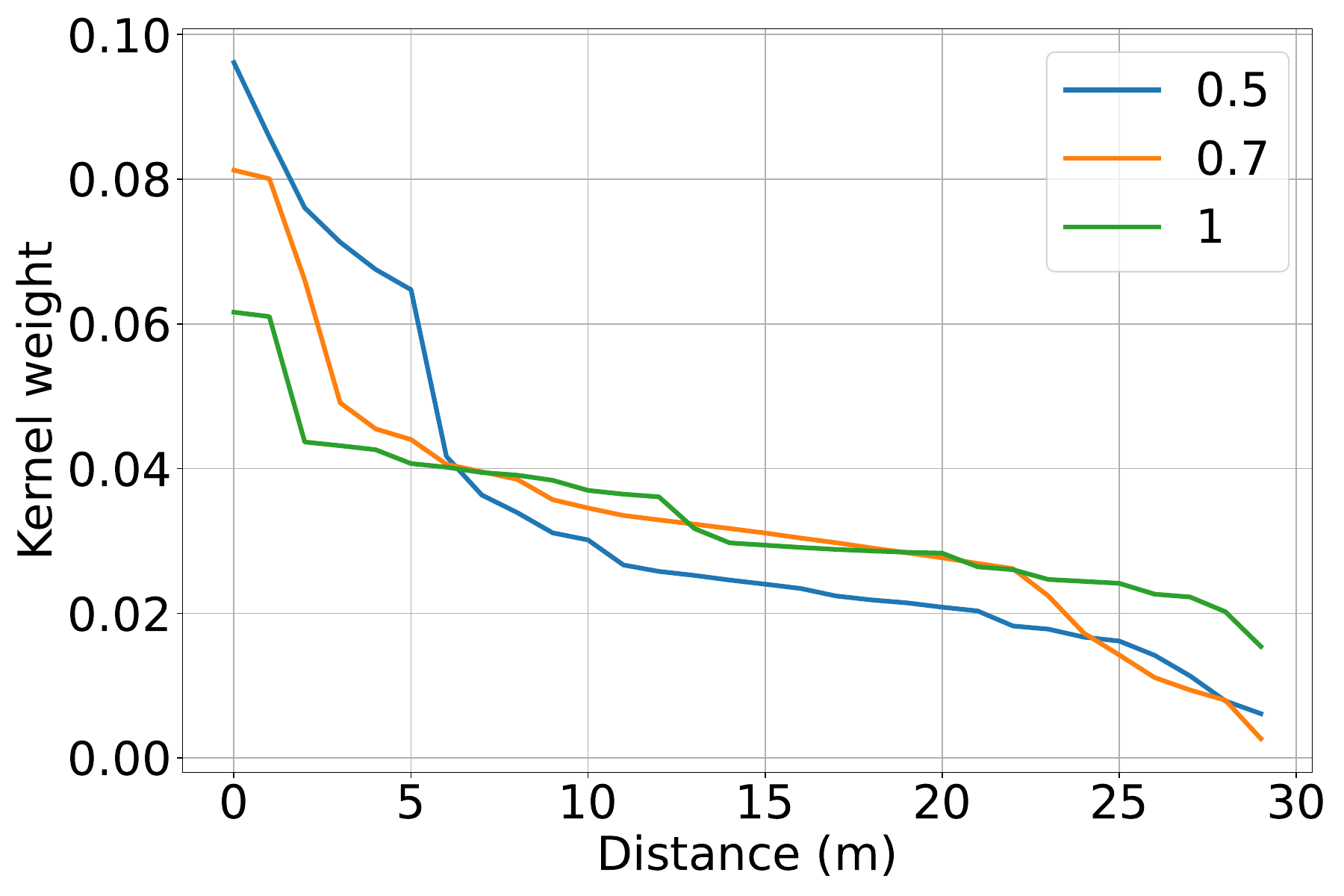}}
    \caption{Simulation results for different speed feedback gain $\beta_0$ values.}
    \label{fig:AV acc beta}
    \label{fig:FD AV acc beta}
\end{figure}

For the speed feedback gain $\beta_0$, we run simulations with $\beta_0$ ranging from 0.5 to 1. From the estimation error in Fig.~\ref{fig:AV acc beta}(a), we see that for all the considered $\beta_0$ values, the looking-ahead model has an estimation lower than 0.4\%, showing that the non-local model gives an accurate estimation of traffic. For the looking-ahead kernel weights shown in Fig.~\ref{fig:AV acc beta}(b), there are also two trends:  a majority of the non-local weights are within 5 meters, and more downstream weights are allocated with the increase of $\beta_0$. For example, for the weights at $x=0$, $\beta_0=0.5$ has the highest value, and $\beta_0=1$ has the lowest value. And at $x=10$ meters,  $\beta_0=1$  corresponds to the highest value while $\beta_0=0.5$ has the lowest weight. For the non-local fundamental diagram, we also find that the $\beta_0$ value also has no significant effect on the fundamental diagram. 

\subsubsection{Looking-ahead gain}

\begin{figure}[!t]
    \centering
    \subcaptionbox{Estimation error}{\includegraphics[width=0.3\linewidth]{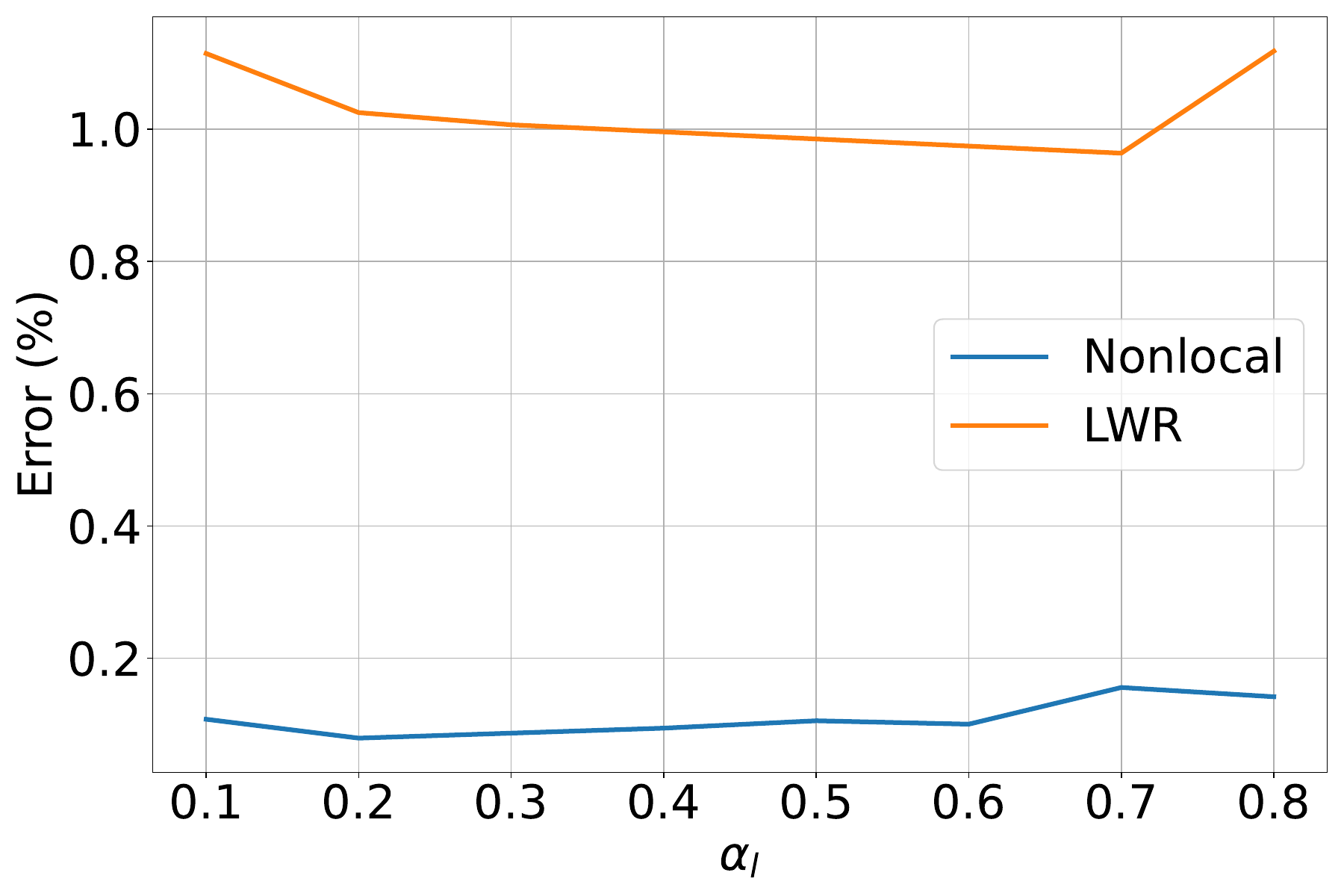}}
    \subcaptionbox{Non-local kernel}{\includegraphics[width=0.3\linewidth]{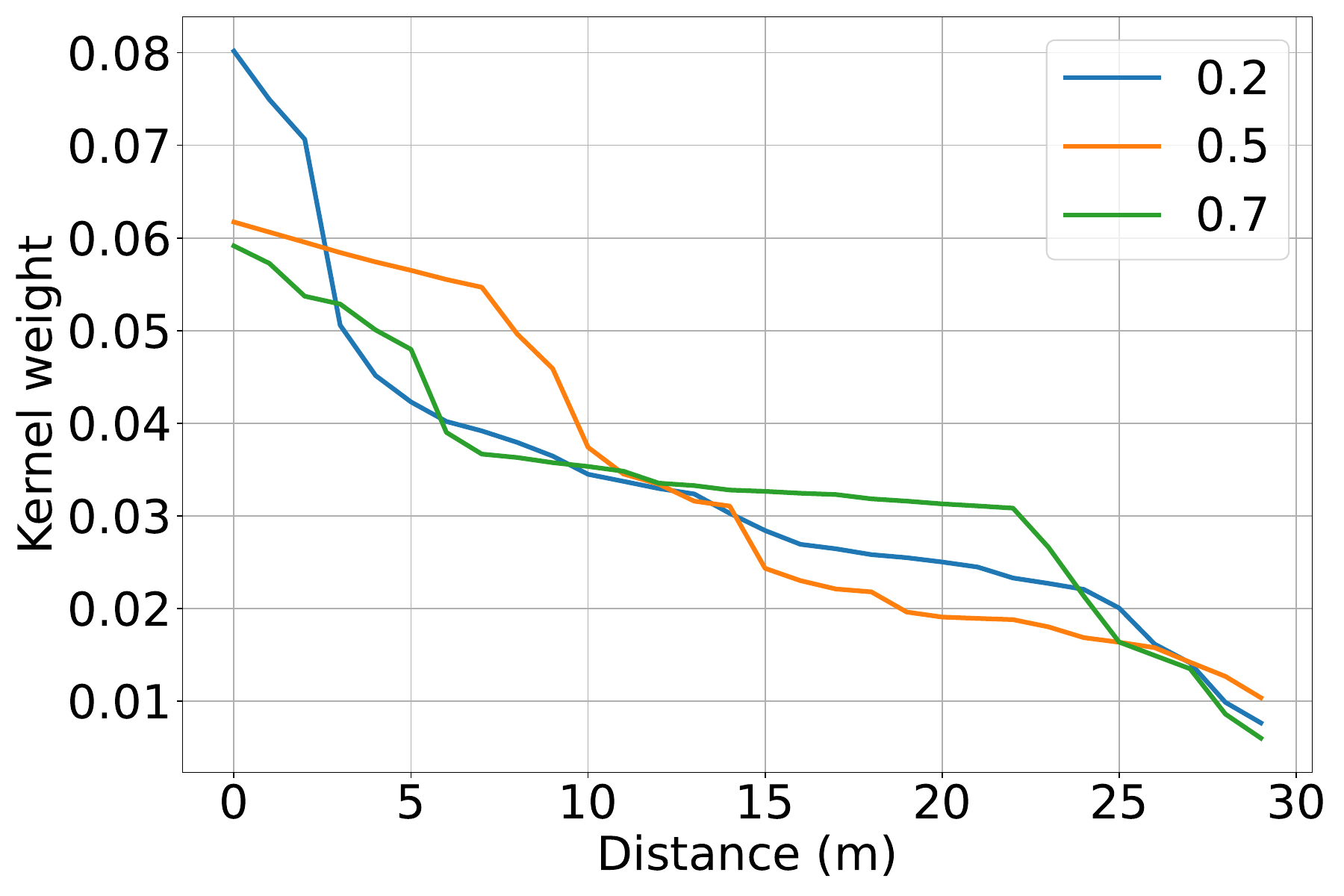}}
    \caption{Simulation results for different looking-ahead gap feedback gain  $\alpha_{-1}$.}
    \label{fig:AV ahead alpha}\label{fig:FD AV ahead alpha}
\end{figure}

To present a clear analysis on how the looking-ahead gain affects macroscopic traffic, we only consider the feedback of two leading vehicles $i=-1,-2$. And there are two looking-ahead gains: $\beta_{-1}$ for speed of the leading vehicle $i=-2$ and $\alpha_{-1}$ for gap of the leading vehicle $i=-1$. We run simulations with $\alpha_{-1}$ ranging from 0.1 to 0.8. The estimation error with varying $\alpha_{-1}$ values is given in Fig.~\ref{fig:AV ahead alpha}(a). For all considered $\alpha_{-1}$ gain values, the looking-ahead model constantly maintains a small error, showing that it provides an accurate description of traffic flow. Fig.~\ref{fig:AV ahead alpha}(b) shows the looking-ahead kernel weights for different $\alpha_{-1}$ values. Unlike the ACC case, the looking-ahead kernel now allocates more weights to downstream traffic and begins to sharply decrease around 5 to 25 meters. For example, for $\alpha_{-1}=0.7$, the kernel remains larger than 0.03 for $x\le 23$ meters, and only begins to decelerate sharply after 23 meters. And for $\alpha_{-1}=0.5$, the kernel weights begin to decrease around 15 meters, with the weights larger than 0.03 for $x\le 15 $ m. With a smaller $\alpha_{-1} = 0.2$, more weights are allocated to the local traffic around $x\le 5$ m. For the non-local fundamental diagram, similar to the ACC case, the looking-ahead gap feedback poses no clear change on the fundamental diagram.

\begin{figure}[!t]
    \centering
    \subcaptionbox{Estimation error}{\includegraphics[width=0.3\linewidth]{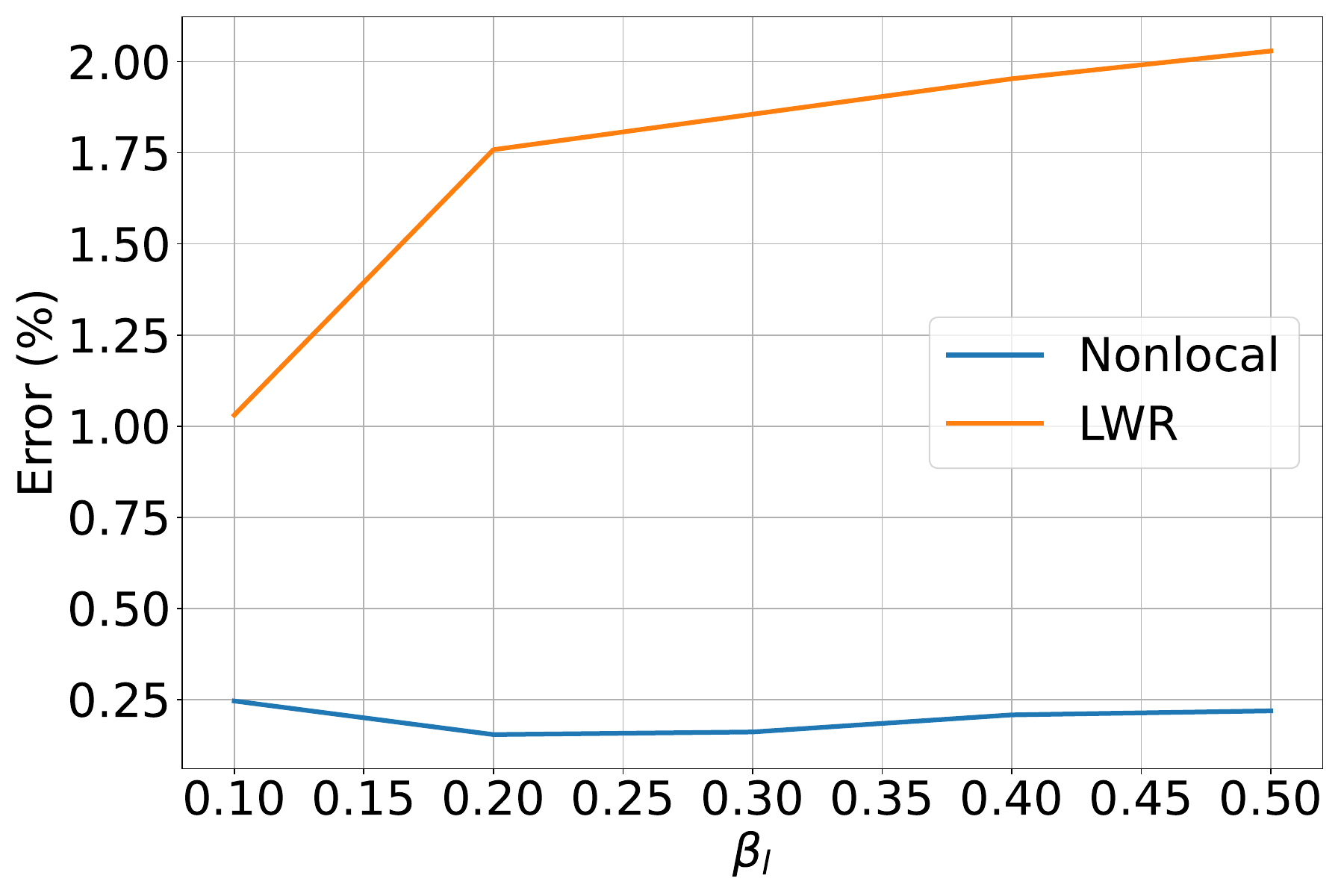}}
    \subcaptionbox{Non-local kernel}{\includegraphics[width=0.3\linewidth]{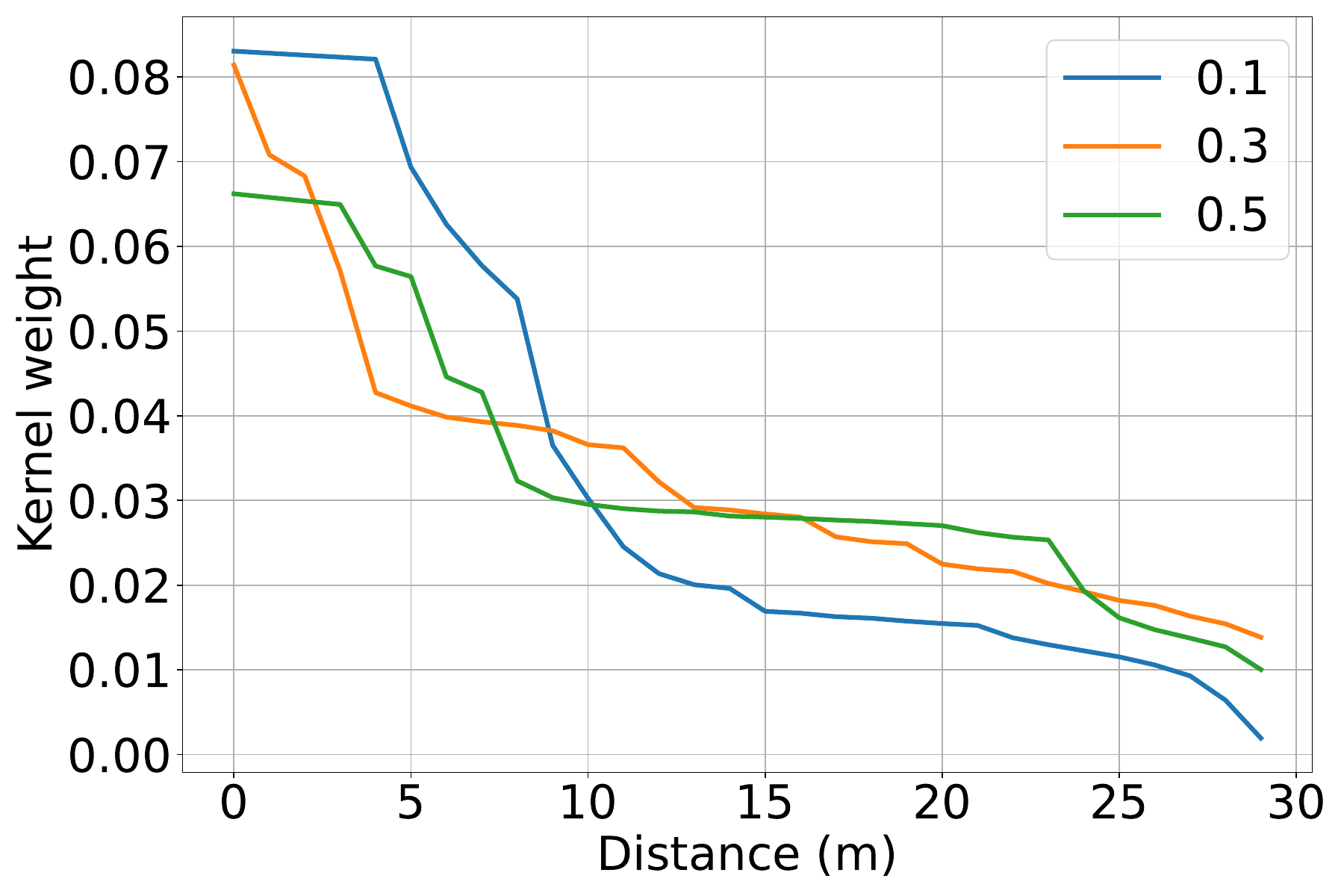}}
    \caption{Simulation results for different looking-ahead speed feedback gain $\beta_{-1}$.}
    \label{fig:AV ahead beta} \label{fig:FD AV ahead beta}
\end{figure}

Fig.~\ref{fig:AV ahead beta} analyzes the effect of looking-ahead speed feedback gain $\beta_{-1}$. As the estimation error in Fig.~\ref{fig:AV ahead beta}(a) shows, the looking-ahead model provides an accurate description of traffic dynamics, with the estimation error constantly lower than 0.25\%. For the looking-ahead weights as shown in Fig.~\ref{fig:AV ahead beta}(b), two trends are observed, similar to the case for $\alpha_{-1}$. The first one is that the kernel  has more  weights for downstream traffic compared with the ACC case. The second one is that with the increase of $\beta_{-1}$, more weights are allocated to the downstream traffic. For example, when $\beta_{-1}=0.5$, the kernel weights have two large decreases, one around 5 meters, and the second around 23 meters. And the weights remain around 0.03 within 10 to 23 meters. With a smaller $\beta_{-1} = 0.3$, the weight at $x=0$ increases, indicating more weights to the local traffic. And the weights have two sharp decelerations around 5 meters and 13 meters, which implies that fewer weights are allocated to downstream traffic. For the fundamental diagram, the the looking-ahead speed gain $\beta_{-1}$ also has only a marginal effect on the fundamental diagram.

\subsubsection{Looking-behind gain}
\begin{figure}[!t]
    \centering
    \subcaptionbox{Estimation error}{\includegraphics[width=0.3\linewidth]{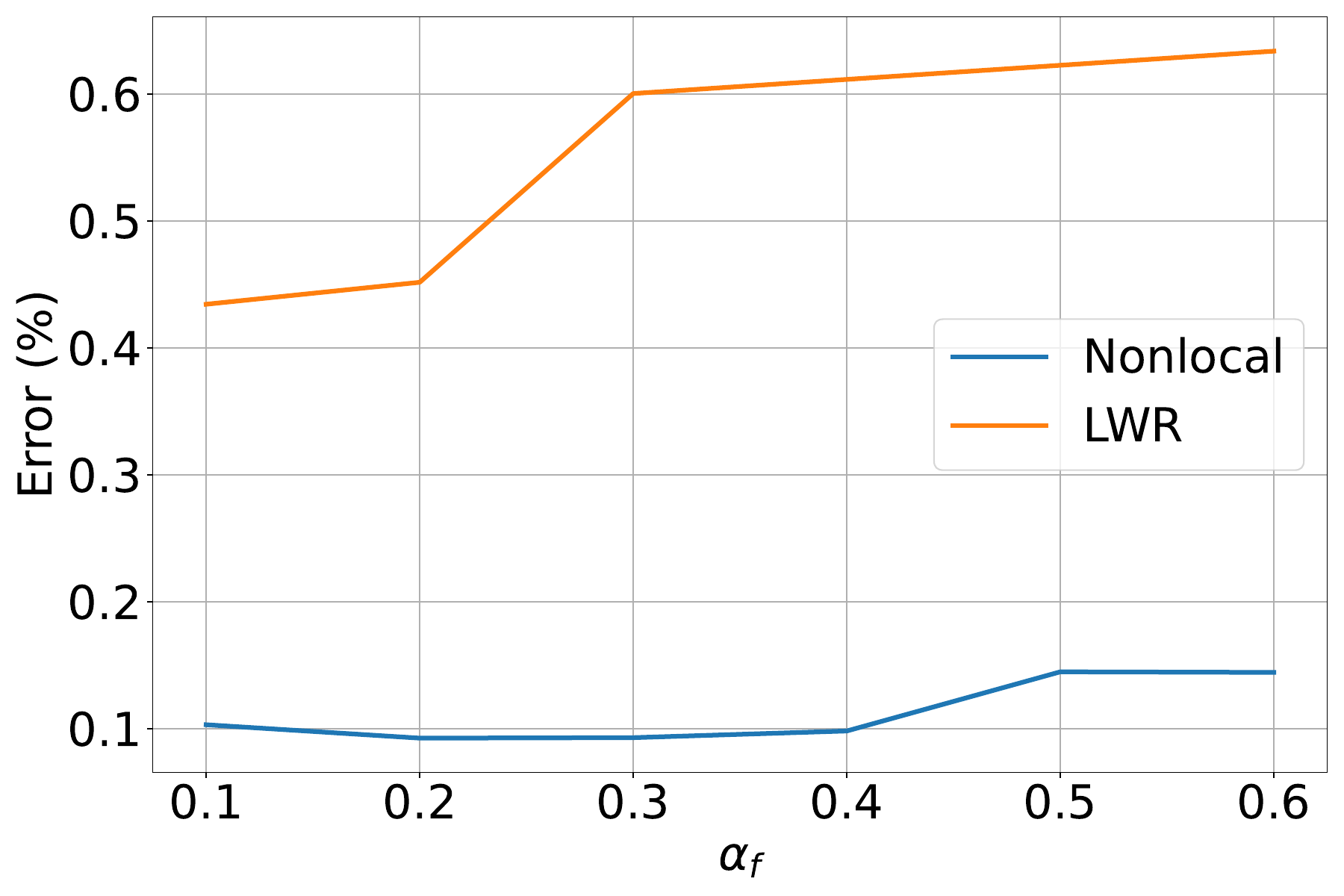}}
    \subcaptionbox{Non-local kernel}{\includegraphics[width=0.3\linewidth]{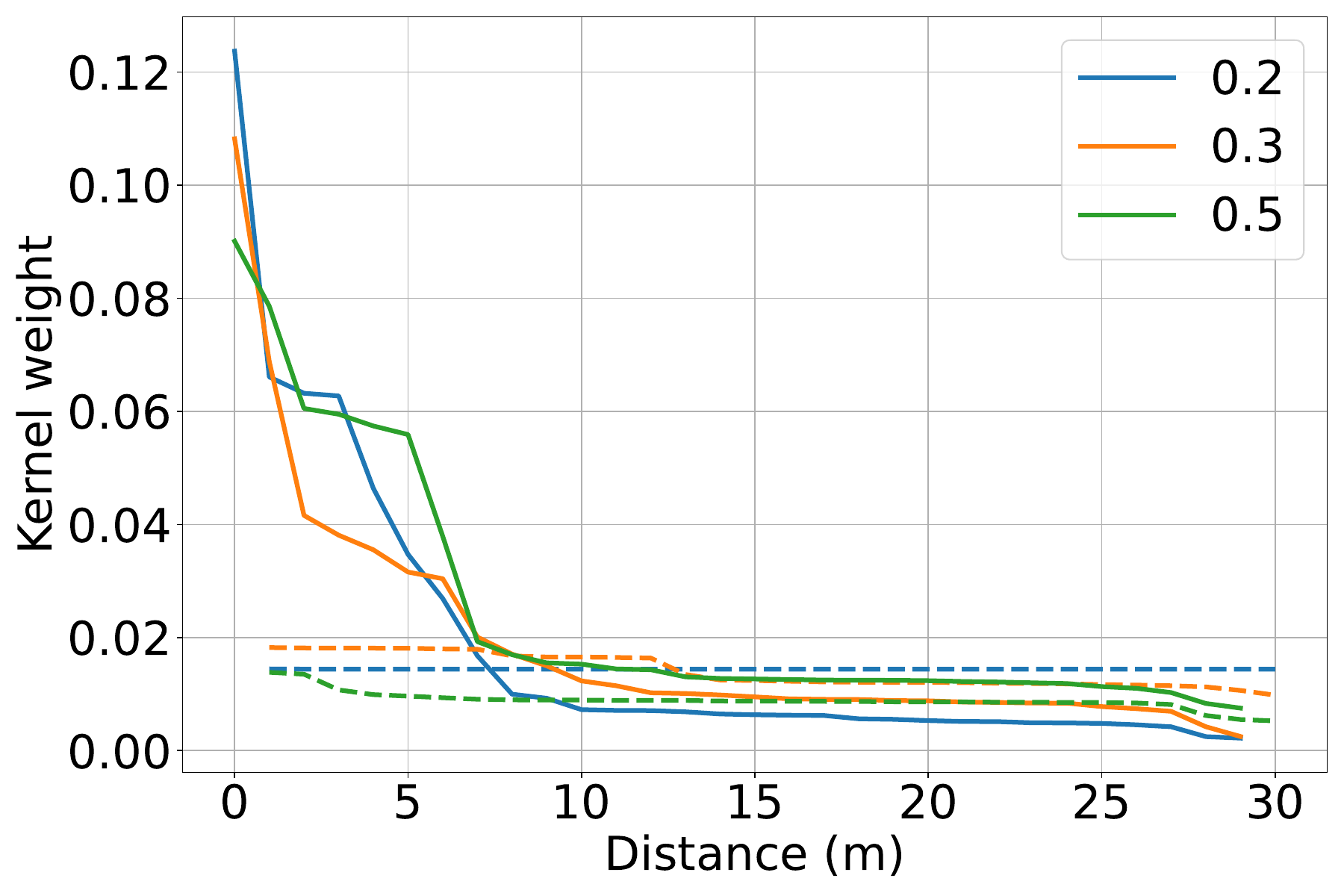}}
    \caption{Simulation results for different looking-behind gap feedback gain $\alpha_{1}$.}
    \label{fig:AV behind alpha}
\end{figure}

For the looking-behind analysis, we consider the case when the CAV includes the speed and gap of the following vehicle $i=1$. There are also two looking-behind gains: $\alpha_1$ and $\beta_1$  for the following vehicle's gap and speed respectively. 
We run the looking-ahead looking-behind non-local model with the non-local density being~\eqref{eq:nonlocal rho look-ahead-behind}. Fig.~\ref{fig:AV behind alpha} analyzes the effect of looking-behind gap feedback gain $\alpha_1$. The estimation error remains lower than 0.2\% as Fig.~\ref{fig:AV behind alpha}(a) shows, and the looking-ahead looking-behind non-local model provides an accurate description of the macroscopic dynamics. For the non-local weights in Fig.~\ref{fig:AV behind alpha}(b), the following trends are identified. First, for the looking-ahead kernel, a majority of the weights are within 5 meters. The looking-ahead kernel weights reduce sharply around 5 meters for all three $\alpha_1$ values. This agrees with the findings in the ACC case.   Second, the looking-behind kernel accounts for a small portion and remains almost constant.

\begin{figure}[!t]
    \centering
    \subcaptionbox{Estimation error}{\includegraphics[width=0.3\linewidth]{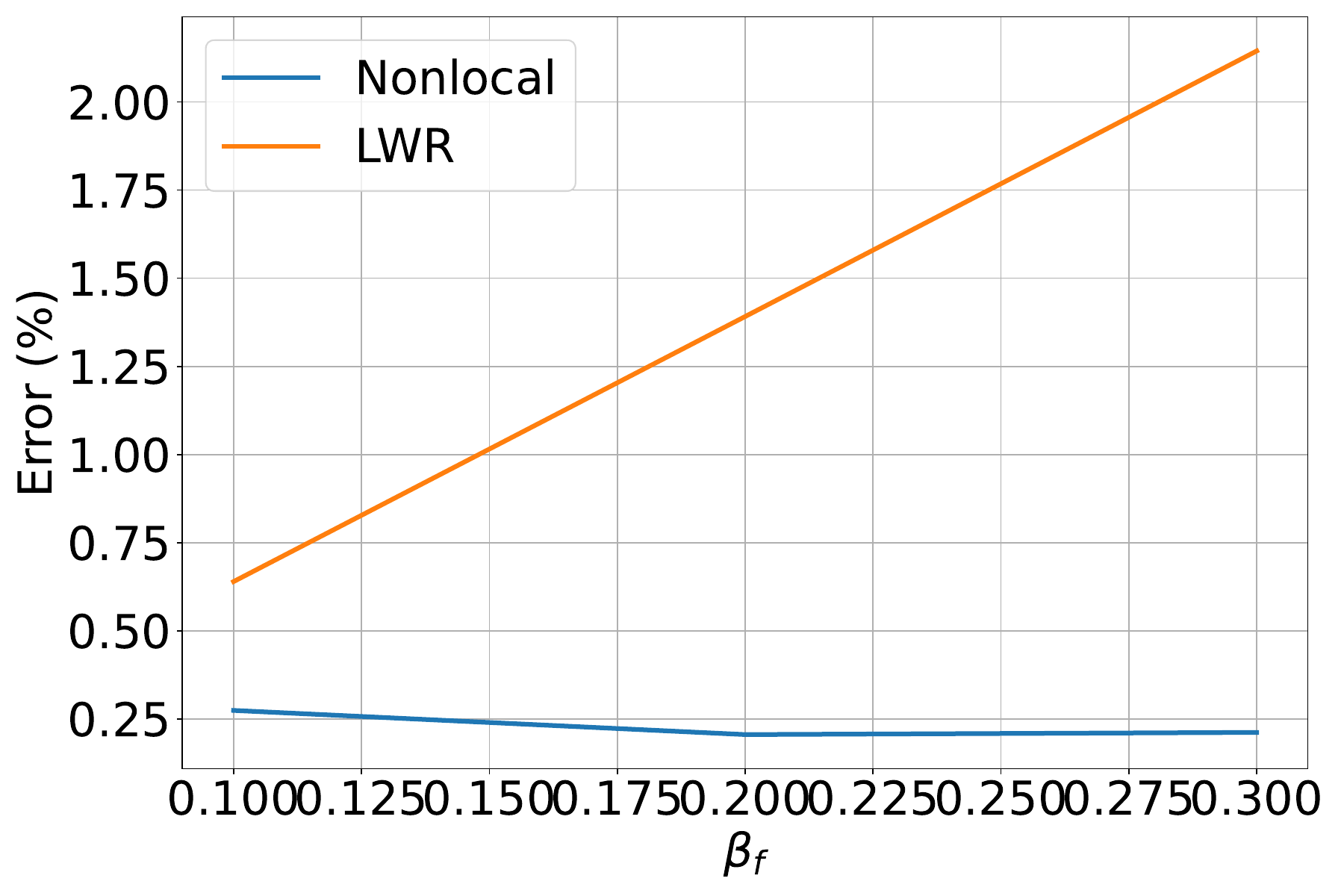}}
    \subcaptionbox{Non-local kernel}{\includegraphics[width=0.3\linewidth]{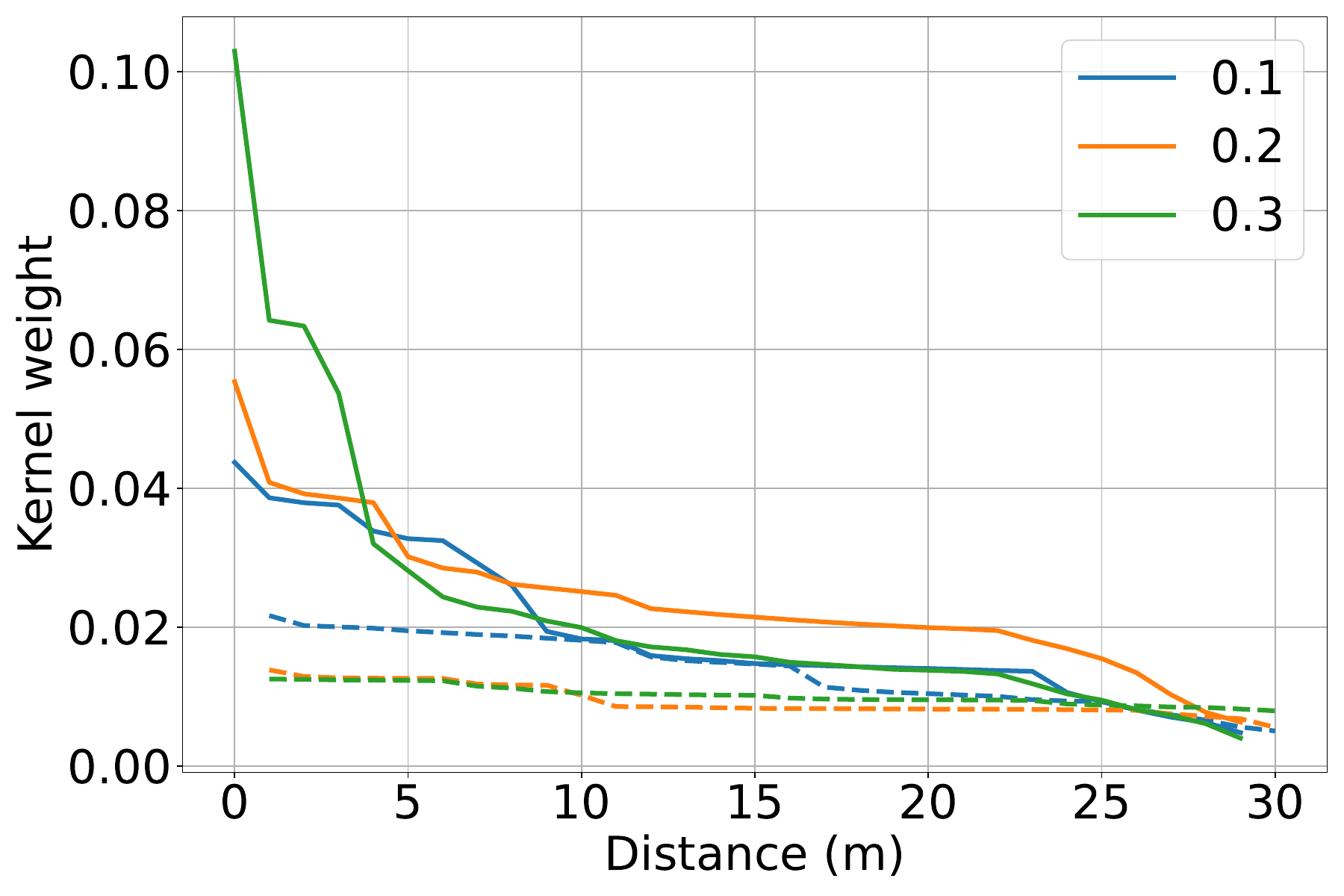}}
    \caption{Simulation results for different looking-behind speed feedback gain $\beta_{1}$.}
    \label{fig:AV behind beta}
\end{figure}

Fig.~\ref{fig:AV behind beta}  analyzes the effect of looking-behind speed feedback gain $\beta_1$. From the estimation error in Fig.~\ref{fig:AV behind beta}(a), the non-local LWR model constantly provides an accurate description of traffic flow dynamics. For the kernel weights, similar to the results for $\alpha_1$, two trends are identified from Fig.~\ref{fig:AV behind beta}(b): The looking-ahead weights decrease sharply after $x\ge 5$ meters, and the looking-behind kernel remains approximately the same. For the non-local fundamental diagram, we also find that the fundamental diagram has only small variations when $\beta_1$ changes.

\subsection{Optimal speed function analysis}\label{subsec:vs}

\begin{figure}[!t]
    \centering
    \subcaptionbox{Estimation error}{\includegraphics[width=0.3\linewidth]{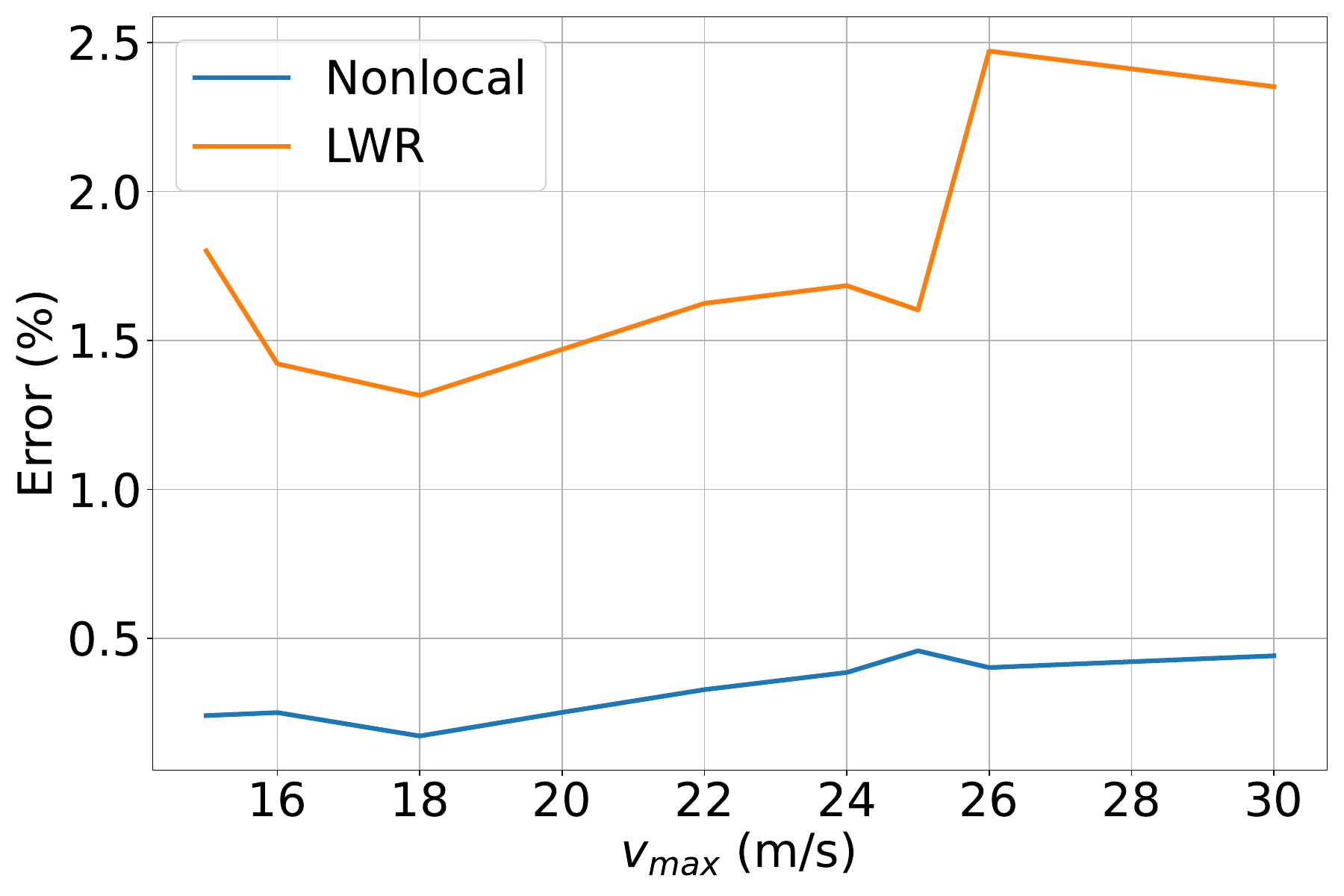}}
    \subcaptionbox{Non-local kernel}{\includegraphics[width=0.3\linewidth]{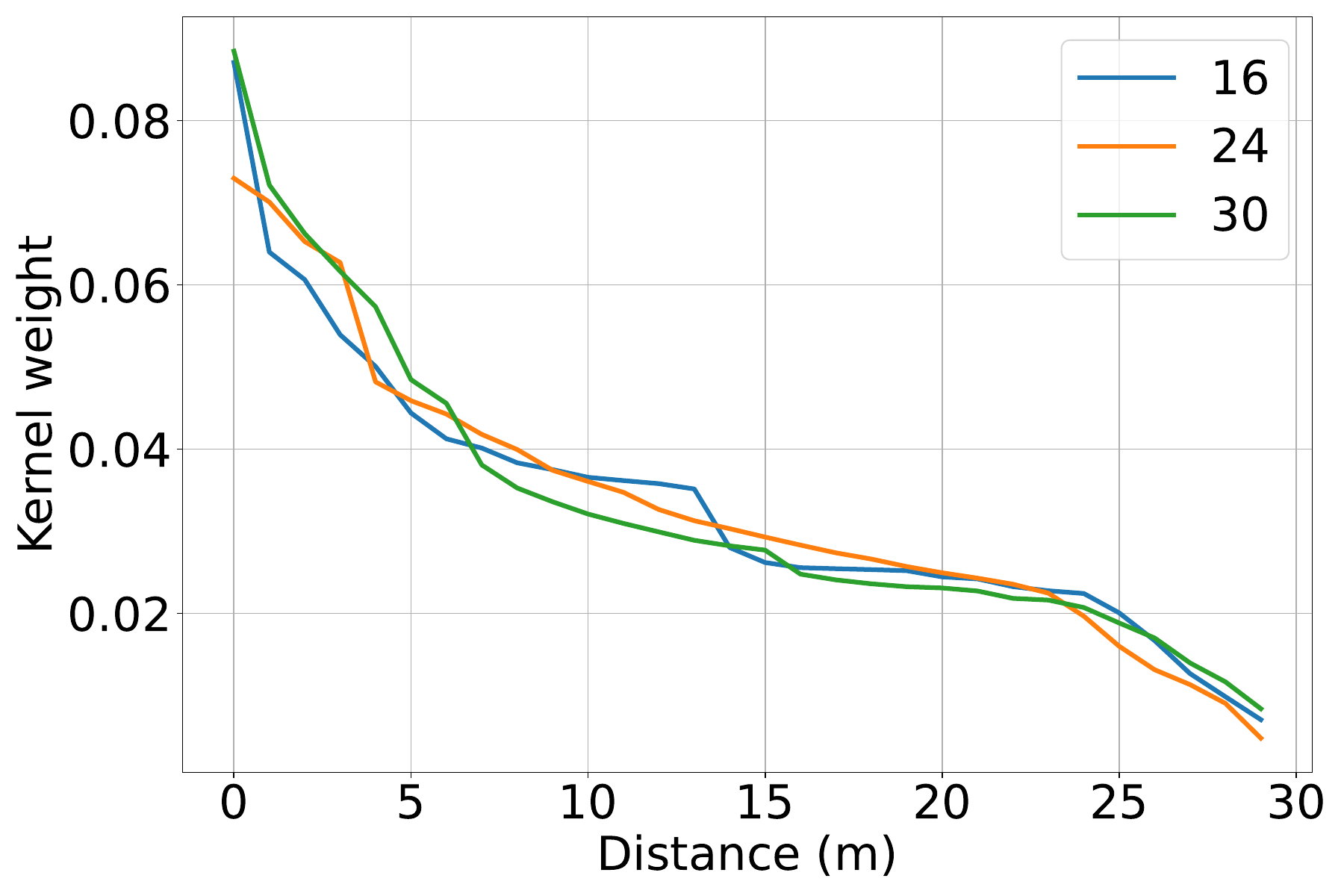}}
    \caption{Simulation results for different maximum speed  $v_{\max}$ values.}
    \label{fig:AV acc vmax}
\end{figure}

\begin{figure}[!t]
    \centering
    \subcaptionbox{Fundamental diagram}{\includegraphics[width=0.3\linewidth]{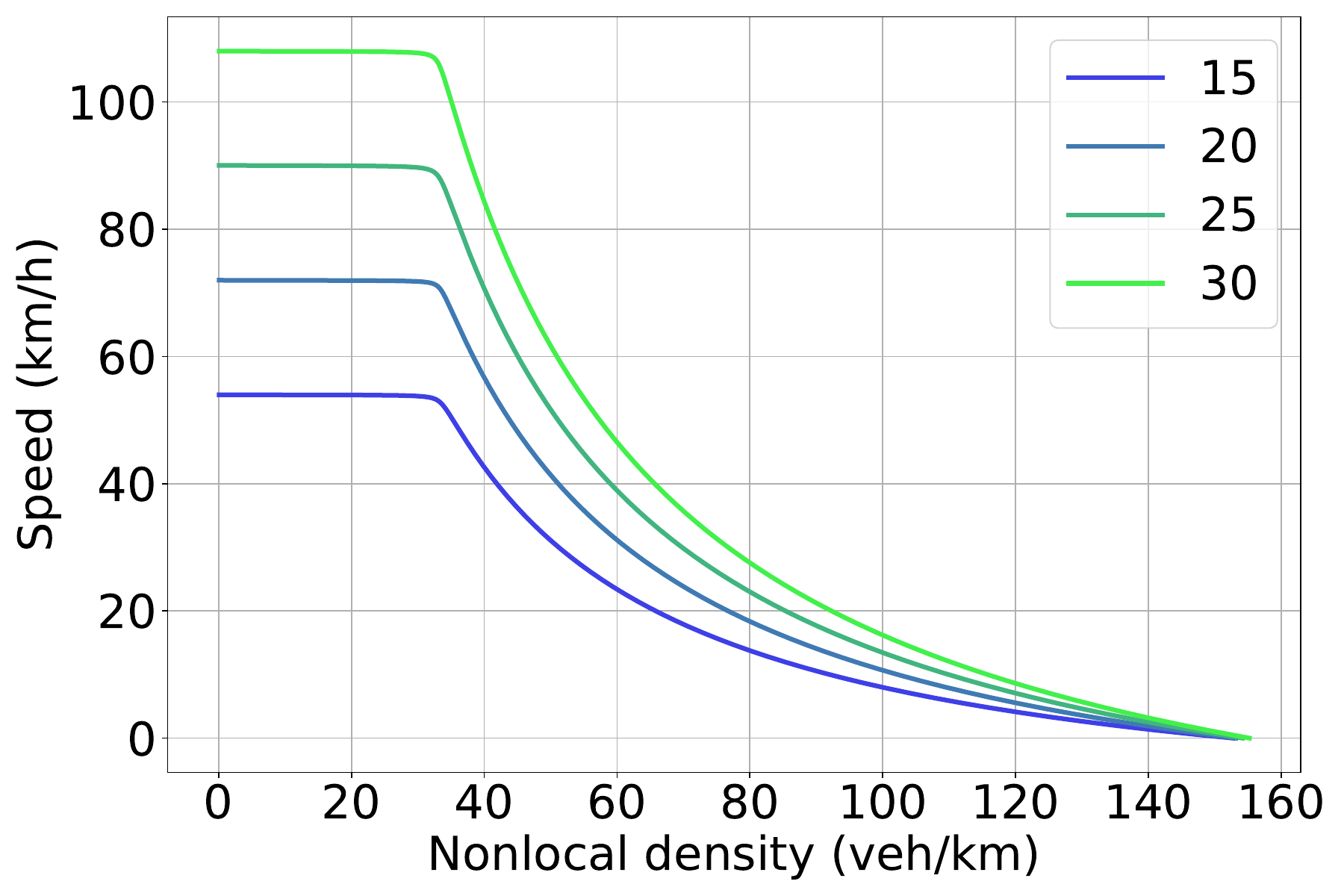}}
    \subcaptionbox{Jam density}{\includegraphics[width=0.3\linewidth]{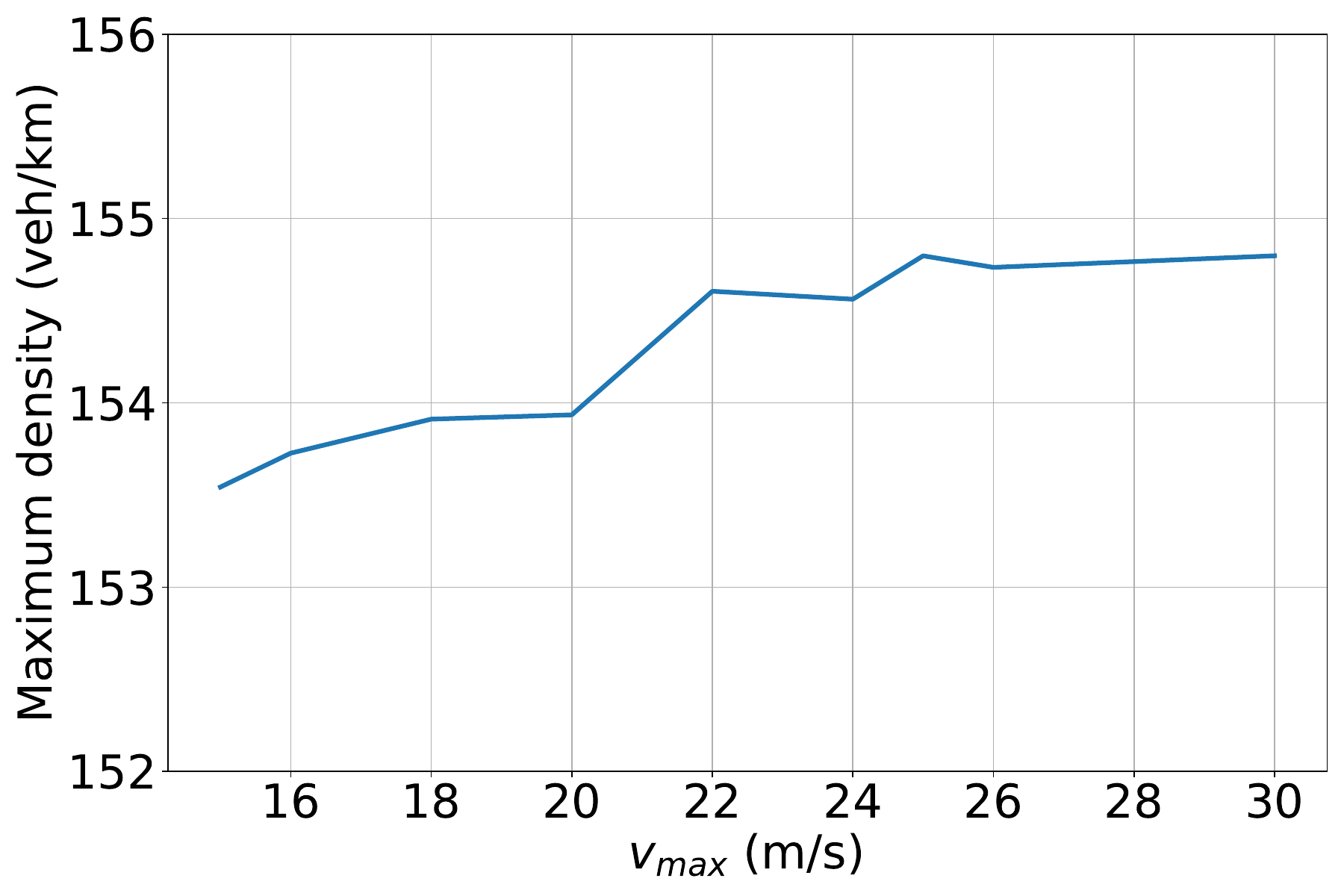}}
    \subcaptionbox{Free-flow speed}{\includegraphics[width=0.3\linewidth]{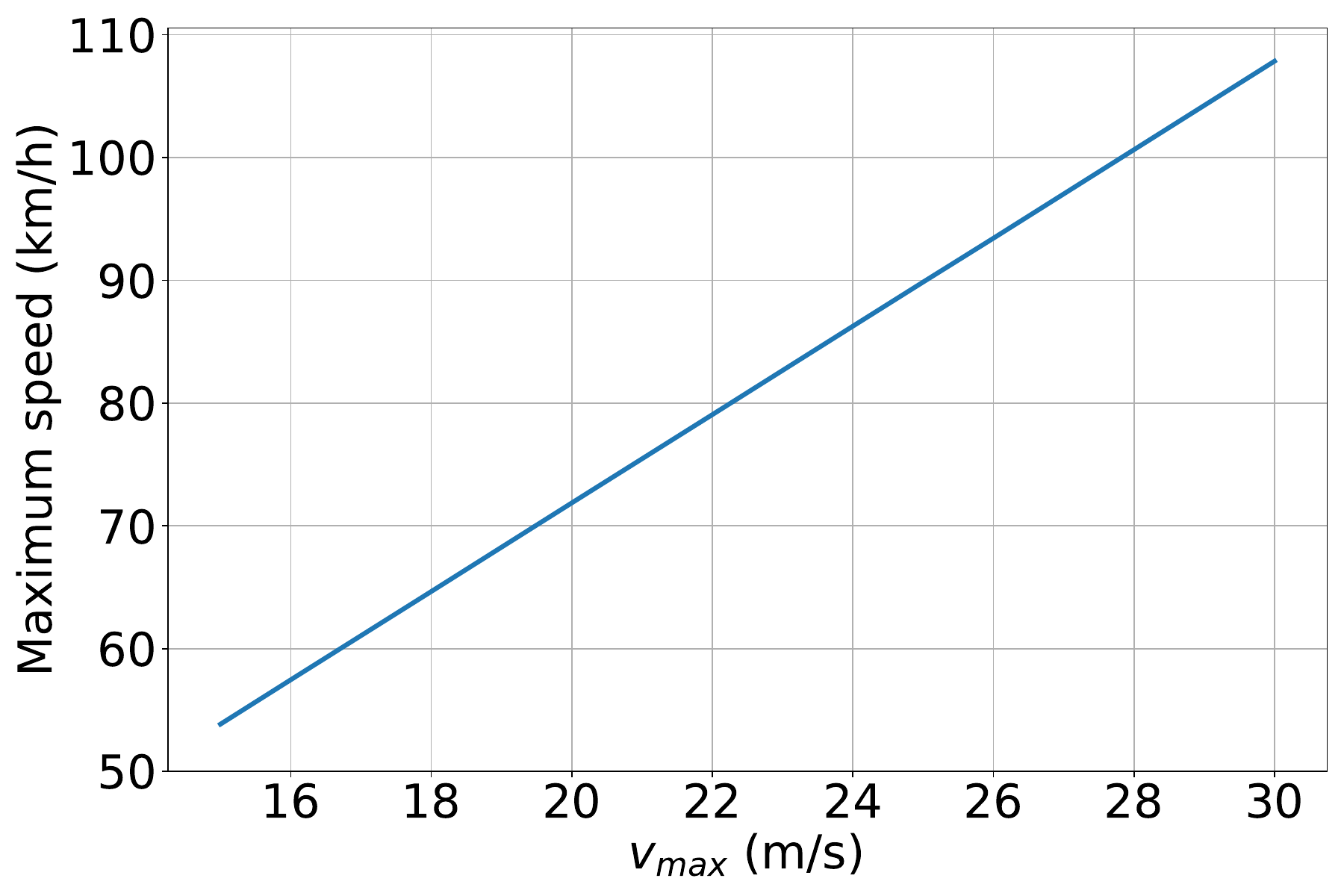}}
    \caption{Non-local fundamental diagram for different maximum speed  $v_{\max}$ values.}
    \label{fig:FD AV acc vmax}
\end{figure}

\begin{figure}[!t]
    \centering
    \subcaptionbox{Estimation error}{\includegraphics[width=0.3\linewidth]{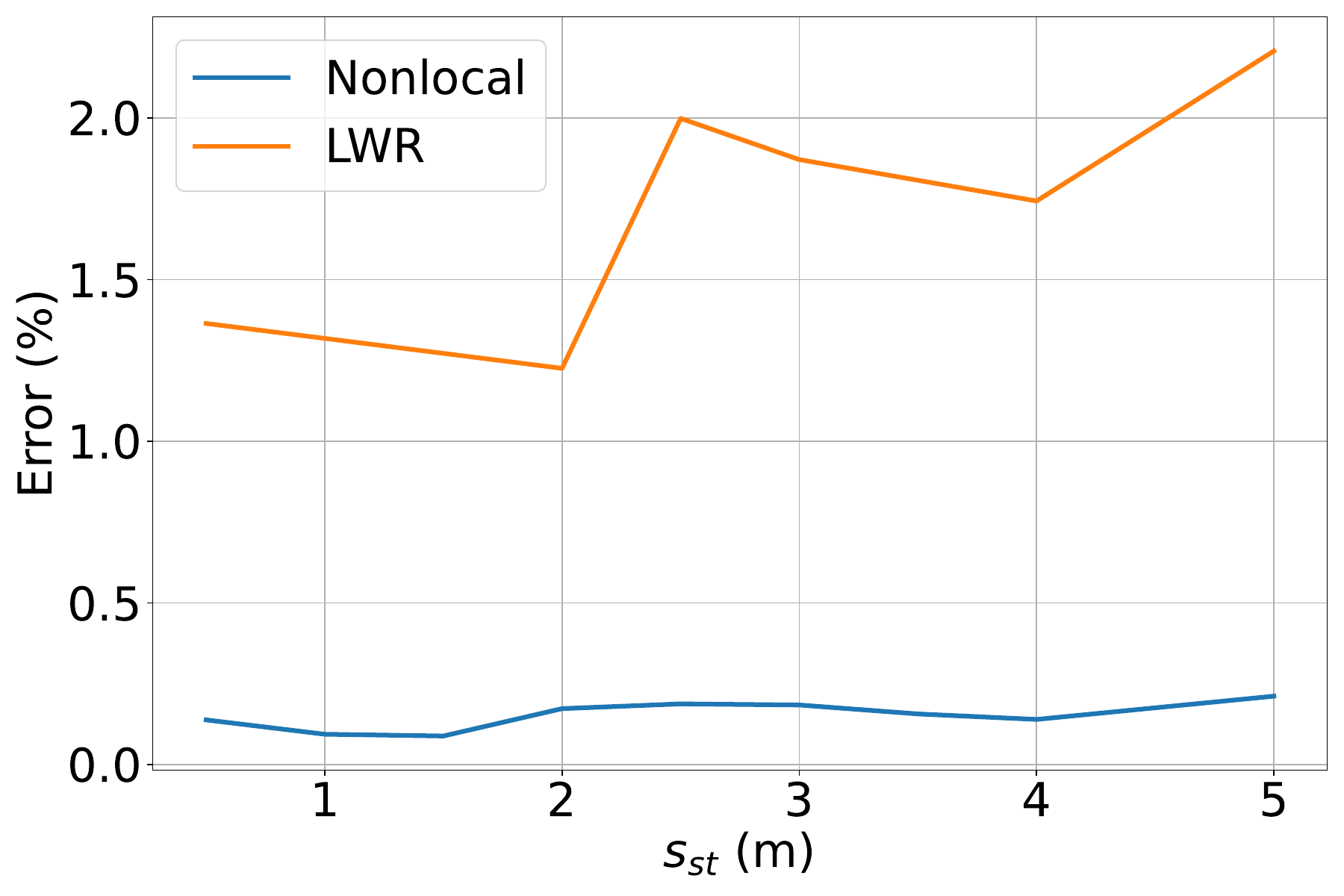}}
    \subcaptionbox{Non-local kernel}{\includegraphics[width=0.3\linewidth]{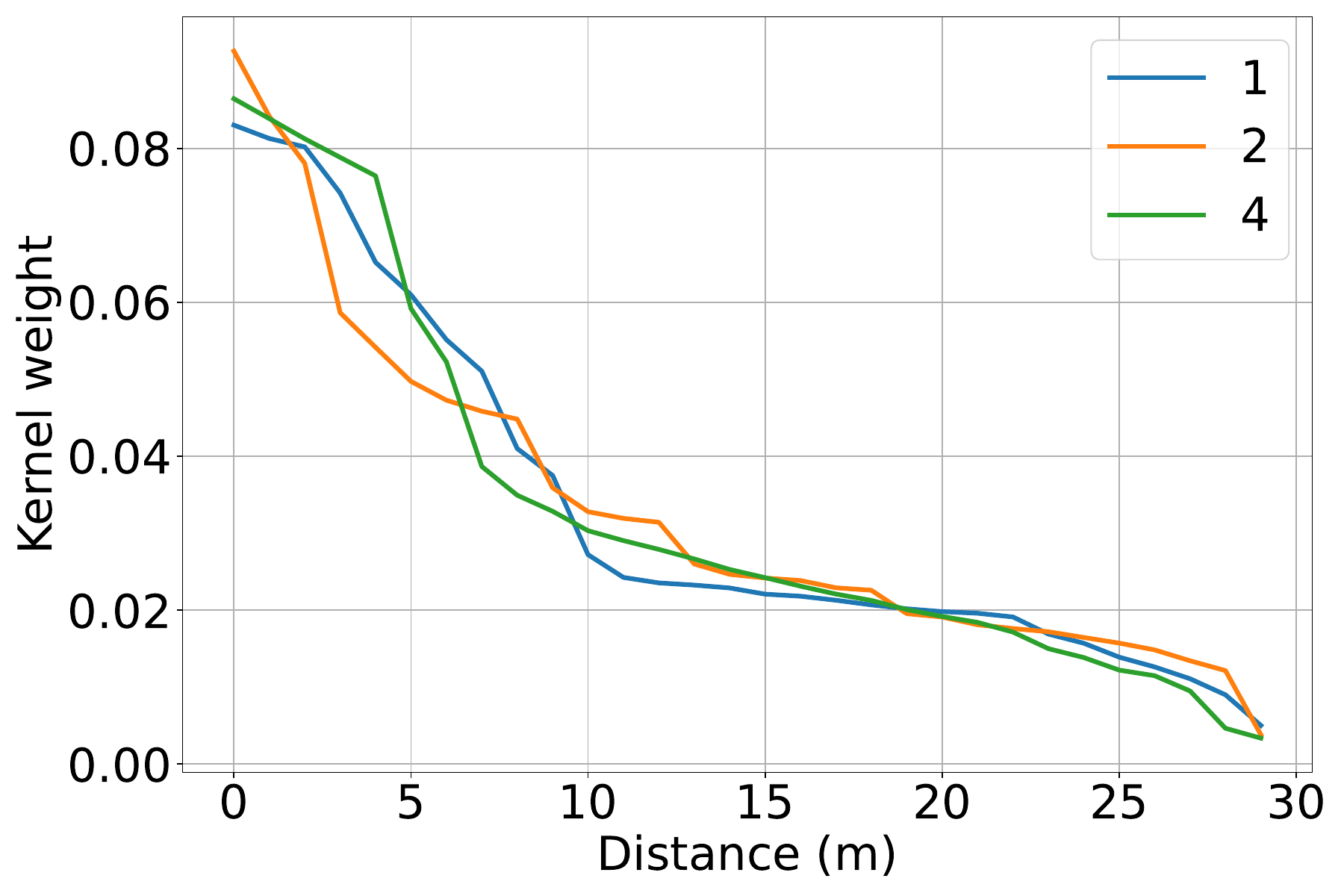}}
    \subcaptionbox{Non-local fundamental diagram}{\includegraphics[width=0.3\linewidth]{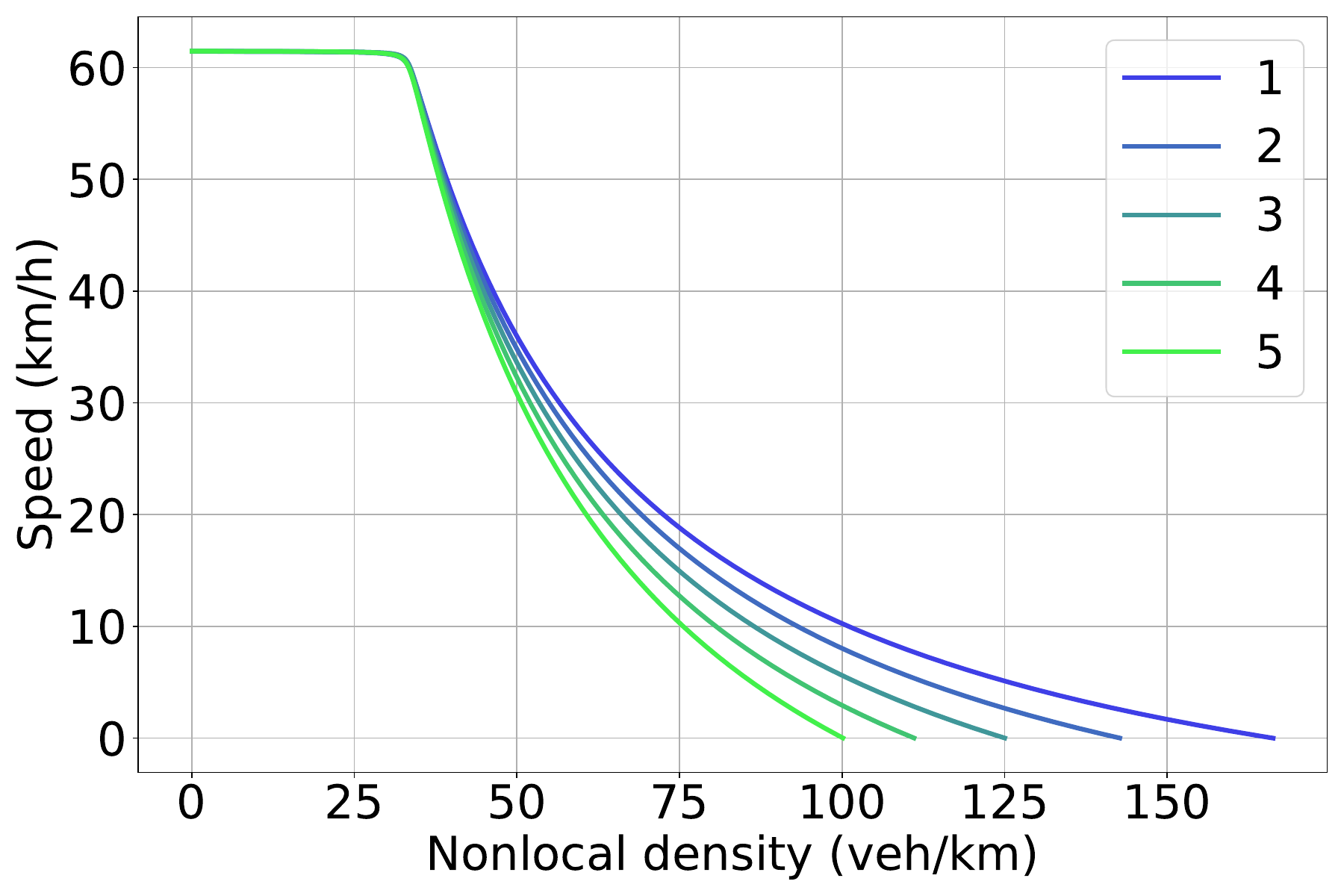}}
    \caption{Simulation results for different stopping gap   $s_{\mathrm{st}}$ values.}
    \label{fig:AV acc sst}
\end{figure}

\begin{figure}[!t]
    \centering
    \subcaptionbox{Estimation error}{\includegraphics[width=0.3\linewidth]{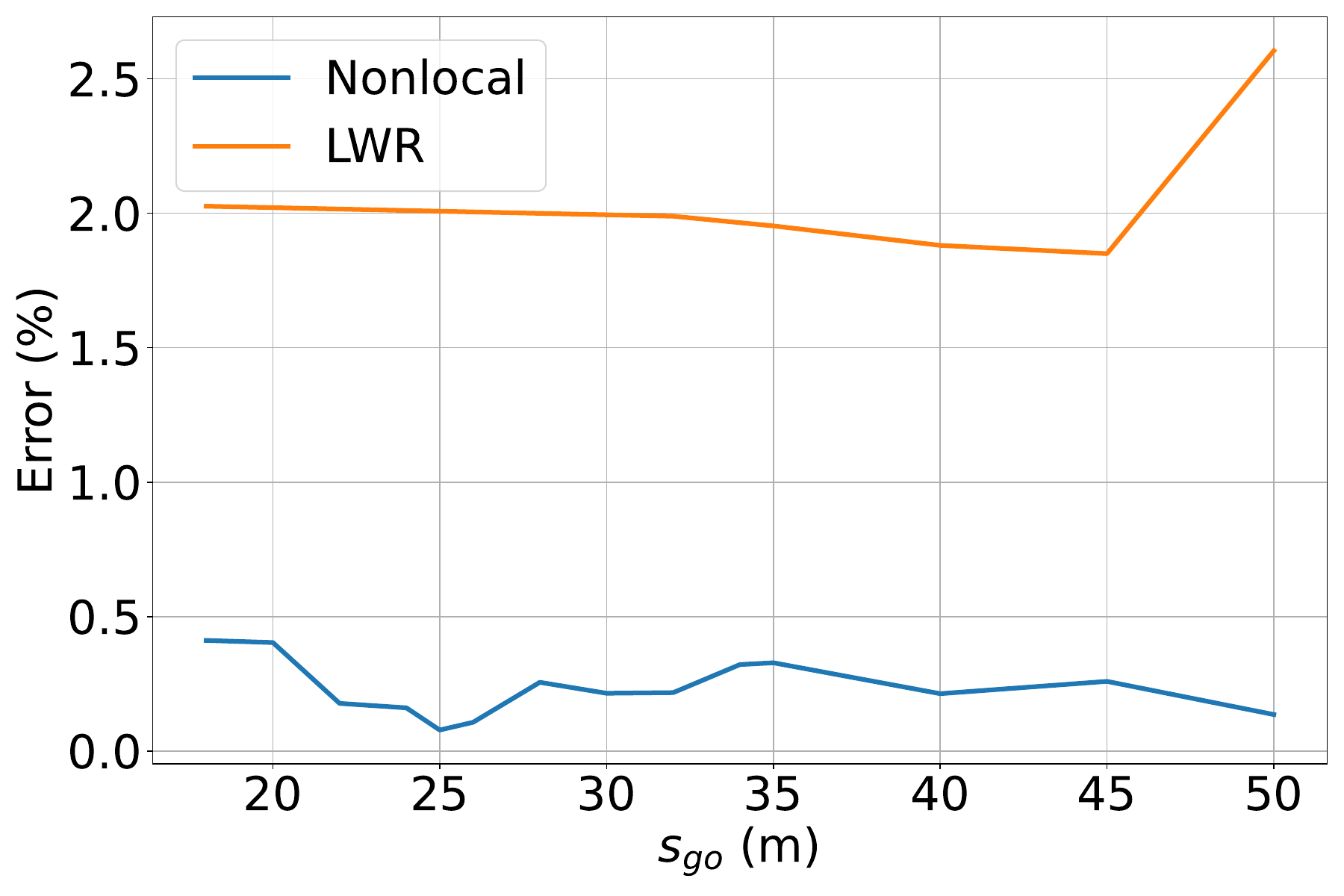}}
    \subcaptionbox{Non-local kernel}{\includegraphics[width=0.3\linewidth]{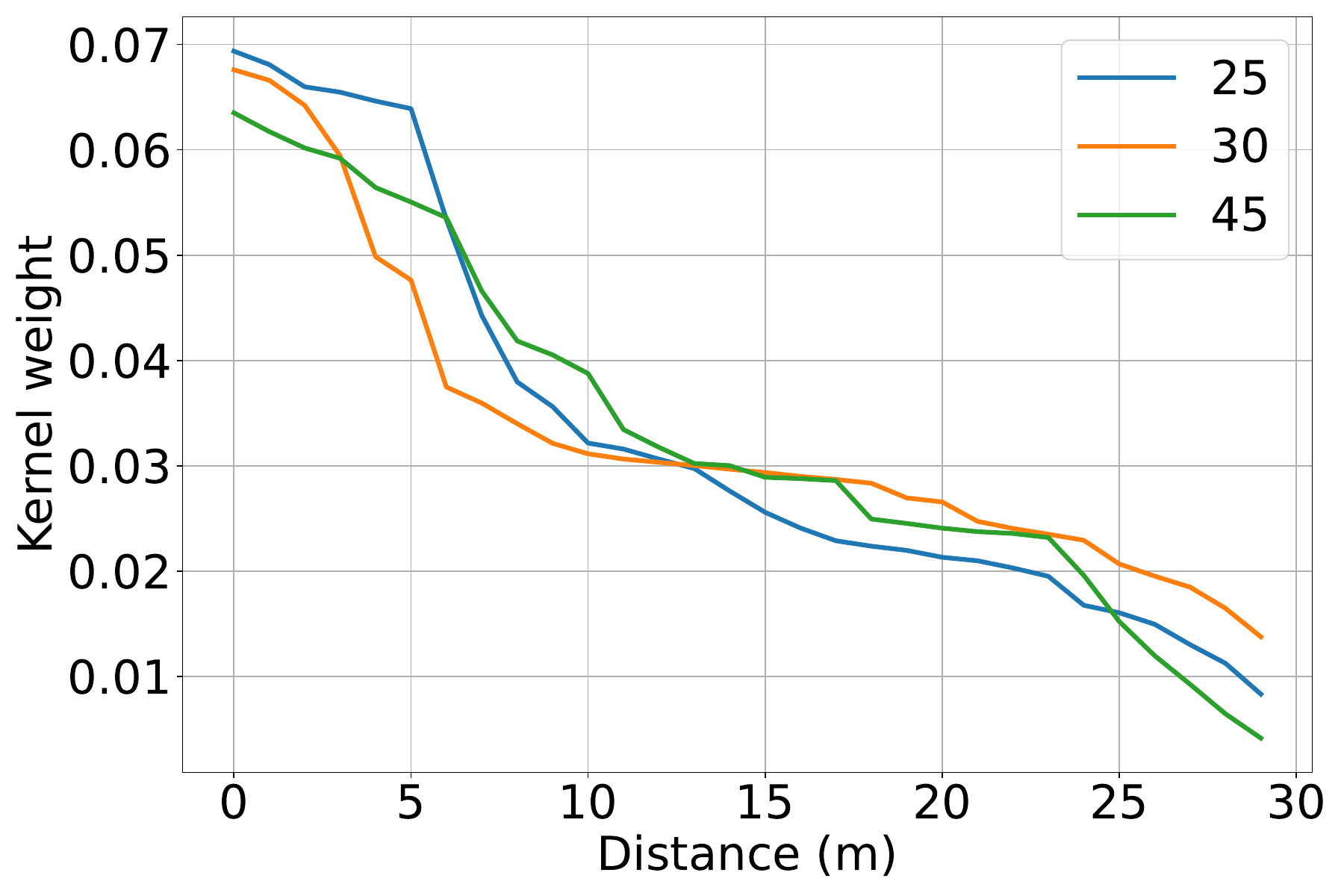}}
    \subcaptionbox{Non-local fundamental diagram}{\includegraphics[width=0.3\linewidth]{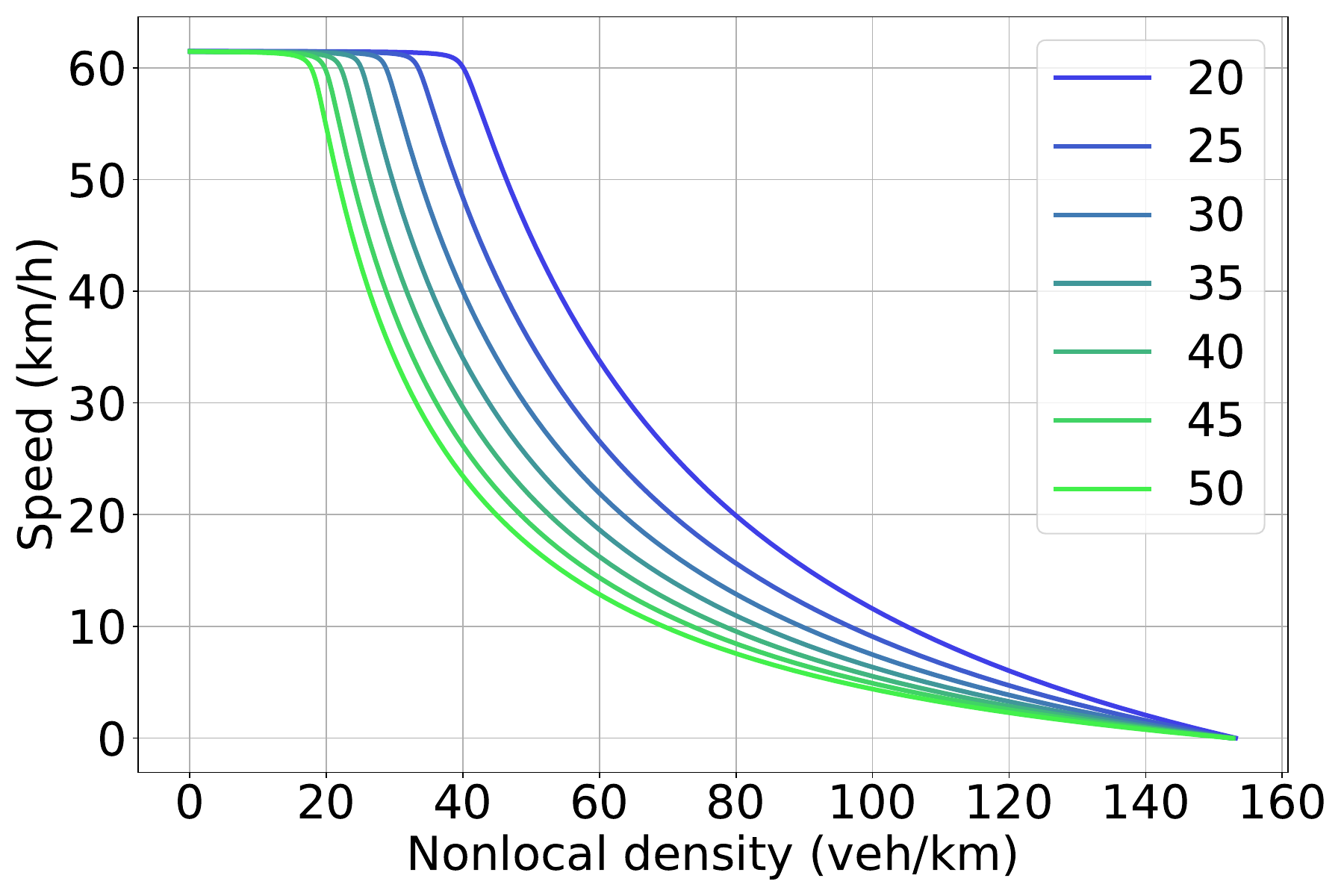}}
    \caption{Simulation results for different free-flow gap   $s_{\mathrm{go}}$ values.}
    \label{fig:AV acc sgo}
\end{figure}

We analyze the effect of the three parameters in the gap-dependent desired speed $V_{\mathrm{opt}}$: $v_{\max}$ in Figs.~\ref{fig:AV acc vmax}-\ref{fig:FD AV acc vmax}, $s_{\mathrm{st}}$ in Fig.~\ref{fig:AV acc sst}, and $s_{\mathrm{go}}$ in Fig.~\ref{fig:AV acc sgo}. For the maximum speed $v_{\max}$, we see from Fig.~\ref{fig:AV acc vmax}(a) that for a wide range of $v_{\max}$ values, the non-local model provides an accurate description, with the estimation error lower than 0.5\%. From the looking-ahead kernels in Fig.~\ref{fig:AV acc vmax}(b), different $v_{\max}$ parameter values correspond to a similar looking-ahead kernel: a majority of the non-local effect is within five meters. From the comparison of fundamental diagrams  in Fig.~\ref{fig:FD AV acc vmax}, we see that the free-flow speed increases linearly as the $v_{\max}$ value increases.  This agrees with our intuitive assumption, i.e., an increase in the microscopic free-speed brings a larger free-flow speed in the macroscopic traffic. The critical density remains approximately the same when $v_{\max}$ varies, and the capacity also increases linearly as $v_{\max}$ increases. We also find from Fig.~\ref{fig:FD AV acc vmax}(a) that when $v_{\max}$ increases, there is a slight increase in the jam density. 

For different $s_{\mathrm{st}}$ values, we see from Fig.~\ref{fig:AV acc sst}(a) that the estimation error for the  non-local model remains  consistently lower than 0.5\%. The non-local kernels in Fig.~\ref{fig:AV acc sst}(b) also show the same two trends as for $v_{\max}$. First, the  kernel is approximately the same for different $s_{\mathrm{st}}$ values, i.e., this parameter has little effect on the non-local kernel. Second, a majority of the non-local effect is within 5 meters.  For the fundamental diagram in Fig.~\ref{fig:AV acc sst}(c), the jam density decreases with a larger stopping gap $s_{\mathrm{st}}$, which agrees with the common expectation.

For different $s_{\mathrm{go}}$ values, the non-local model also has an accurate description of traffic flow, as Fig.~\ref{fig:AV acc sgo}(a) shows. The looking-ahead weights also present the two trends as in the case for $v_{\max}$ and $s_{\mathrm{st}}$.   The kernel weights are approximately the same for different $s_{\mathrm{go}}$ values, and the 5 meter range accounts for the major part. For the fundamental diagram, as Fig.~\ref{fig:AV acc sgo}(c) shows, the free-flow speed and jam density remain approximately the same when $s_{\mathrm{go}}$ increases.  The critical density  decreases with a larger $s_{\mathrm{go}}$. To summarize, the three parameters in $V_{\mathrm{opt}}$ have little effect on the non-local kernels, but they mainly affect the fundamental diagram.

\begin{figure}[!t]
    \centering
    \subcaptionbox{Estimation error}{\includegraphics[width=0.3\linewidth]{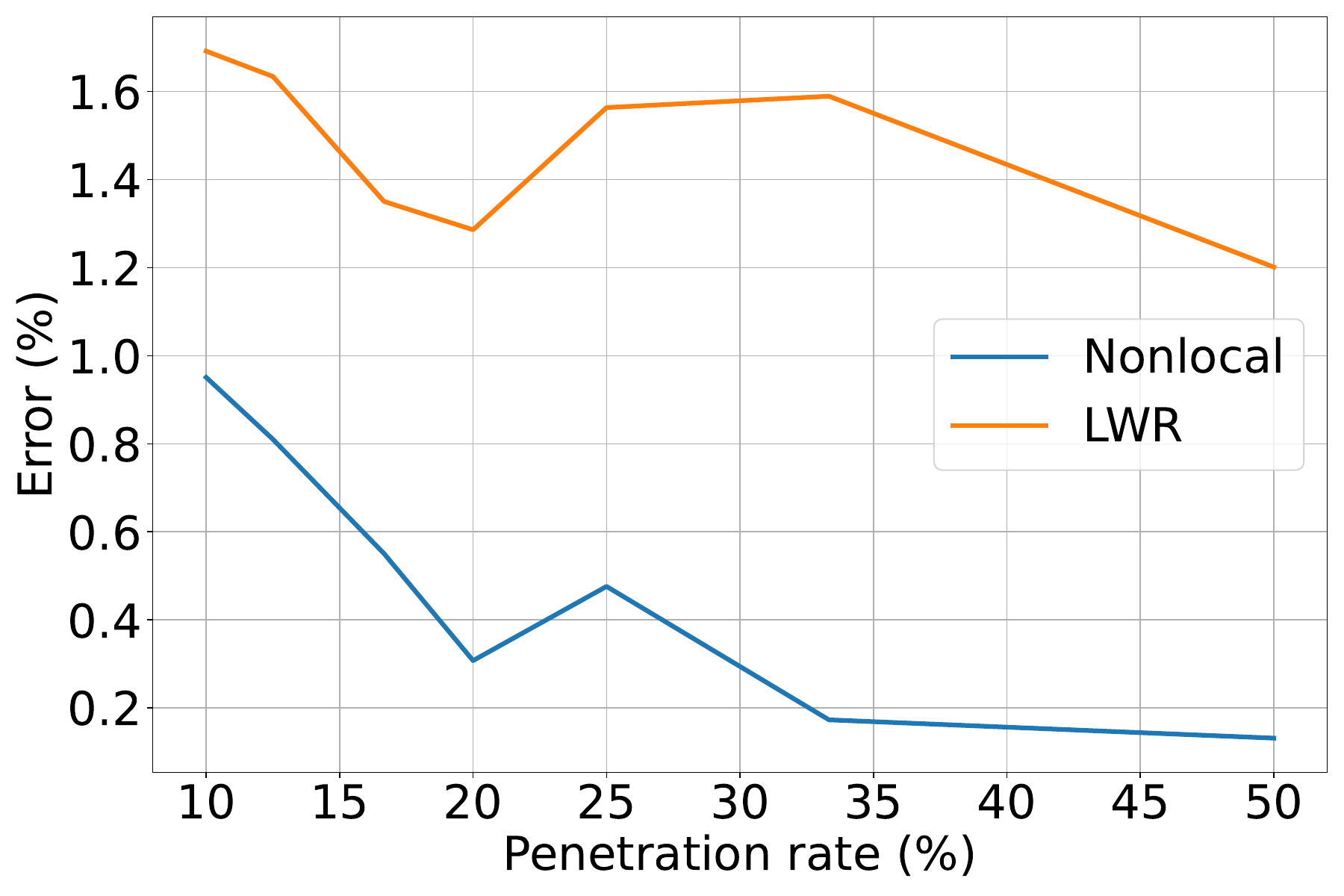}}
    \subcaptionbox{Nonlocal kernel}{\includegraphics[width=0.3\linewidth]{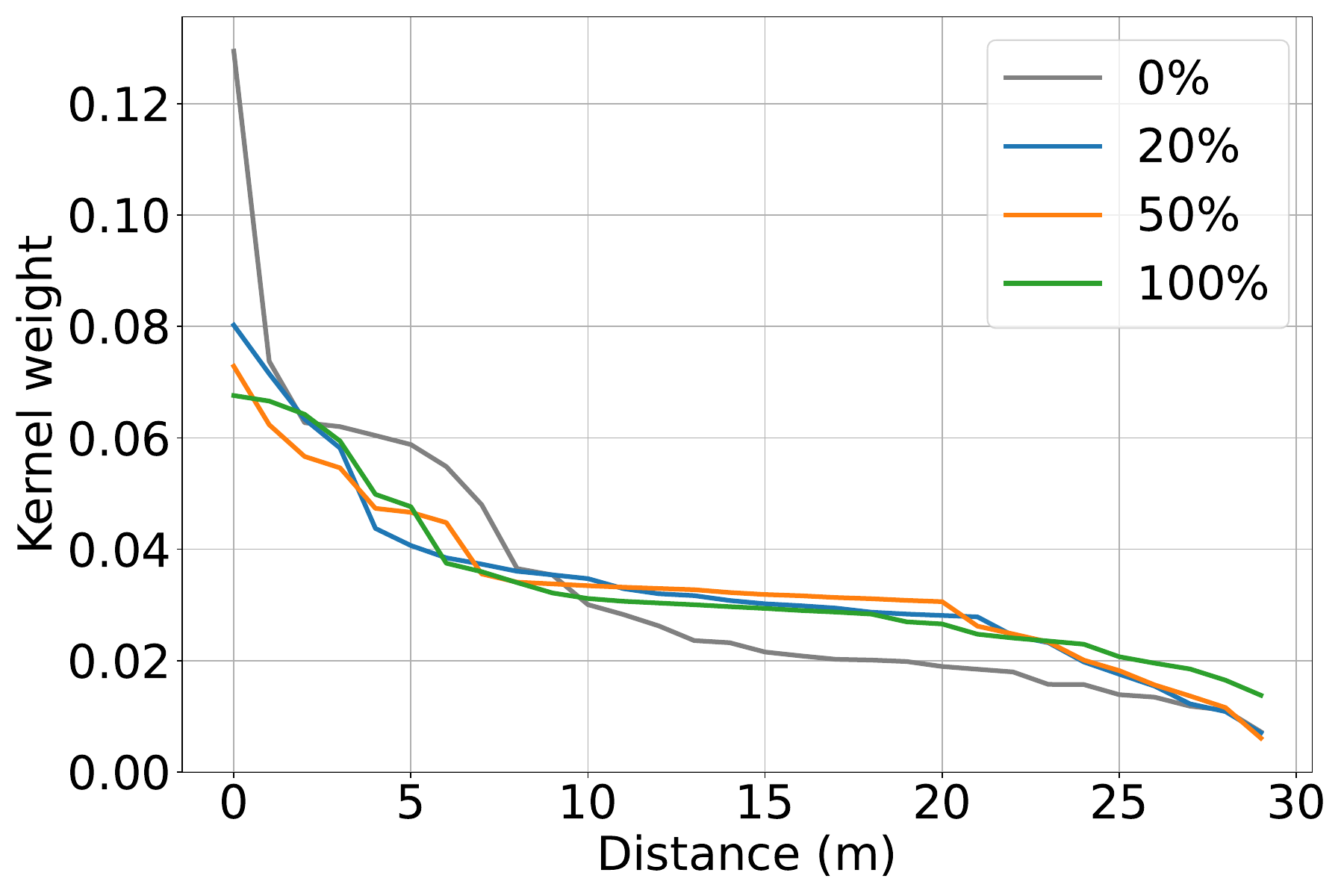}}
    \subcaptionbox{Fundamental diagram}{\includegraphics[width=0.3\linewidth]{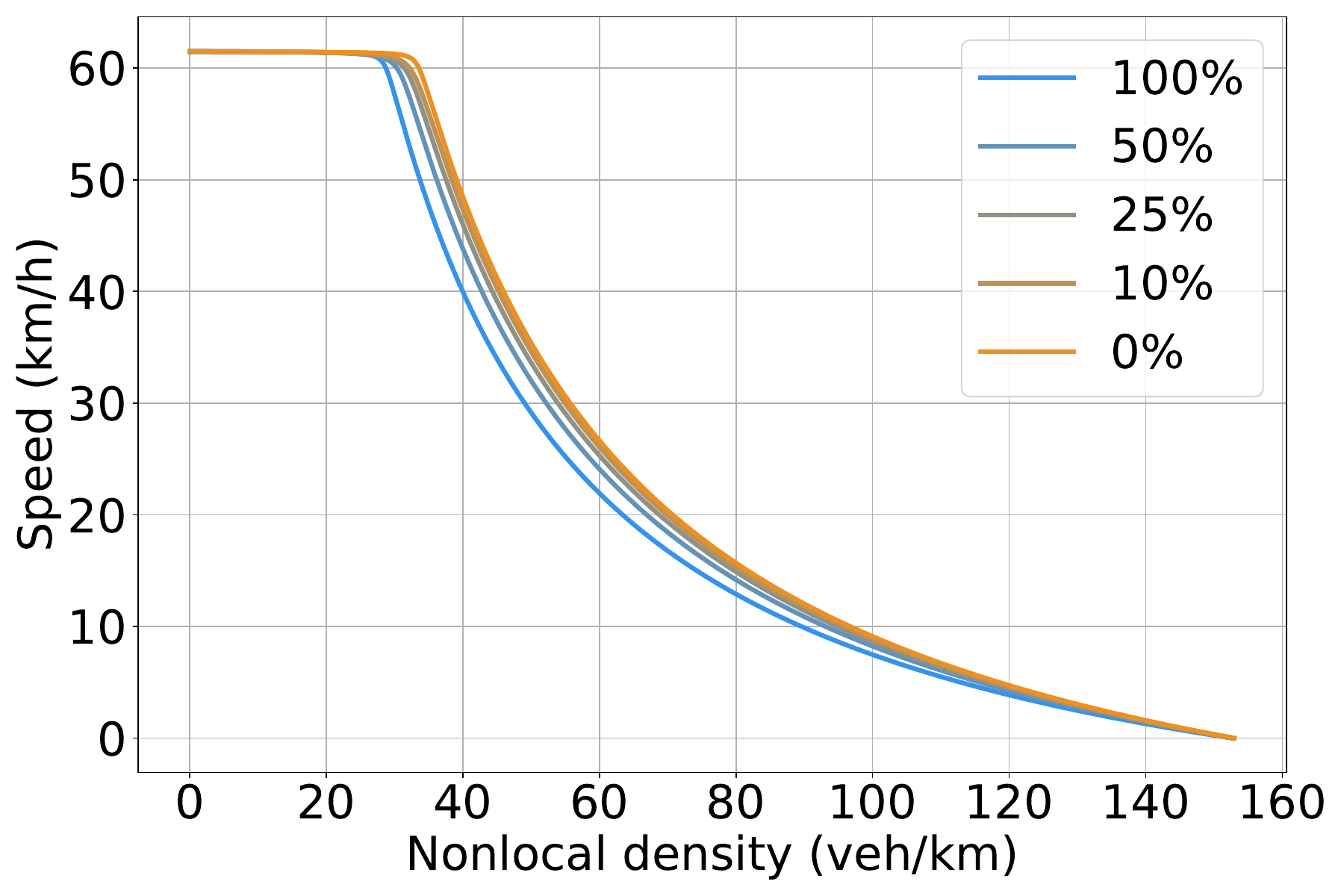}}
    \caption{Simulation results for mixed traffic. The CAV takes a different $s_{\mathrm{go}}$ value.}
    \label{fig:mix sgo}
\end{figure}

\begin{figure}[!t]
    \centering
    \subcaptionbox{Estimation error}{\includegraphics[width=0.3\linewidth]{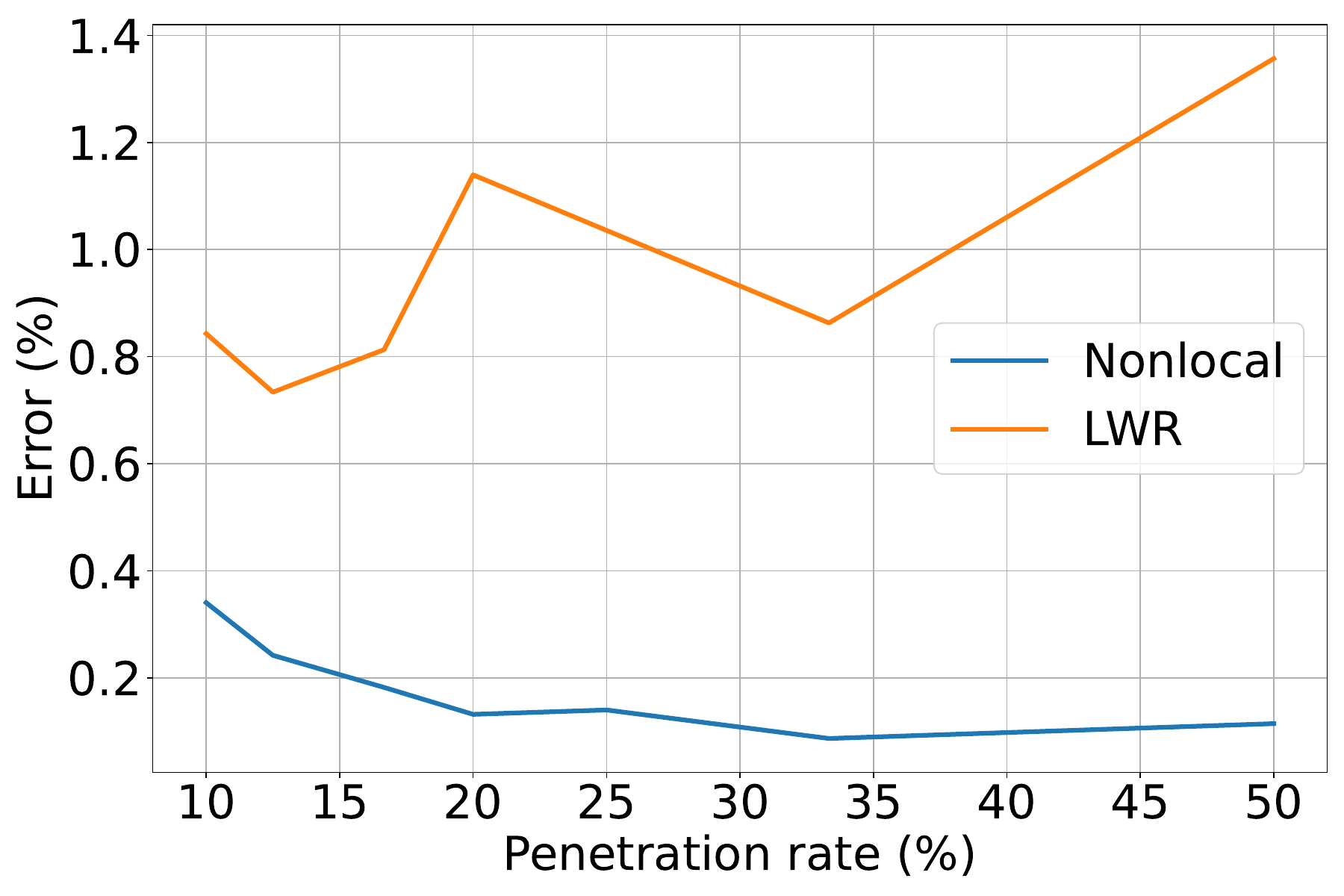}}
    \subcaptionbox{Nonlocal kernel}{\includegraphics[width=0.3\linewidth]{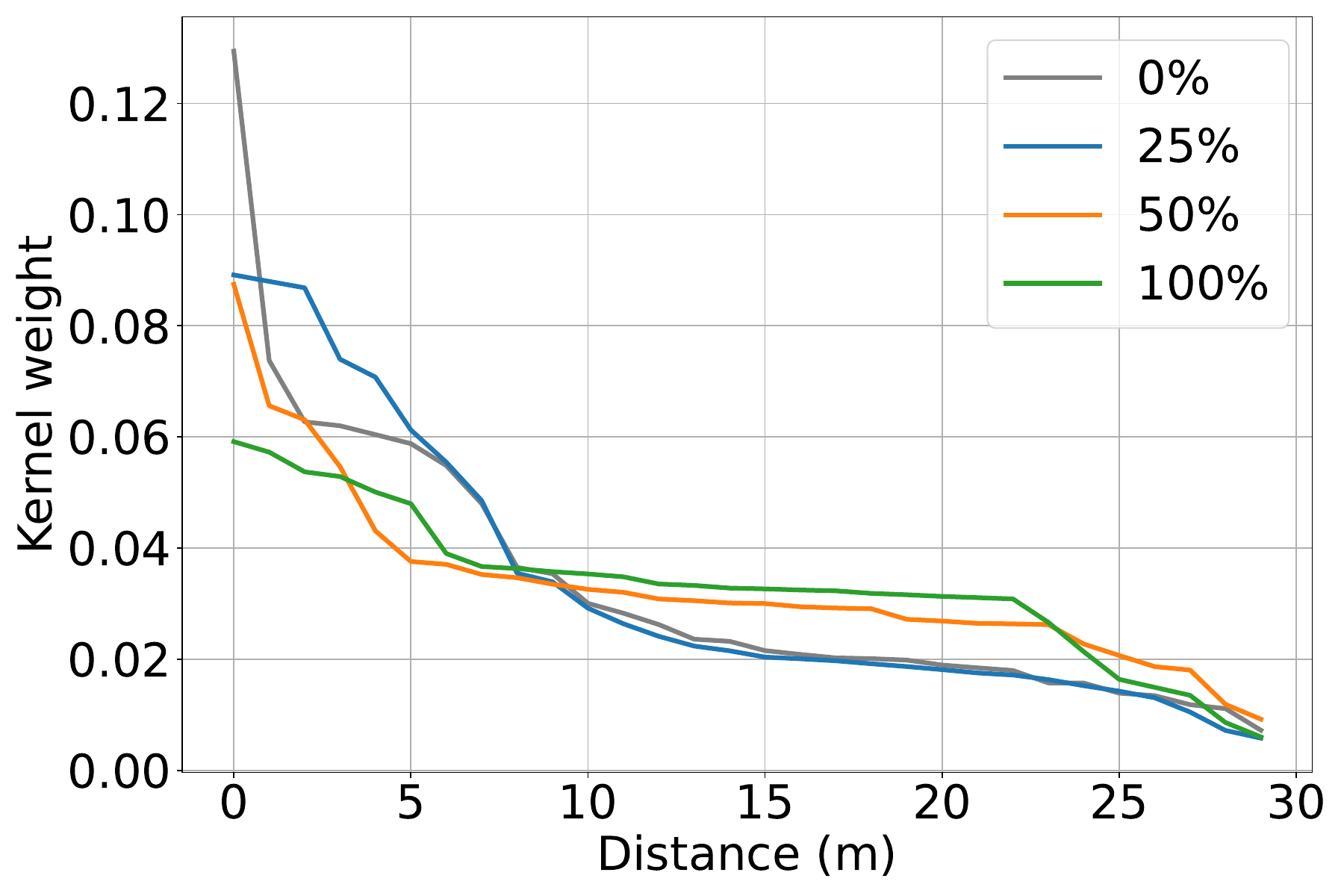}}
    \subcaptionbox{Fundamental diagram}{\includegraphics[width=0.3\linewidth]{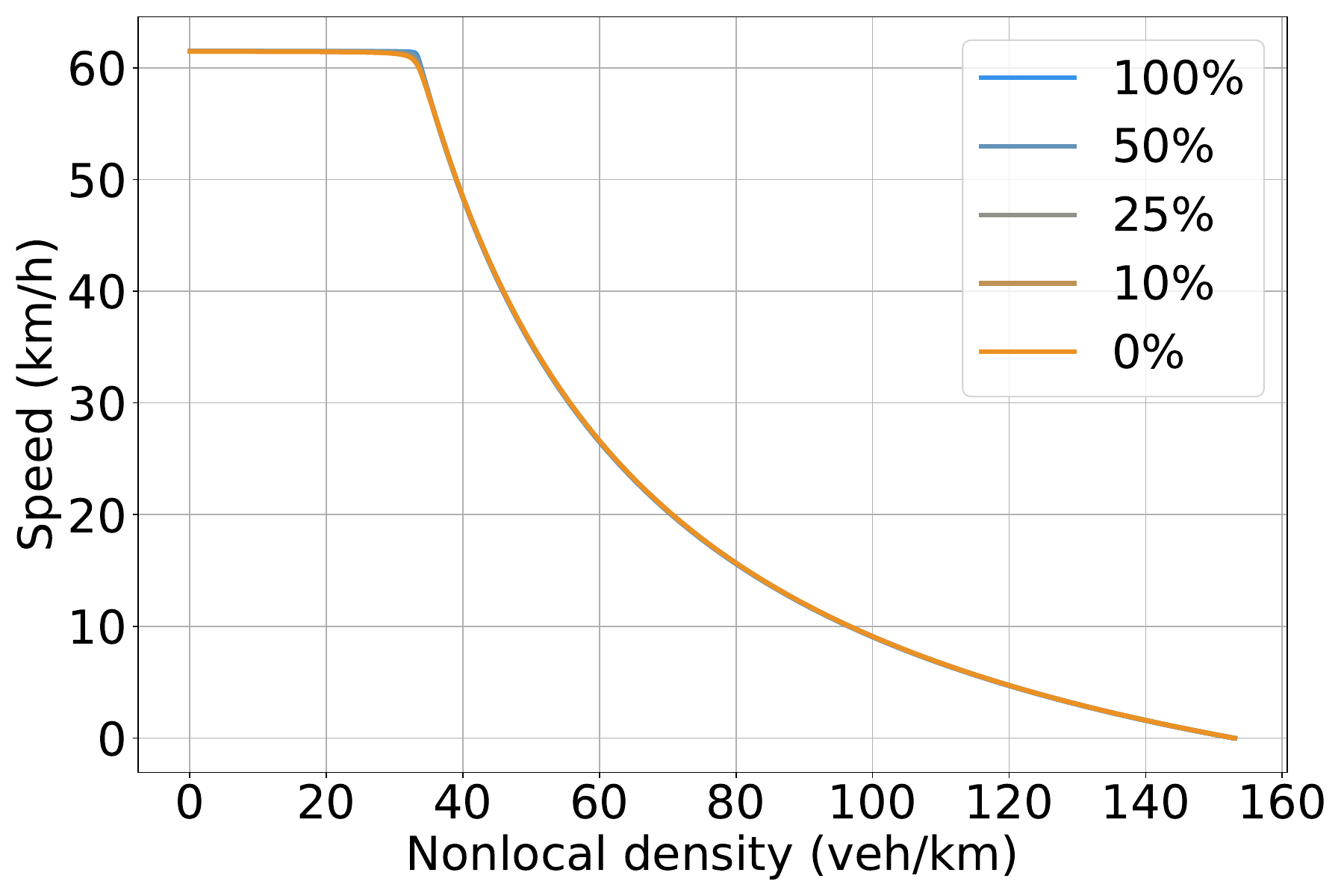}}
    \caption{Simulation results for mixed traffic. The CAV uses the same desired speed function as HVs but takes the looking-ahead spacing gain as $\alpha_{-1} = 0.7$.}
    \label{fig:mix alphal}
\end{figure}

\begin{figure}[!t]
    \centering
    \subcaptionbox{Estimation error}{\includegraphics[width=0.3\linewidth]{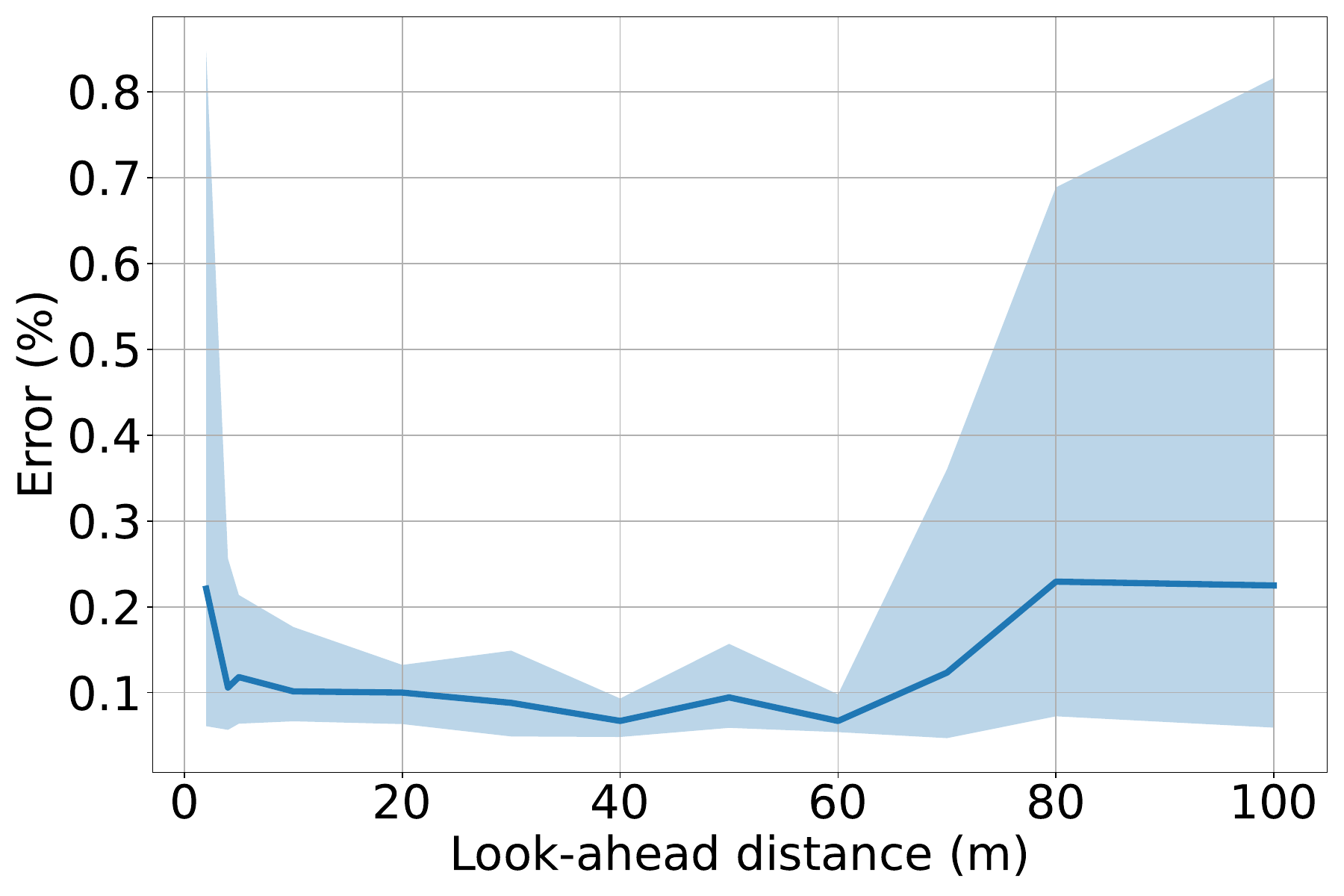}}\\
    \subcaptionbox{Looking-ahead kernel, $\eta_a=20$ m}{\includegraphics[width=0.3\linewidth]{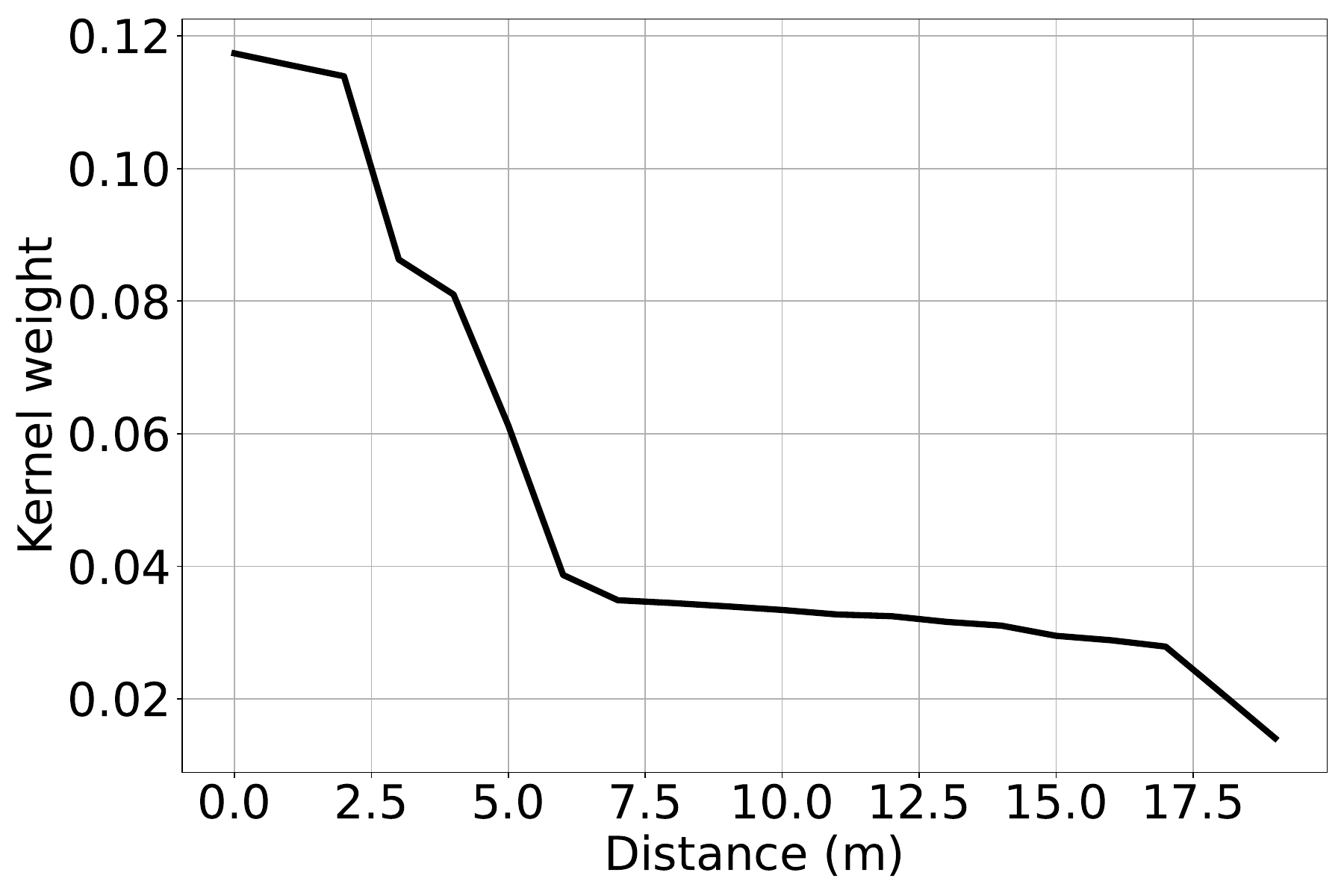}}
    \subcaptionbox{Looking-ahead kernel, $\eta_a=40$ m}{\includegraphics[width=0.3\linewidth]{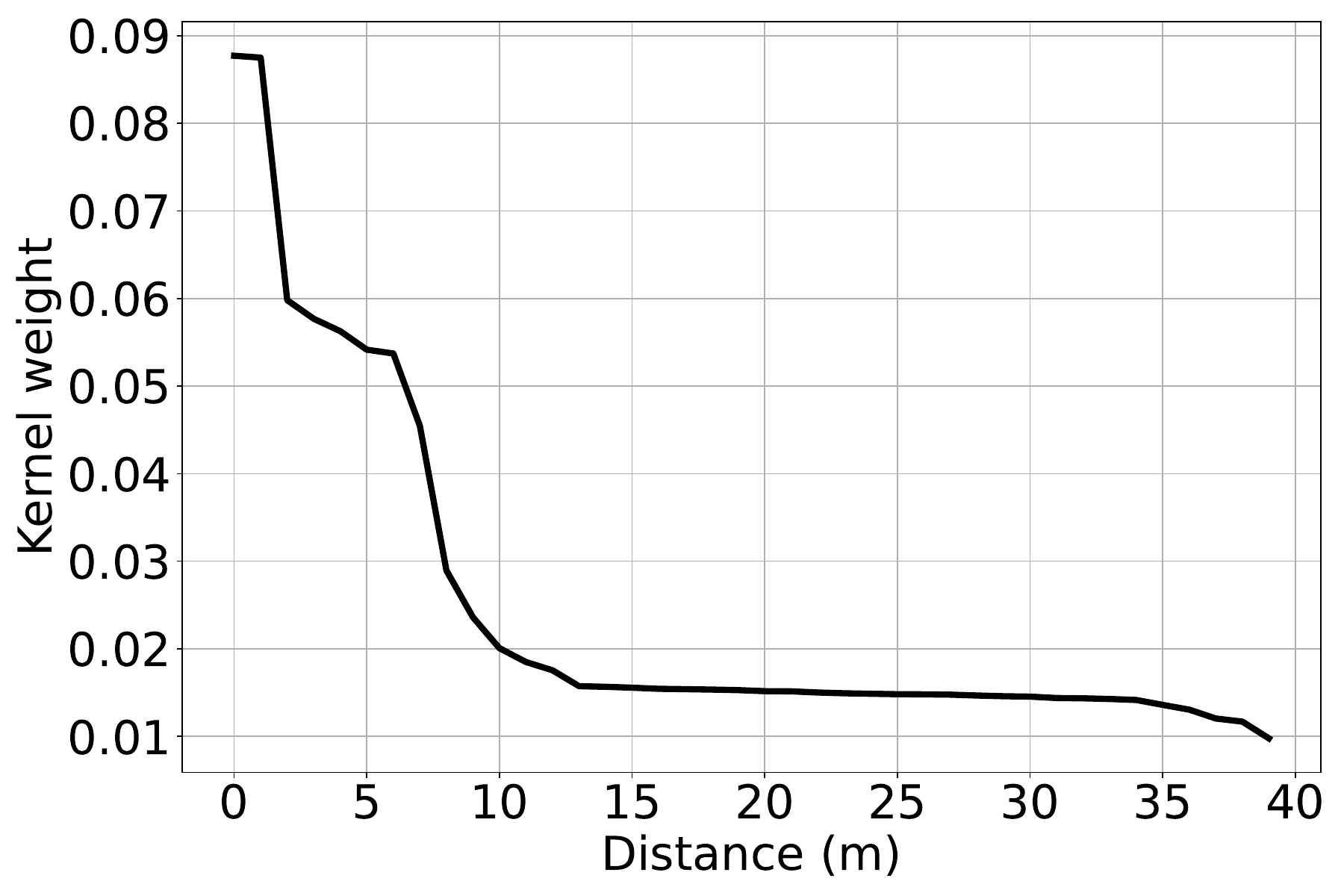}}
    \subcaptionbox{Looking-ahead kernel, $\eta_a=55$ m}{\includegraphics[width=0.3\linewidth]{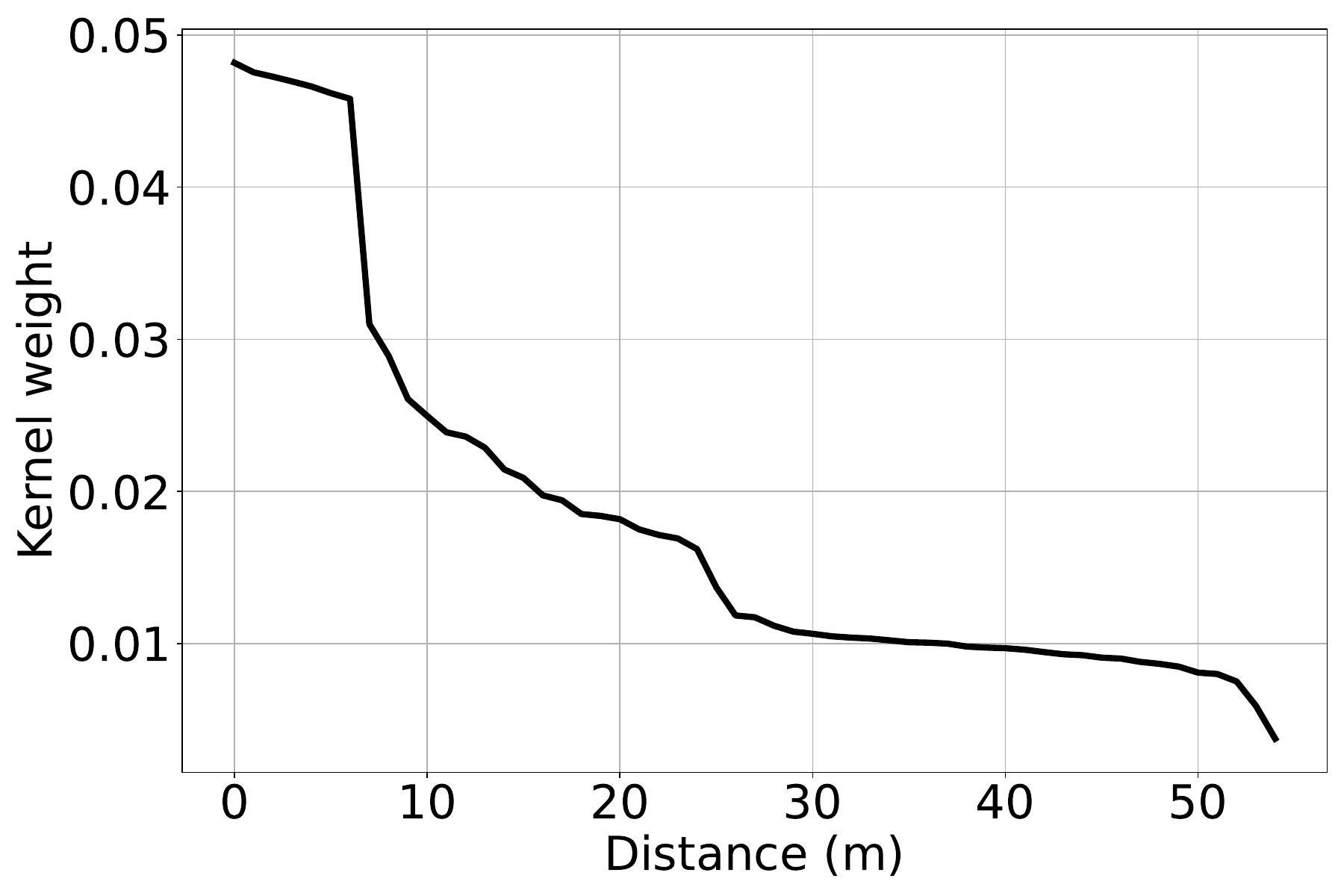}}\\
    \subcaptionbox{Fundamental diagram, $\eta_a=20$ m}{\includegraphics[width=0.3\linewidth]{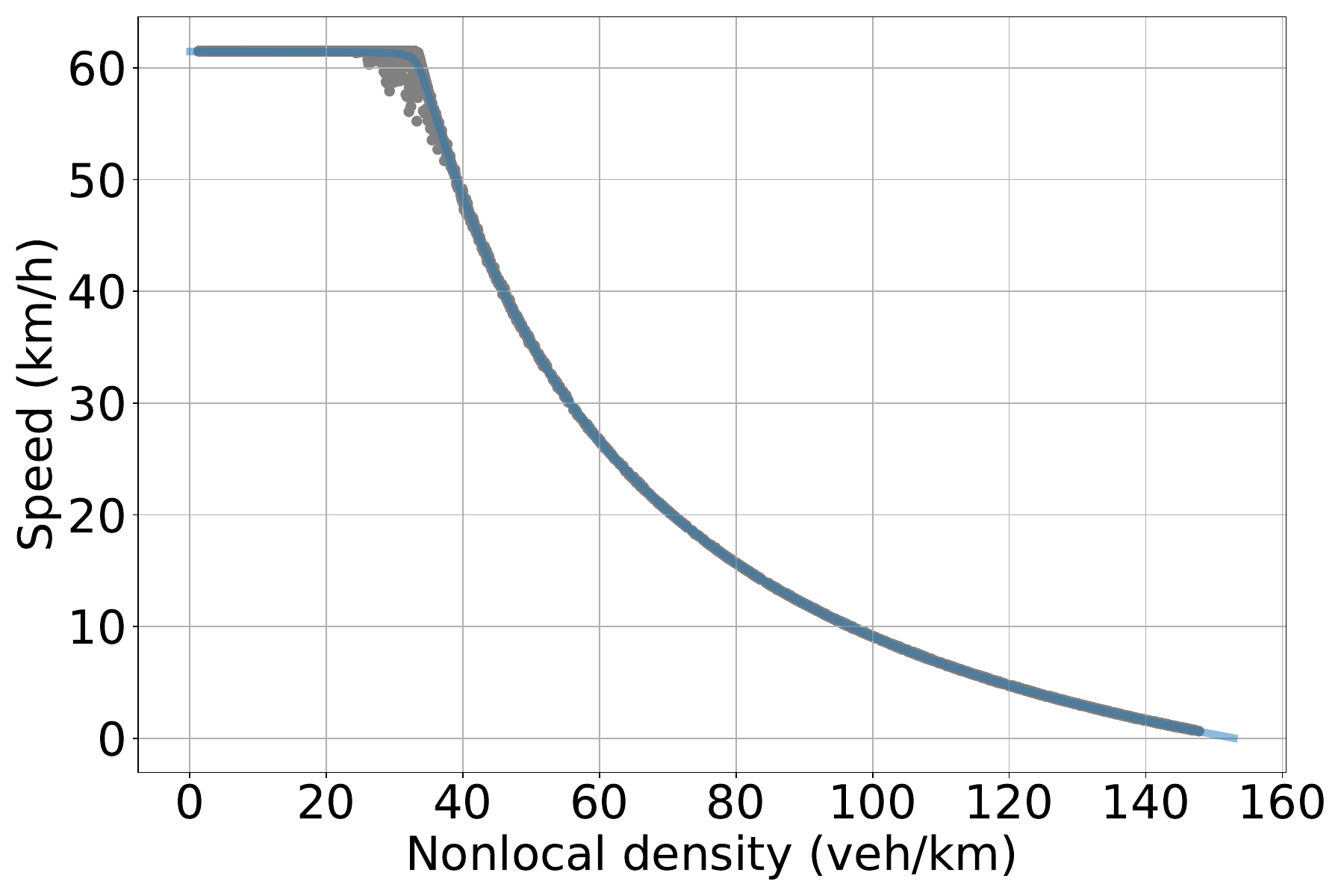}}
    \subcaptionbox{Fundamental diagram, $\eta_a=40$ m}{\includegraphics[width=0.3\linewidth]{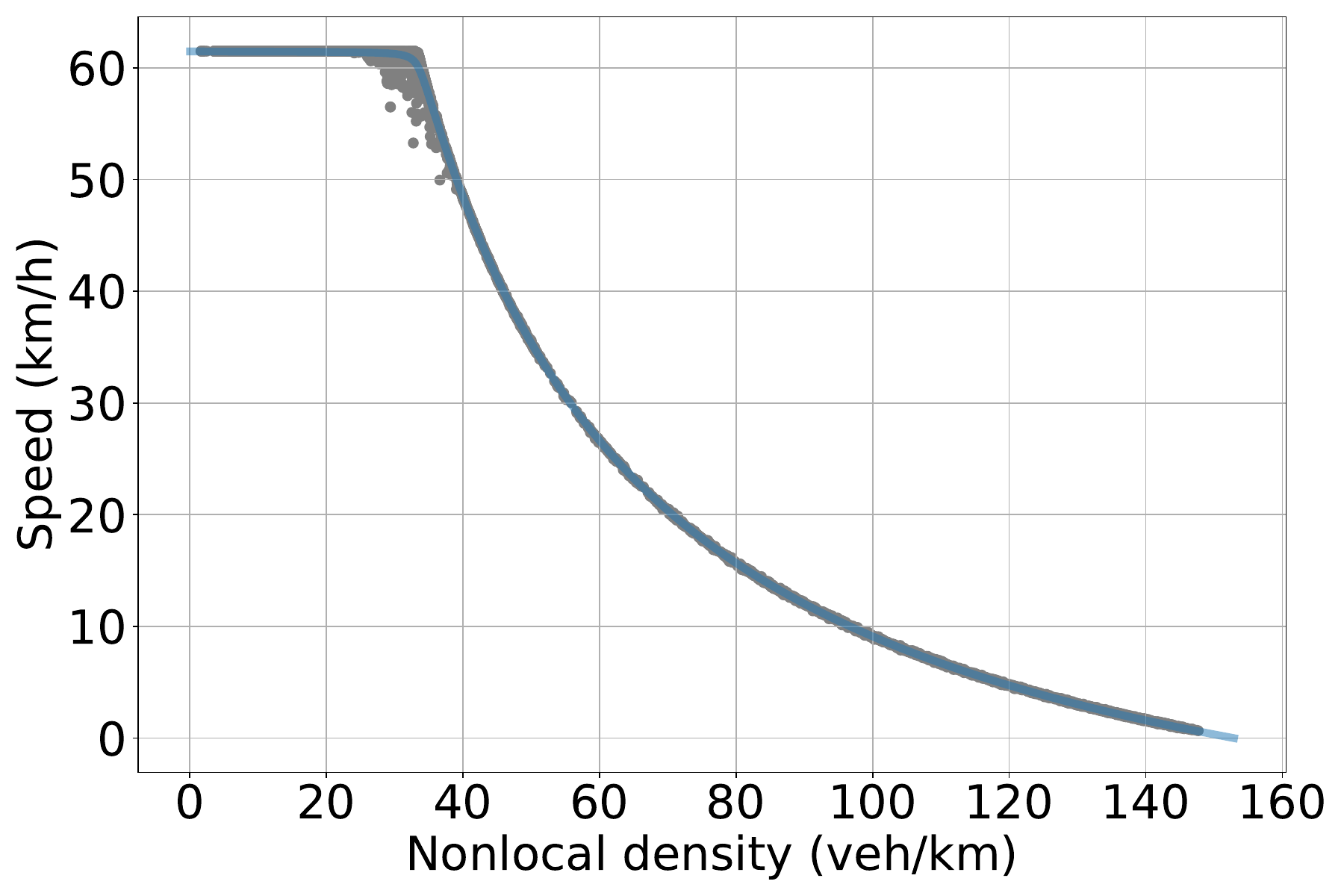}}
    \subcaptionbox{Fundamental diagram, $\eta_a=55$ m}{\includegraphics[width=0.3\linewidth]{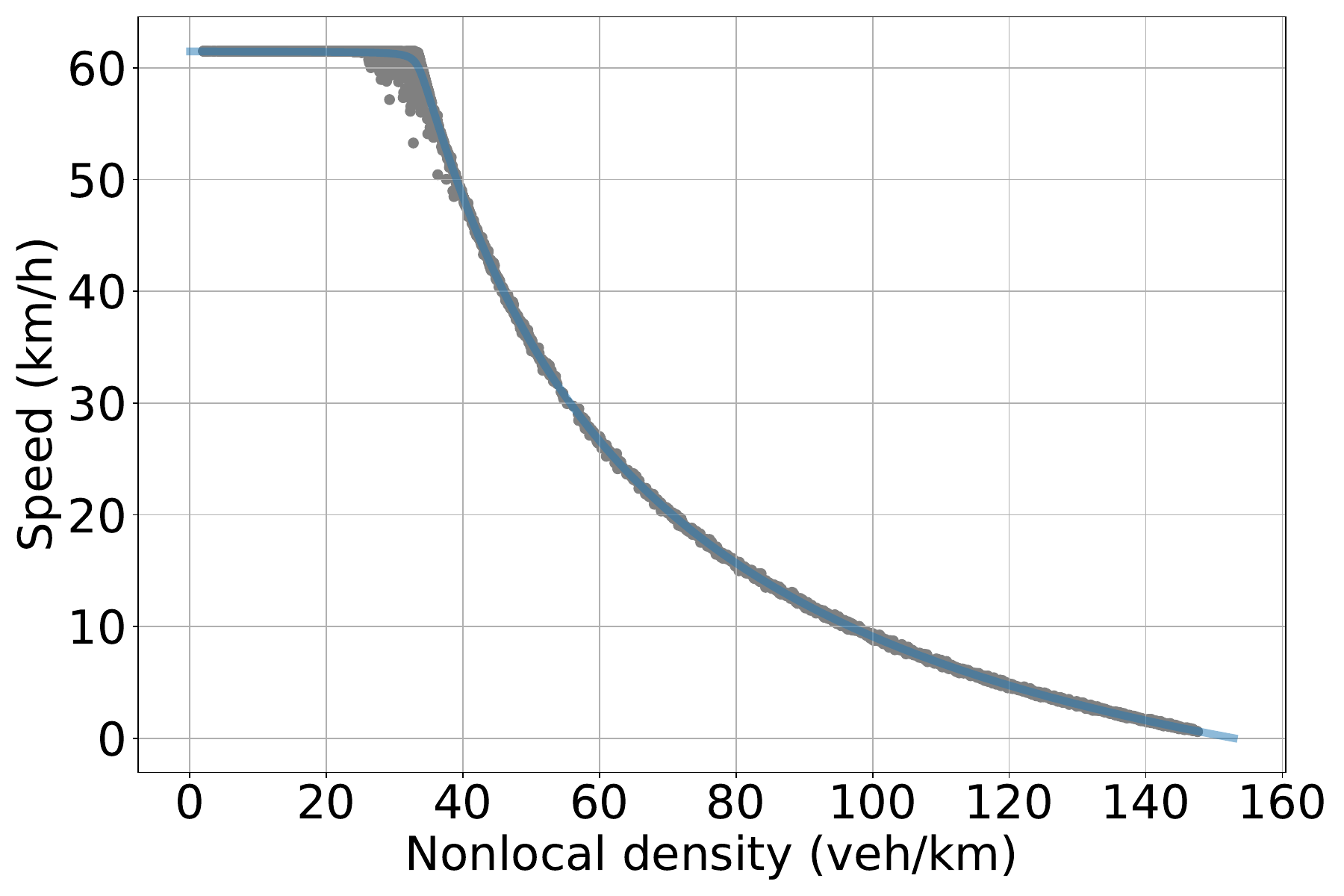}}
    \caption{Simulation results with different looking-ahead kernel length $\eta_a$.}
    \label{fig:kernel length acc}
\end{figure}

\begin{figure}[!t]
    \centering
    \subcaptionbox{Estimation error}{\includegraphics[width=0.3\linewidth]{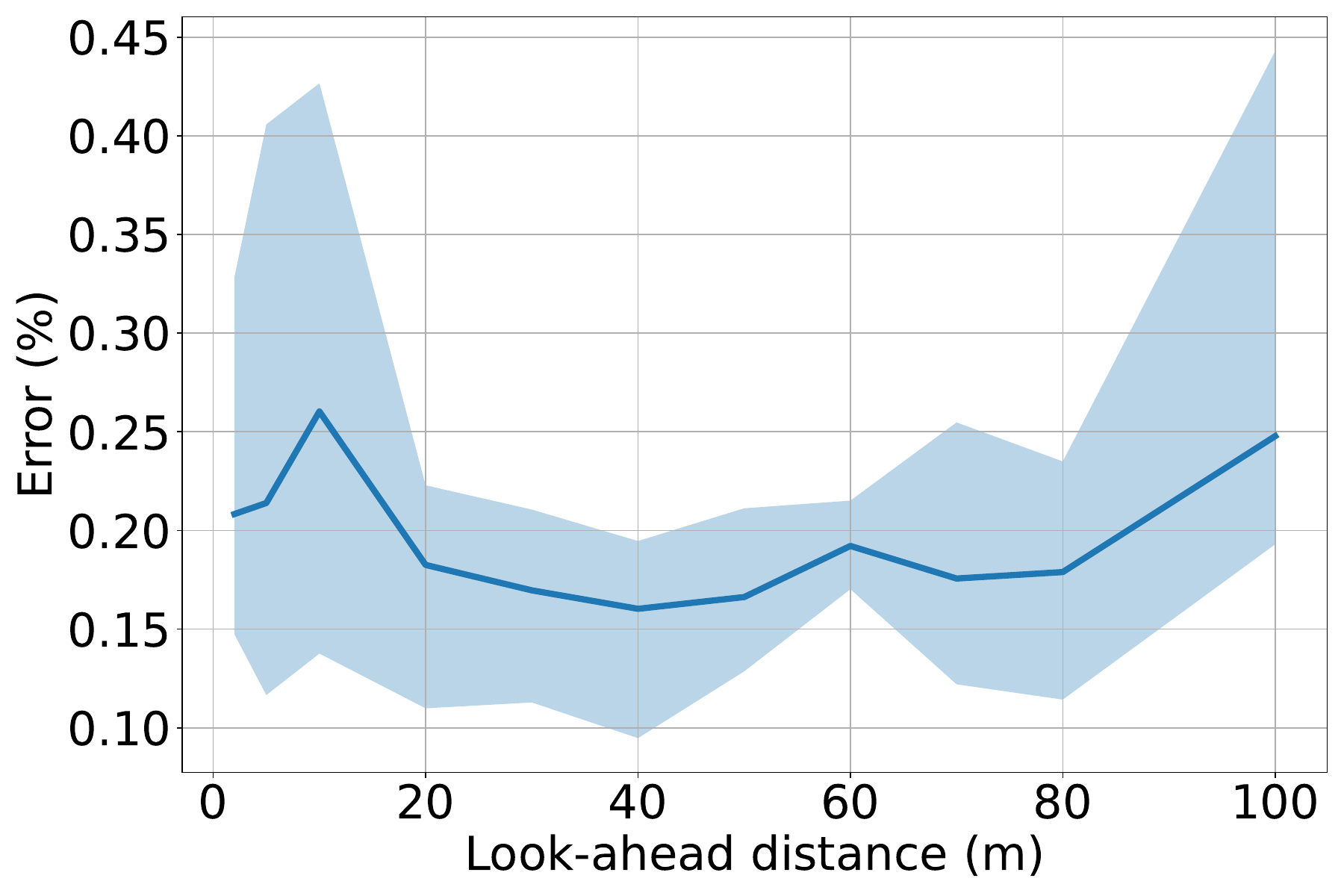}}\\
    \subcaptionbox{Nonlocal kernel, $\eta_b=20$ m}{\includegraphics[width=0.3\linewidth]{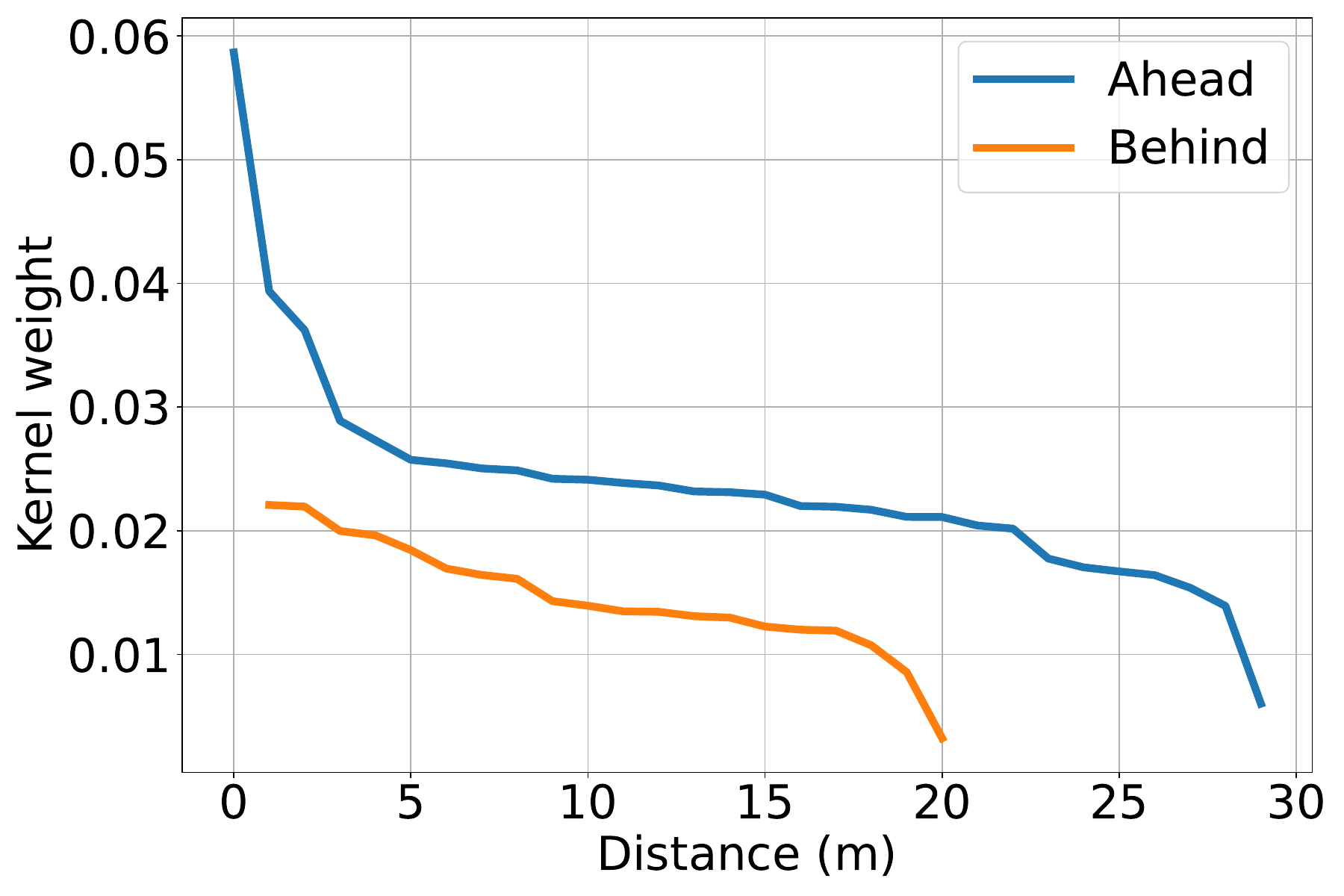}}
    \subcaptionbox{Nonlocal kernel, $\eta_b=40$ m}{\includegraphics[width=0.3\linewidth]{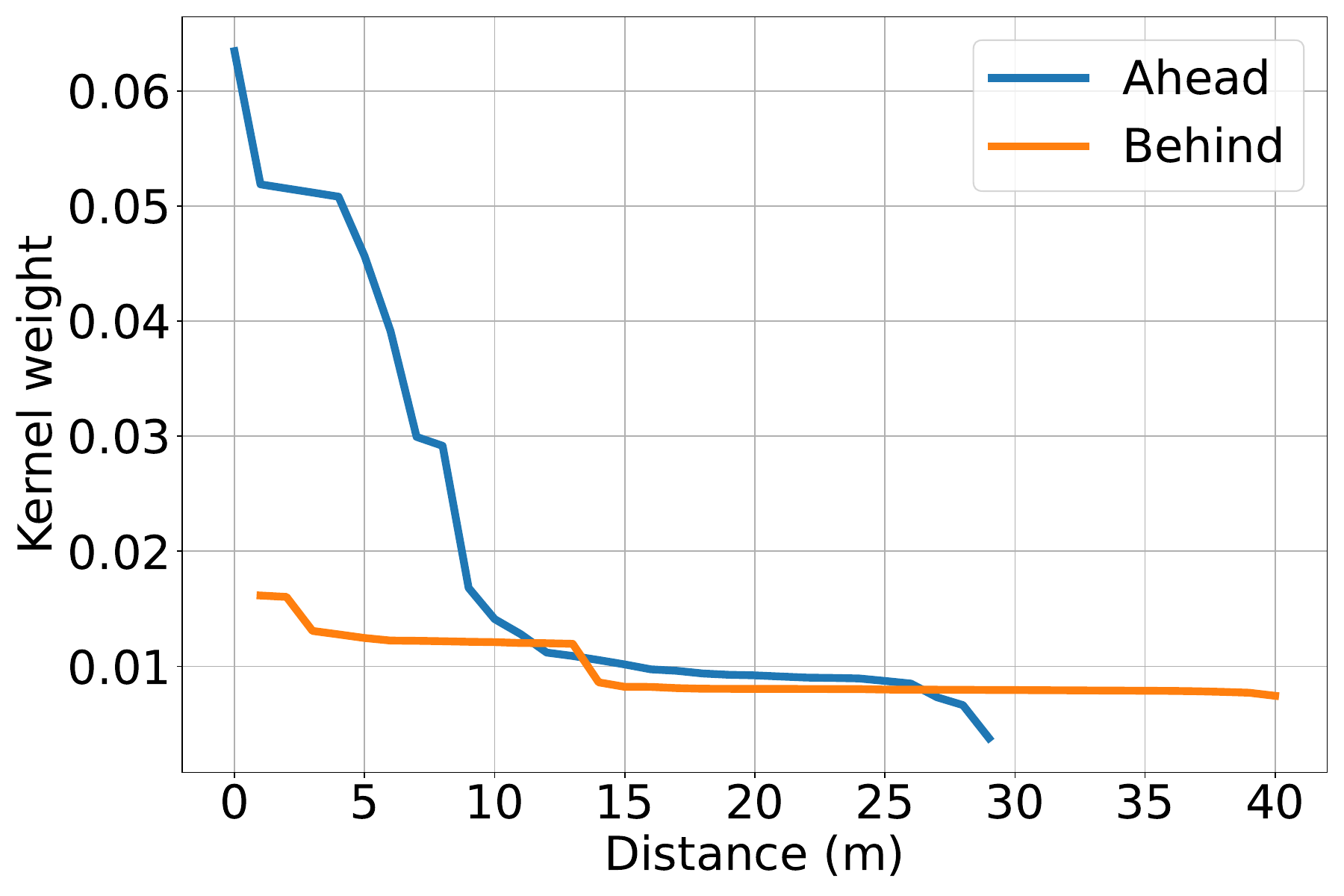}}
    \subcaptionbox{Nonlocal kernel, $\eta_b=50$ m}{\includegraphics[width=0.3\linewidth]{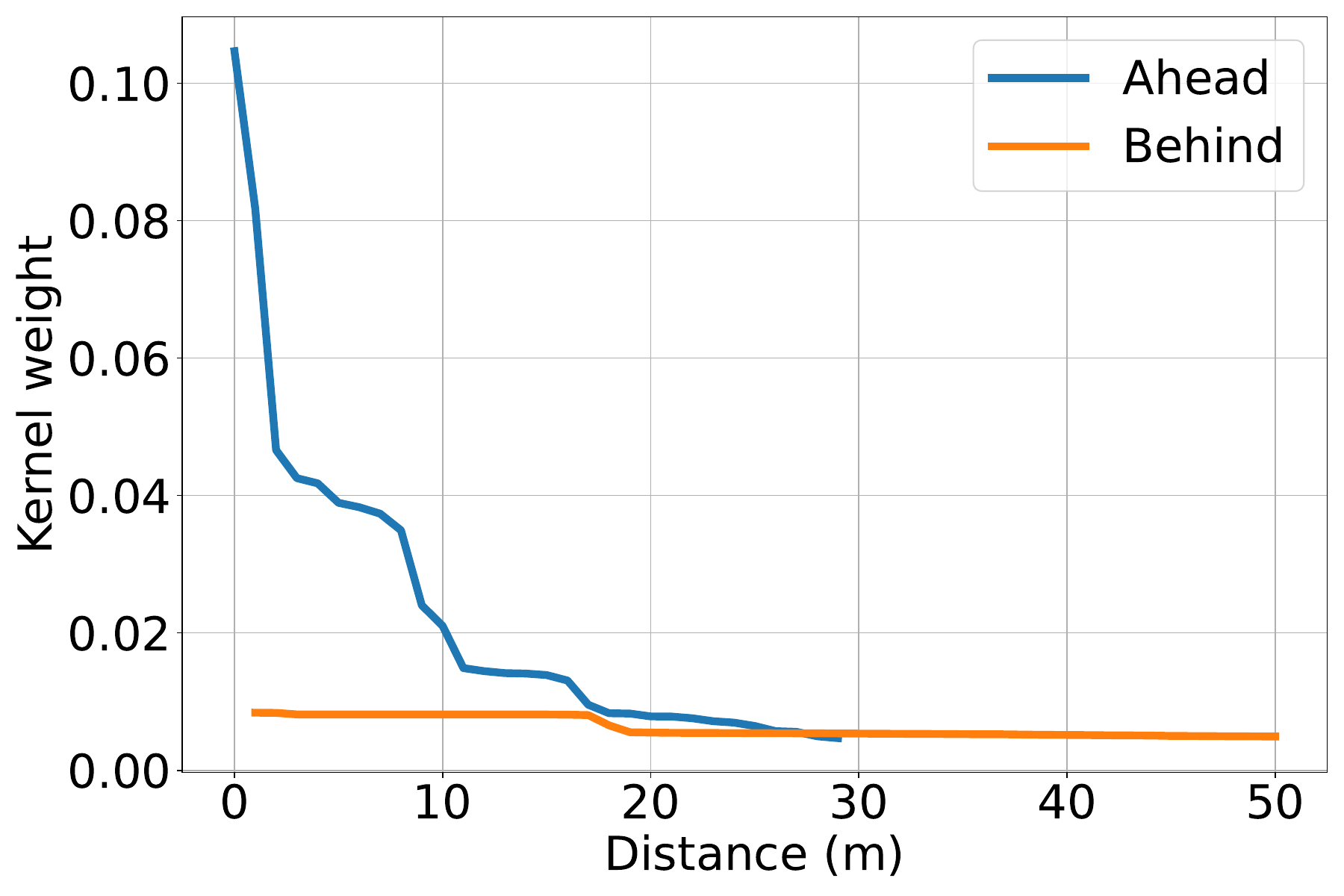}}\\
    \subcaptionbox{Fundamental diagram, $\eta_b=20$ m}{\includegraphics[width=0.3\linewidth]{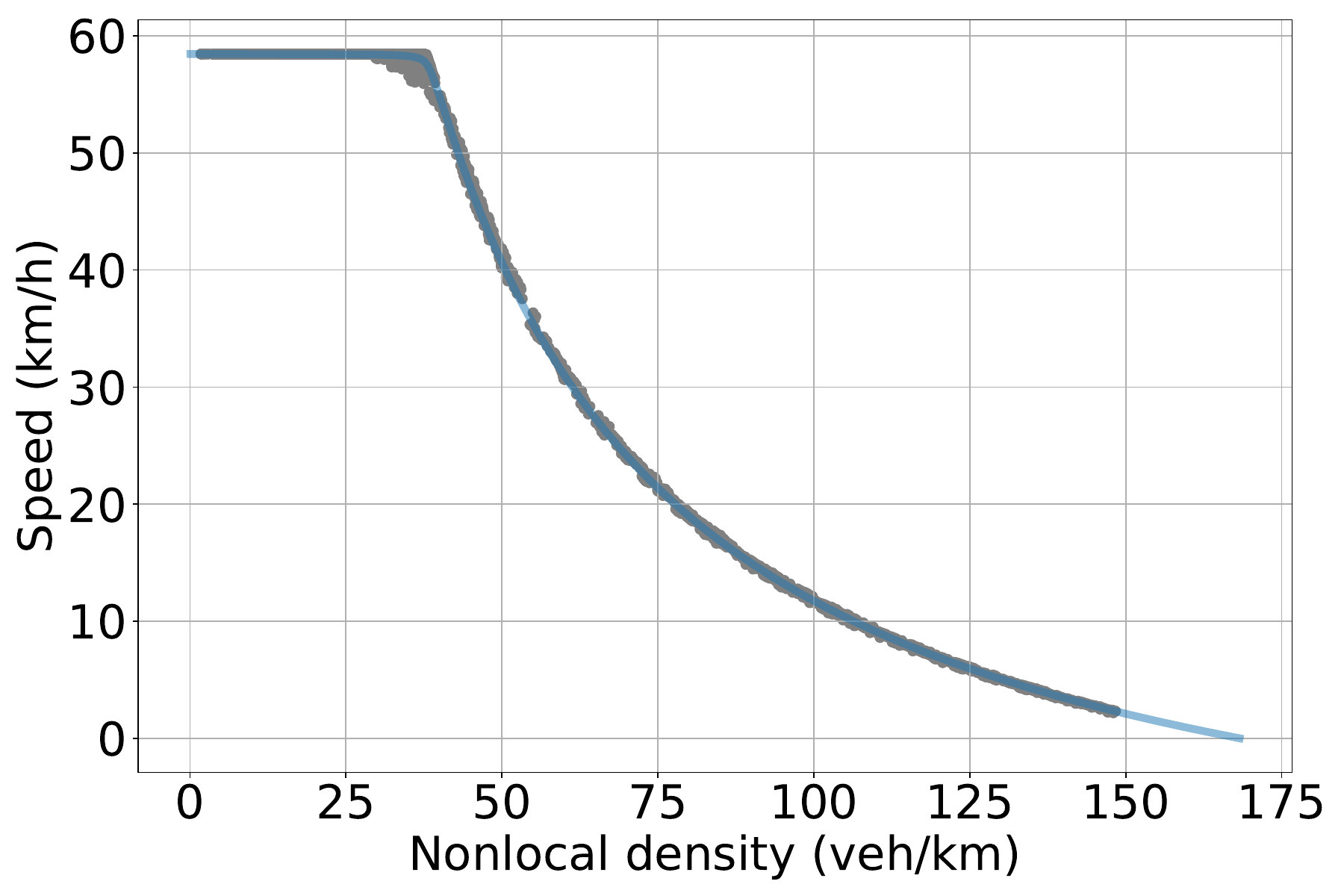}}
    \subcaptionbox{Fundamental diagram, $\eta_b=40$ m}{\includegraphics[width=0.3\linewidth]{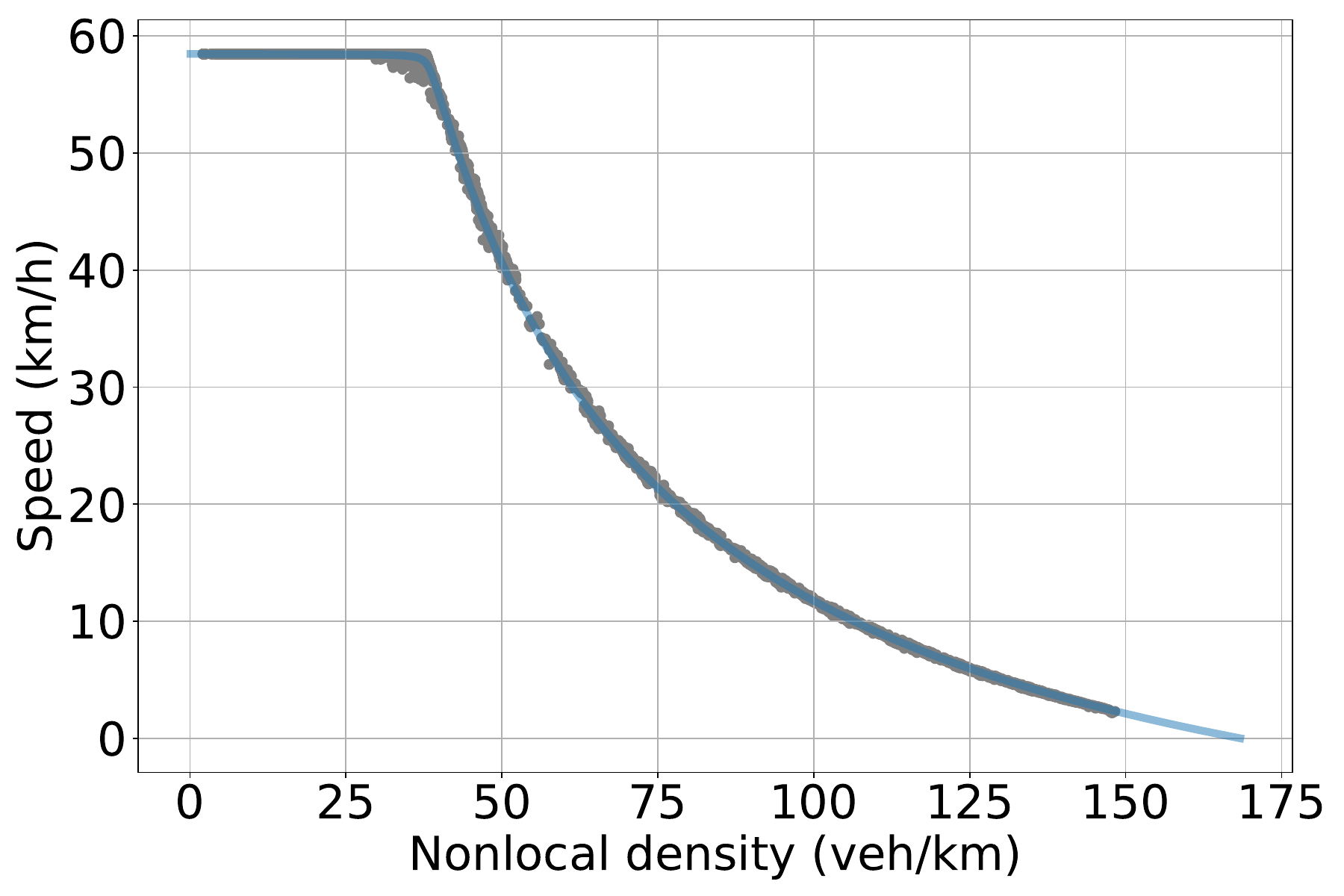}}
    \subcaptionbox{Fundamental diagram, $\eta_b=50$ m}{\includegraphics[width=0.3\linewidth]{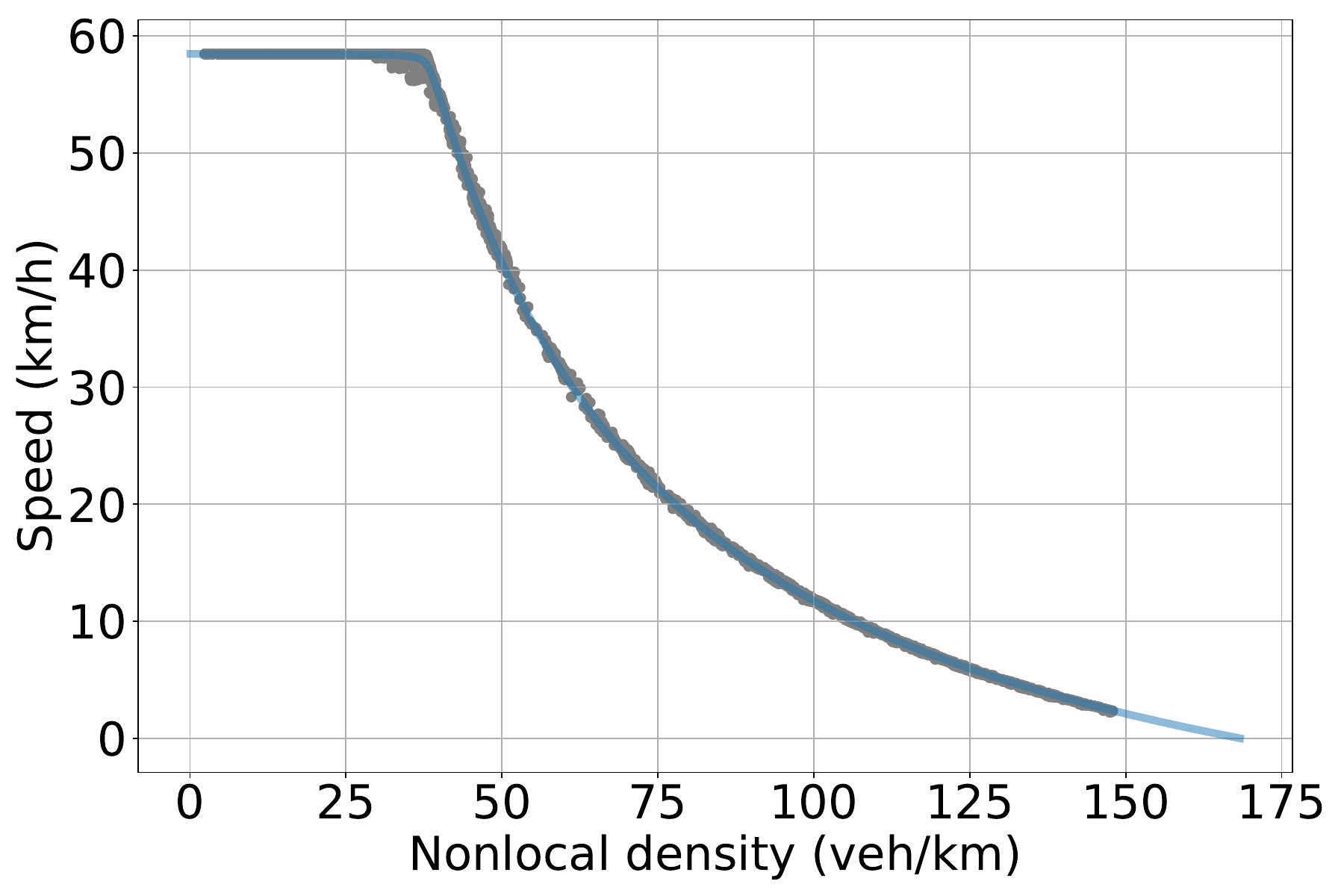}}
    \caption{Simulation results with different looking-behind kernel length $\eta_b$. We fix the looking-ahead kernel length as $\eta=30$ m.}
    \label{fig:kernel length nudging}
\end{figure}

\begin{figure}[!t]
    \centering
    \subcaptionbox{Car-following~\eqref{eq:micro CF}}{\includegraphics[width=0.3\linewidth]{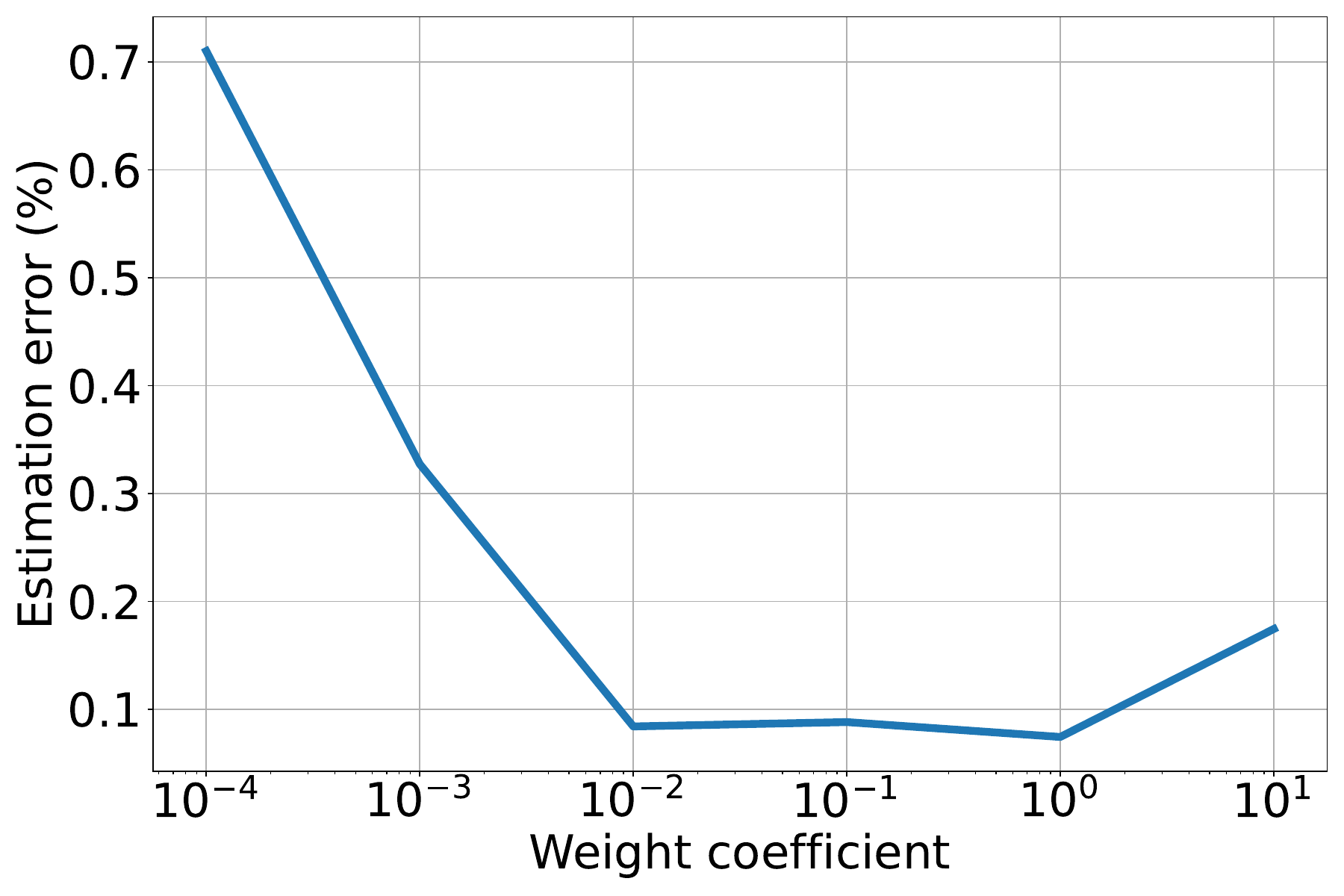}}
    \subcaptionbox{Look-ahead~\eqref{eq:micro look ahead}}{\includegraphics[width=0.3\linewidth]{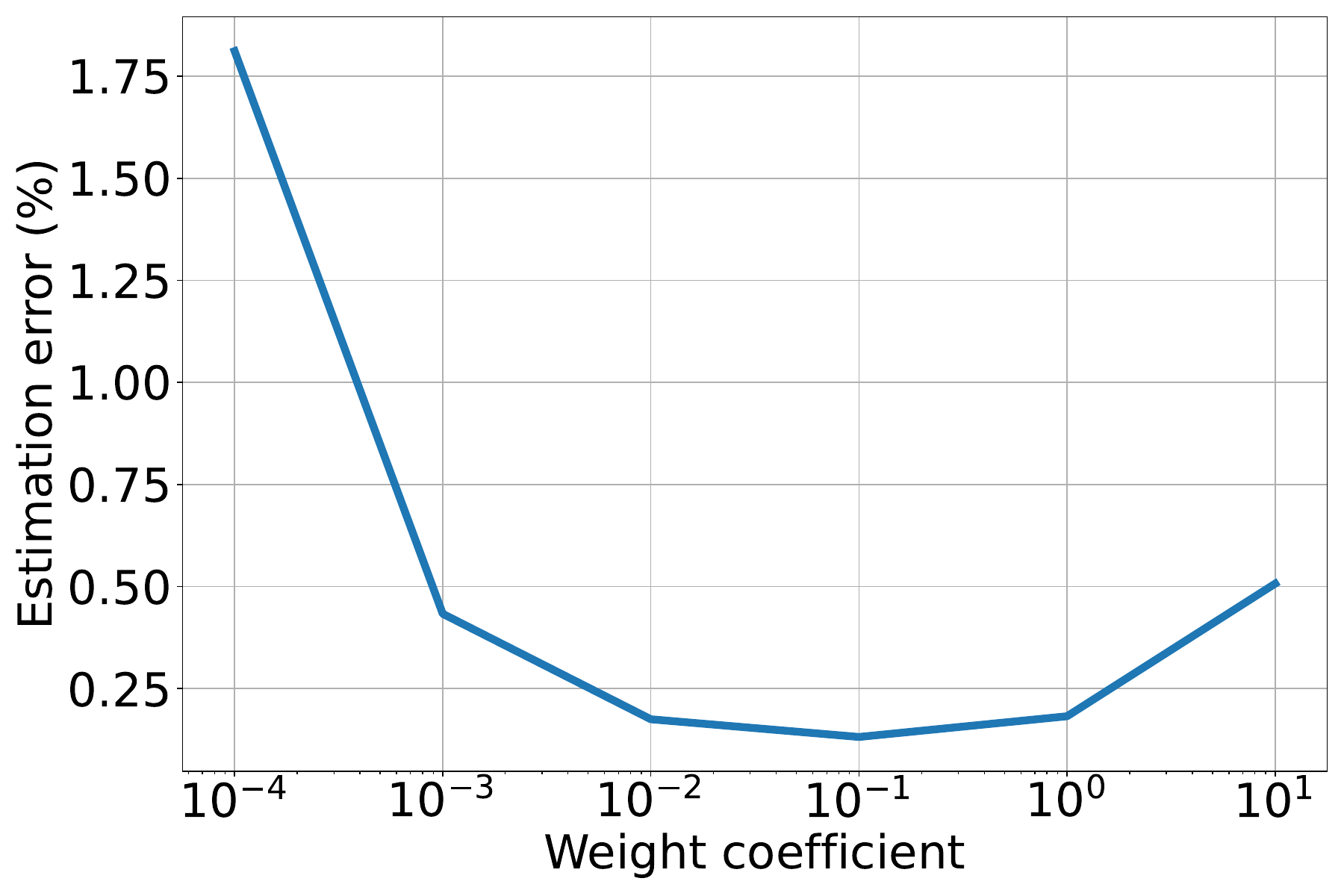}}
    \subcaptionbox{Look-behind~\eqref{eq:micro nudging pos}}{\includegraphics[width=0.3\linewidth]{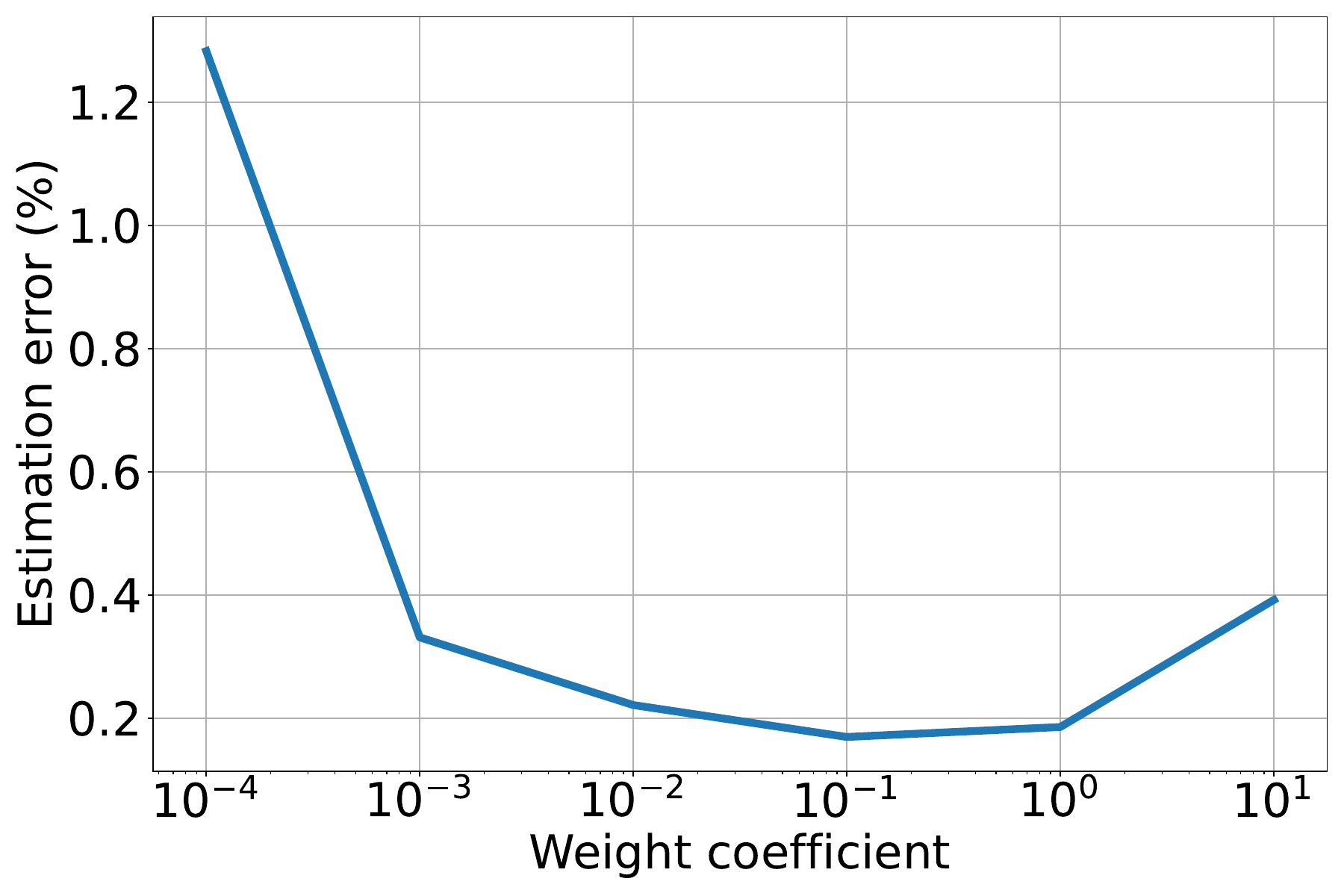}}
    \caption{Estimation error with varying weight coefficients $c_d$ of the data loss~\eqref{eq:loss data}.}
    \label{fig:data loss weight}
\end{figure}

\subsection{Mixed traffic analysis}\label{subsec:mixed}

In this part, we analyze mixed traffic scenarios. We consider two types of vehicles, one being HVs and another being CAVs. We assume that the CAVs are evenly distributed within the vehicle chain. For the HVs, we use the car-following model~\eqref{eq:micro CF} with parameters as calibrated in Table~\ref{tab:micro parameter}. 

We first consider the case where the CAV is controlled by ACC. To present a clear analysis, we assume that the CAV takes the same controller gain as HVs, but takes a different desired speed function $V_{\mathrm{opt}}$. We run simulations where the CAV takes a free-traffic gap $s_{\mathrm{go}} = 30$ m. Fig.~\ref{fig:mix sgo} gives the simulation results under different penetration rates. In Fig.~\ref{fig:mix sgo}(a), we give the estimation error got by local LWR model and looking-ahead non-local LWR model. The results show that the proposed approach remains an accurate description of traffic flow dynamics. In Fig.~\ref{fig:mix sgo}(b), we compare the looking-ahead kernel weights under different penetration rates. Similar to the pure CAV traffic case, the penetration rate has only a small effect on the kernel. For different penetration rates, the kernels share a same trend. A majority of the kernel weight is within 5 meters. Fig.~\ref{fig:mix sgo}(c) plots the non-local fundamental diagram with different penetration rates.

Now we consider the case where CAVs use the same desired speed function, but have different controller gains. We run simulations with the looking-ahead gap gain $\alpha_{-1} = 0.7$ and give the results in Fig.\ref{fig:mix alphal}.  From the estimation error results in Fig.\ref{fig:mix alphal}(a), we see that the proposed modeling approach with non-local LWR physics regularization remains an estimation error lower than 0.5\% for a wide range of penetration rate. For the non-local kernel shown in Fig.~\ref{fig:mix alphal}(b), we find that with more looking-ahead CAVs, more weights are allocated to downstream traffic. For example, for the weight value at $x=0$, full HV traffic (grey line) has the highest value, and full CAV traffic (green line)  corresponds to the lowest value. But for the downstream traffic after 10 meters, the weights for HV traffic and  mixed traffic with $p=25\%$ quickly decrease; while for the $p=50\%$ and $p=100\%$ case, the weights remains approximately the same between 10 meters and 20 meters. Fig.~\ref{fig:mix alphal}(c) compares the non-local fundamental diagram for different penetration rates. As the results show, deploying CAVs that use the same desired speed function but different controller gains has little effect on the non-local fundamental diagram. This finding is consistent with the pure CAV traffic results.

\subsection{Sensitivity analysis}\label{subsec:sensitivity}

In this part, we conduct more analysis on how the  hyper-parameters affect the modeling results. We first analyze how the non-local kernel length affects the results. We run simulations with the kernel length $\eta_a$ varying from 2 meters to 100 meters. Fig.~\ref{fig:kernel length acc} gives the simulation results corresponding to the car-following model~\eqref{eq:micro CF} with varying looking-ahead kernel length $\eta_a$ values. From the estimation error in Fig.~\ref{fig:kernel length acc}(a), for a wide range of kernel length choices from 5 meters to 60 meters, the estimation constantly  remains accurate. We further compare the learned kernel weights corresponding to different kernel length values in Fig.~\ref{fig:kernel length acc}(b), Fig.~\ref{fig:kernel length acc}(c), and Fig.~\ref{fig:kernel length acc}(d). We see for different settings of kernel length, the learned kernels share the same trend: a majority of the looking-ahead effect happens with in 5 meters. In the third row of Fig.~\ref{fig:kernel length acc}, we plot the learned non-local fundamental diagram for different kernel length values. As the results show, for all the three looking-ahead values, the non-local fundamental diagrams present only a small variation between speed and non-local density.

In Fig.~\ref{fig:kernel length nudging}, we analyze how the looking-behind kernel length value affects the modeling result. We fix the looking-ahead kernel length as $\eta_a = 30$ m and run simulations with the looking-behind kernel length  $\eta_b$ varying between 2 meters to 100 meters. As the estimation error in Fig.~\ref{fig:kernel length nudging}(a) shows, for a wide choice of the $\eta_b$ value between 20 meters to 60 meters, the non-local flow model accurately describe traffic dynamics, with the estimation error lower than 0.2\%.  From the learned kernel weights in the second row of Fig.~\ref{fig:kernel length nudging}, we see that there is a uniform trend of the looking-behind kernel with different kernel length.  The looking-behind kernel weights are relatively small, compared with the looking-ahead kernel weight values. For example, even at the $x=1$ meter location, the looking-behind weight is lower than 0.03, whereas the looking-ahead weight is around  0.06 to 0.1.   
Fig.~\ref{fig:kernel length nudging}(e), Fig.~\ref{fig:kernel length nudging}(f), and Fig.~\ref{fig:kernel length nudging}(g) give the speed vs non-local density scatter plot. As the figures show, the non-local fundamental diagram provides an accurate description of macroscopic traffic when vehicles have looking-behind behaviors.

Now we analyze how the  weight coefficient $c_d$ for the data loss in~\eqref{eq:loss data} affects the modeling result. We run simulations with the weight coefficient varying from $10^{-4}$ to $10$. In Fig.~\ref{fig:data loss weight}, we give the estimation error under different coefficient values. We consider all the three cruising behaviors: normal car-following as in~\eqref{eq:micro CF}, car-following with looking-ahead behaviors as in~\eqref{eq:micro look ahead}, and car-following behaviors with looking-behind behaviors as in~\eqref{eq:micro nudging pos}. From the estimation error in Fig.~\ref{fig:data loss weight}, we see that for all the three cases, the estimation error gets its lowest value between $c_d = 10^-2$ and $c_d = 1$. This implies that for different microscopic behaviors, there exists a uniformly wide range of $c_d$ that gives an accurate description of flow dynamics, which avoids the necessity to fine-tune this training parameter when the microscopic cruising model changes.

\section{Conclusion}

{
In this paper, we design a hybrid modeling approach to directly learn  non-local flow models from looking-ahead looking-behind CAV motion models. We analyze how the CAV control designs, including the looking-ahead/looking-behind feedback gains and the desired-speed function,  affect macroscopic flow dynamics. We first design and conduct simulator experiments to analyze looking-ahead and looking-behind behaviors in human drivers' cruising strategies, which are then used as a baseline for analyzing CAV controller design. 
Results show that the looking-ahead and looking-behind gains mainly affect the non-local kernel weights, while the optimal speed function mainly affects the non-local fundamental diagram. 
 While this paper focuses on the longitudinal dynamics, the proposed modeling method can be extended for lateral dynamics, provided that a suitable macroscopic model with lane-change is available. Another future direction is to deploy CAV controllers for vehicle test and collect trajectory data of the mixed traffic. 

}

\bibliographystyle{plain}
\bibliography{ref.bib}

\end{document}